\documentclass[twoside,12pt]{article}
\usepackage{epsfig}
\usepackage{pst-all}
\usepackage{capt-of}
\usepackage{multicol}
\usepackage{float}
\usepackage{subcaption}

\def\Journal#1#2#3#4{{#1} {#2} (#4) #3 }

\def\NPA{{\em Nucl. Phys.} A}

\def\PLB{{\em Phys. Lett.} B}

\def\PRL{\em Phys. Rev. Lett.}

\def\PREP{\em Phys. Rep.}

\def\PRD{{\em Phys. Rev.} D}
\def\PRC{{\em Phys. Rev.} C}

\newcommand{\be}{\begin{equation}}
\newcommand{\ee}{\end{equation}}
\newcommand{\bea}{\begin{eqnarray}}
\newcommand{\eea}{\end{eqnarray}}

\begin{document}
\title{\vspace{1cm} The Neutron-Rich Edge of the Nuclear Landscape.\\ Experiment and Theory}
\author{Fr\'ed\'eric Nowacki,$^{1,2}$ Alexandre  Obertelli,$^3$ and Alfredo  Poves$^4$ \\
\\
$^1$Universit\'e de Strasbourg, IPHC, 23 rue du Loess, 67037 Strasbourg, France\\
$^2$CNRS, UMR7178, 67037 Strasbourg, France\\
$^3$Institut f\"ur Kernphysik,
Technische Universit\"at Darmstadt, Germany\\
$^4$Departamento de Fisica Teorica and IFT UAM-CSIC\\
Universidad Autonoma de Madrid, Spain}
\maketitle

 \section{Introduction and rationale} 
  The foundations of the microscopic description of the nuclear dynamics lay in the nuclear shell
  model (or independent particle model (IPM)) proposed by Mayer and Jensen in 1949 \cite{mayer1949,jensen1949}, to
  explain the experimental evidence of the existence of nuclear magic numbers, somehow reminiscent of the atomic numbers of the
  noble gases. In the IPM, magic nuclei play a pivotal role, even if the applicability of the pure IPM is limited to these and to their close neighbours
  with one proton or neutron plus or minus. Even more, the IPM fails badly in its predictions for the excitation spectra of doubly magic nuclei. 
  Paradoxically, the vulnerability of the nuclear magic numbers and the overwhelming dominance  of the nuclear correlations, was put under the rug by
  the expediency of going semiclassical, adopting the mean field approach and breaking  the rotational symmetry and the particle number conservation
  to incorporate the quadrupole-quadrupole and pairing correlations respectively: the Bohr-Mottelson unified model \cite{bohr1953}. Microscopic  
  calculations in the laboratory frame were out of reach except for the  lightest nuclei, belonging to the $p$-shell. However, experimenters were
  getting intriguing results which should have produced a crisis in the IPM paradigm. And the first example took place at the neutron-rich edge
  and in a nucleus only a neutron short of magicity, precisely the theme of this review.  Wilkinson and Alburger \cite{wilkinson1959} found that the ground
  state of $^{11}$Be had total angular momentum and parity $J^{\pi}=\frac{1}{2}^+$ instead of   the IPM prediction $\frac{1}{2}^-$. Talmi and Unna \cite{talmi1960} gave a very clever 
  (even if partly wrong) explanation of this fact in terms of what we should call nowadays monopole shell evolution. They missed the crucial role
  of the correlations in the inversion of normal and intruder configurations and by the same token the discovery of the first Island of Inversion (IoI)
  at N=8 far from stability, fifteen years before the mass measurements at CERN by C. Thibault and what would later be the Orsay group of the ISOLDE collaboration \cite{thibault1975}, in the N=20 region. Soon after,
  the experimental studies of the excitation spectrum of doubly magic $^{16}$O   \cite{carter1964,chevallier1967}  discovered the presence at very
  low excitation energy of 0$^+$ states, which were interpreted as 4p-4h (four particles four holes) and 8p-8h states \cite{brown1966,gerace1967,zuker1968, zuker1969,gerace1969}.
  Indeed to have a first excited state of 4p-4h nature in a doubly magic nucleus is seriously at odds with the predictions of the IPM.
  
\begin{figure}[H]
\begin{center}
\includegraphics[width=1.0\textwidth]{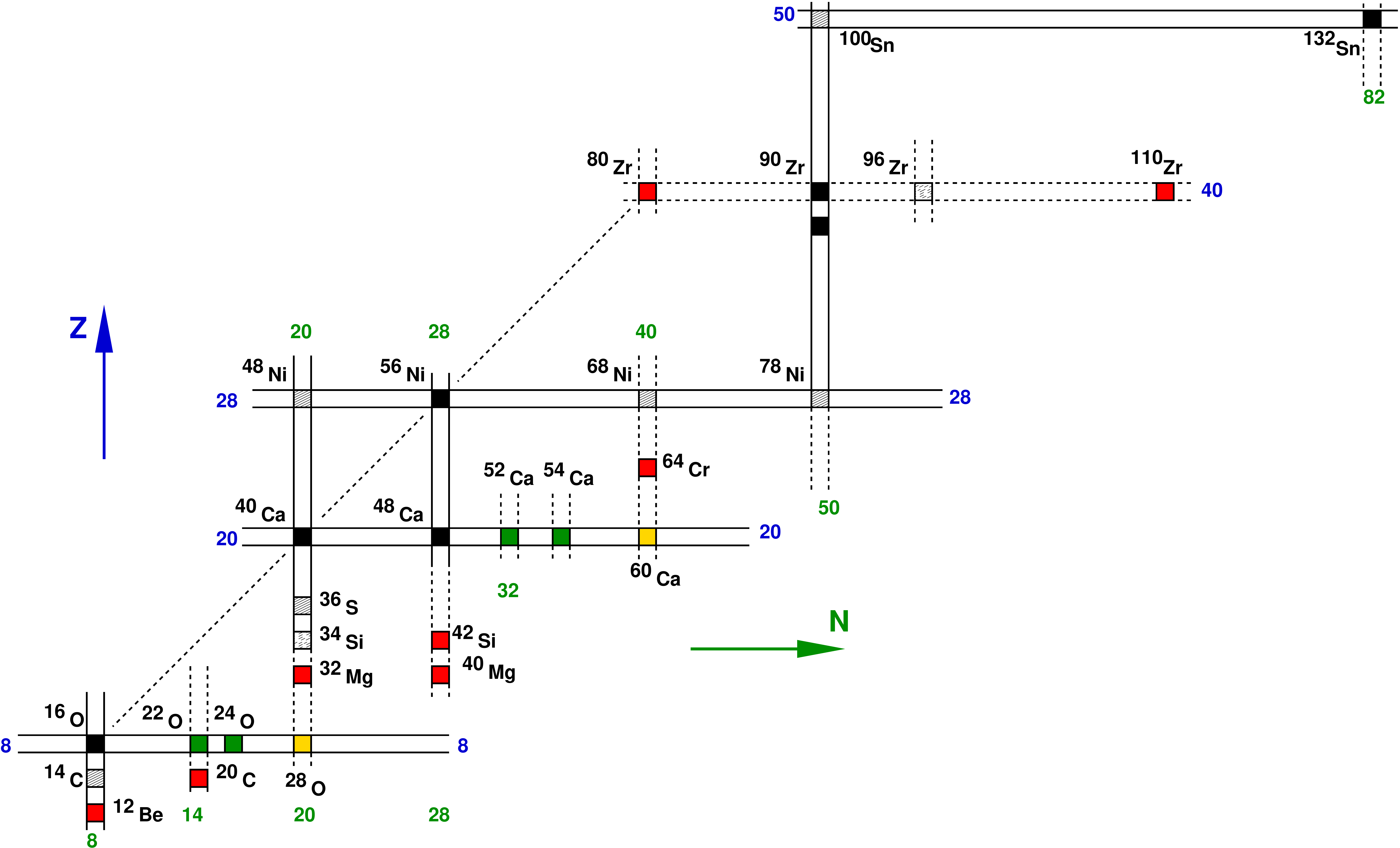} 
\caption{Landscape of the light and mid-mass part  of the Segr\'e chart. Classical doubly magic nuclei (black), new doubly magic
isotopes (grey), expected semi-magic turned deformed (red), new local doubly magic (green) and very neutron-rich nuclei whose structure is under debate (yellow) are highlighted. Most cases are discussed in the review.
\label{landscape-ini}}
\end{center}
\end{figure}

  \medskip
  \noindent
  Jumping ahead half a century, we shall review the experimental methods that make it possible to discover new manifestations of the many-body nuclear dynamics, the facilities where these experiments can be performed, and the observables which convey the relevant information. As the field has grown enormously and we cannot cover all the sectors of the Segr\'e chart, we have chosen to concentrate in the exotic very neutron-rich regions, mainly those depicted in Fig. \ref{landscape-ini}, although we shall touch upon as well the physics of the very neutron rich nuclei beyond N=82.

\noindent 
Medium-mass and heavy neutron-rich nuclei play as well an important role in the synthesis of elements in the cosmos. In particular, violent events of the universe such as supernovae explosions and merging neutron stars, where $\sim$ MeV neutrons are produced at high density, are sites for the so-called {\it r process}, which efficiently produces nuclei from a balanced equilibrium between neutron capture and $\beta$ decay \cite{burbridge1957,kajino2019}. The structure of these radioactive nuclei impact the resulting abundance of elements \cite{horowitz2019,martin2016}.
  With regret, we shall not deal with the nuclei close to the proton drip line whose flagship is the N=Z isotope $^{100}$Sn, shown to have the largest Gamow-Teller matrix element measured up to date with  log{\bf ft}=2.5 \cite{hinke2012,lubos2019}. We will not discuss the phenomenon of shape coexistence in neutron-deficient nuclei \cite{andreyev2000,bouchez2003}. We shall neither deal with the mass frontier, namely the superheavy nuclei now synthesized up to Z=118 \cite{hofmann2000,morita2010,oganessian2012}.

\medskip
  \noindent
The paper is organized as follows: Section 2 describes the
  experimental facilities and methods which make it possible to produce and measure the properties of the extreme neutron rich nuclei. In Section 3, we develop the theoretical framework that will accompany us along the review; the shell-model approach with large scale configuration interaction (mixing) SM-CI, with special emphasis in the competition between the spherical mean field and
  the nuclear correlations (mainly pairing and quadrupole-quadrupole). The symmetry properties of the latter are treated in detail as they will show to be of great heuristic value. In Section 4 we explore in detail the Islands of Inversion (IoI's) at N=20 and N=28. In Section 5 we make a side excursion into the heavier Calcium and Potassium isotopes, to discuss current issues on shell evolution and new magic numbers far from stability. In Section 6 we visit the N=40 Island of Inversion. The protagonist of Section 7 is the doubly magic nucleus $^{78}$Ni, its shape coexistence and the prospect of a new IoI at N=50 below Z=28. Section 8 leads us to examine the N=70 and N=82 neutron closures approaching the neutron drip line. Section 9 contains the conclusions and our outlook.

\medskip
\noindent
Inevitably, this review is bound to contain a very large number of acronyms, hence we have included a Glossary at the end, to help its reading. 
  
\section{The experimental view}
\subsection{Production of very neutron-rich isotopes, techniques and facilities} 
With the coming-soon operation of several new-generation large scale Radioactive-Ion Beam (RIB) facilities ~\cite{motobayashi2014}, the Facility for Antiproton and Ion Research (FAIR)~\cite{fair} in Europe, the Facility for Rare Isotope beams (FRIB)~\cite{frib} in the US and the Heavy Ion Accelerator Facility (HIAF)~\cite{hiaf} and Raon~\cite{raon} in Asia to quote the most important ones, together with the existing facilities under regular upgrades, the nuclear structure community will reach in the next years unprecedented means to address experimentally the nuclear many-body problem.

\subsubsection{The historical seeds to RI physics} 
The very successful story of RIB production and physics started in the 1950s in Copenhagen where O. Kofoed-Hansen and K. O. Nielsen extracted in a vacuum tube radioactive isotopes produced from the fission of Uranium~\cite{kofoed1951}. The fission was induced by neutrons obtained from the breakup of deuterons accelerated by the cyclotron of the Niels Bohr Institute. Early 1960s, R. Bernas, R. Klapisch and collaborators from Orsay, France, took up the idea and developed a method to produce isotopes from fragmentation reactions~\cite{bernas1965,klapisch1965} which will become the Isotopic Separation On-Line method developed and used at CERN. ISOLDE, an acronym for Isotope Separator On Line DEvice, was originally proposed at the 600 MeV Proton Synchrocyclotron in 1964. The first experiments started there in 1967. The technical aspects of on-line mass separation were discussed for the first time at the conference series on Electromagnetic Isotope Separators (EMIS) at Aarhus, Denmark in 1965. The first scientific conference on the topic was organised the following year \cite{lysekil1966}. The first ISOL studies focused on the production and study of neutron-rich alkali Li and Na isotopes from the interaction of 10.5 GeV~\cite{klapisch1968} and 24 GeV~\cite{klapisch1969,klapisch1972} proton beams with heavy targets at the CERN proton synchrotron~\cite{adams1960}. The target-like reaction residues were caught in heated graphite foils from which they could diffuse rather quickly (of the order of milliseconds). Alkali elements were produced by surface ionization and were subsequently mass analyzed. From these nearly ten years of development resulted the first striking evidence of shell evolution: the mass excess of the neutron-rich $^{31,32}$Na isotopes showed that these nuclei were more tightly bound than expected at the time~\cite{thibault1975}.
The possible mass contamination from other elements, the selectivity of the ionization process and the diffusion time necessary to extract isotopes were limiting factors, at the time of this pioneer work. These difficulties were partly lifted with the use of in-flight separators. The first used was a gas-filled magnetic separator at the graphite reactor of Oak Ridge, USA \cite{cohen1958}. Applying deflection in magnetic and electric fields to fission products was pioneered in 1964 at the research reactor in Garching, Germany \cite{ewald1964}.
ISOL beams could later on be used for reaction studies. In 1989, energetic beams of radioactive nuclei were also obtained using the ISOL method: by coupling the two cyclotrons of Louvain-la-Neuve, Belgium, through an on-line mass separator, intense beams of post-accelerated $^{13}$N were obtained. The intensity and good beam quality made it possible to measure the astrophysical important proton-capture reaction on $^{13}$N \cite{decrock}.
A Further important development is the Ion Guide Separation On Line technique (IGISOL) where the production target and its extraction electrodes are embedded in a gas volume  \cite{arje1987}. The primary reaction products are slowed down in the volume with a sufficient probability not to be neutralized. The IGISOL technique allows a fast mass separation, while the extraction is relatively independent on the volatility of the recoils. The method has been continuously improved and used at the University of Jyv{\"a}skyl{\"a}, Finland \cite{aysto2014}.

\medskip
\noindent
A competitive alternative RIB production method was developed in the late 1970s. This was made possible by the development of high-energy heavy-ion beams and it came with in-beam fragmentation in inverse-kinematics where projectile-like reaction products fly in the direction of the beam in a narrow momentum cone centered on the beam velocity~\cite{butler1977,goldhaber1978,symons1979}. The method was first applied to produce neutron-rich He isotopes by I.~Tanihata and collaborators at the Lawrence Berkeley National Laboratory, US, from the fragmentation of a 800 MeV/nucleon $^{11}$B primary beam~\cite{tanihata1985_1}. The reaction cross section of the produced He isotopes onto light Be, C and Al secondary targets was then measured. These first in-beam measurements with RIBs, together with the following measurements for Li isotopes by the same team~\cite{tanihata1985_2}, showed a surprising enhancement of the reaction cross section for $^{6,8}$He and $^{11}$Li along their respective isotopic chains, demonstrating a large matter radius of these dripline nuclei. This observation was soon after interpreted as the development of a hereafter-called neutron halo\footnote{The term \emph{neutron halo} was first mentioned in the literature in 1973 to qualify the excess of neutrons at the surface of heavy stable nuclei concluded from the measurement of antiproton annihilation after capture at low energy~\cite{bugg1973}.} in these loosely bound neutron-rich systems~\cite{hansen1987}.

\medskip
\noindent
The experimental discoveries of shell evolution and neutron halos, in 1975 and 1985, respectively, were two historical seeds for the development of RIB physics as we know it today. They triggered numerous studies and allowed one to reveal new nuclear phenomena, while they were made possible by constant efforts and improvement over the past decades in the production of radioactive isotopes. Several large-scale facilities dedicated to radioactive isotopes produced from heavy-ion beams were developed in the early 80s: the first research experiments at the National Superconducting Cyclotron Laboratory (NSCL), US, took place in 1982 with beams accelerated by the K500 cyclotron. Soon after, higher energies were reached with the K1200 cyclotron. A major upgrade of the facility was the coupling of the two cyclotrons, which started operations in 2001. In parallel, the Grand Acc\'el\'erateur National d'Ions Lourds (GANIL), France, was developed and the first physics experiment was carried in 1983 \cite{detraz1989}. A major upgrade of GANIL took place in 2001 with the completion of the SPIRAL facility which was based on the re-acceleration of ISOL beams by the CIME cyclotron up to 10 MeV/nucleon \cite{villari2003}. Nuclear structure experiments at relativistic energies were made possible at the Gesellschaft f\"ur Schwerionenforschung (GSI), Germany, where experiments with high-energy primary beams accelerated with the SIS18 synchrotron \cite{blasche1985,bohne1992} and RIB separated by the FRS separator \cite{geissel1992} have been performed since 1990. These three facilities, together with the Radioactive Isotope Beam Factory (RIBF) facility of RIKEN since 2007 (see Sec.~\ref{ribf} for more details), have strongly contributed to the field of RI physics over the past decades. Major upgrades qualified as new-generation facilities are currently being undertaken at the sites of these three major facilities, as it will be shortly described below (section~\ref{UpcomingFacilities}). Other facilities with the capabilities to produce RIB, such as the TRI-University Meson Facility (TRIUMF)~\cite{TRIUMF}, Canada, and the Research Center for Nuclear Physics (RCNP)~\cite{RCNP}, Japan, have also contributed to explore nuclei away from stability, to a lower extend.
We propose here to review the state-of-the-art of RIB production from both ISOL and in-flight techniques. To this purpose, we focus on the two world-class facilities which are leading the field today in terms of variety and intensities of RI beams: ISOLDE at CERN and the Radioactive Ion Beam Factory (RIBF), Japan.  

\subsubsection{ISOLDE at CERN: the state-of-the-art ISOL-beam production}
\begin{figure}[!]
\centering
\includegraphics[trim=0cm 0cm 0cm 0cm,clip,width=14cm]{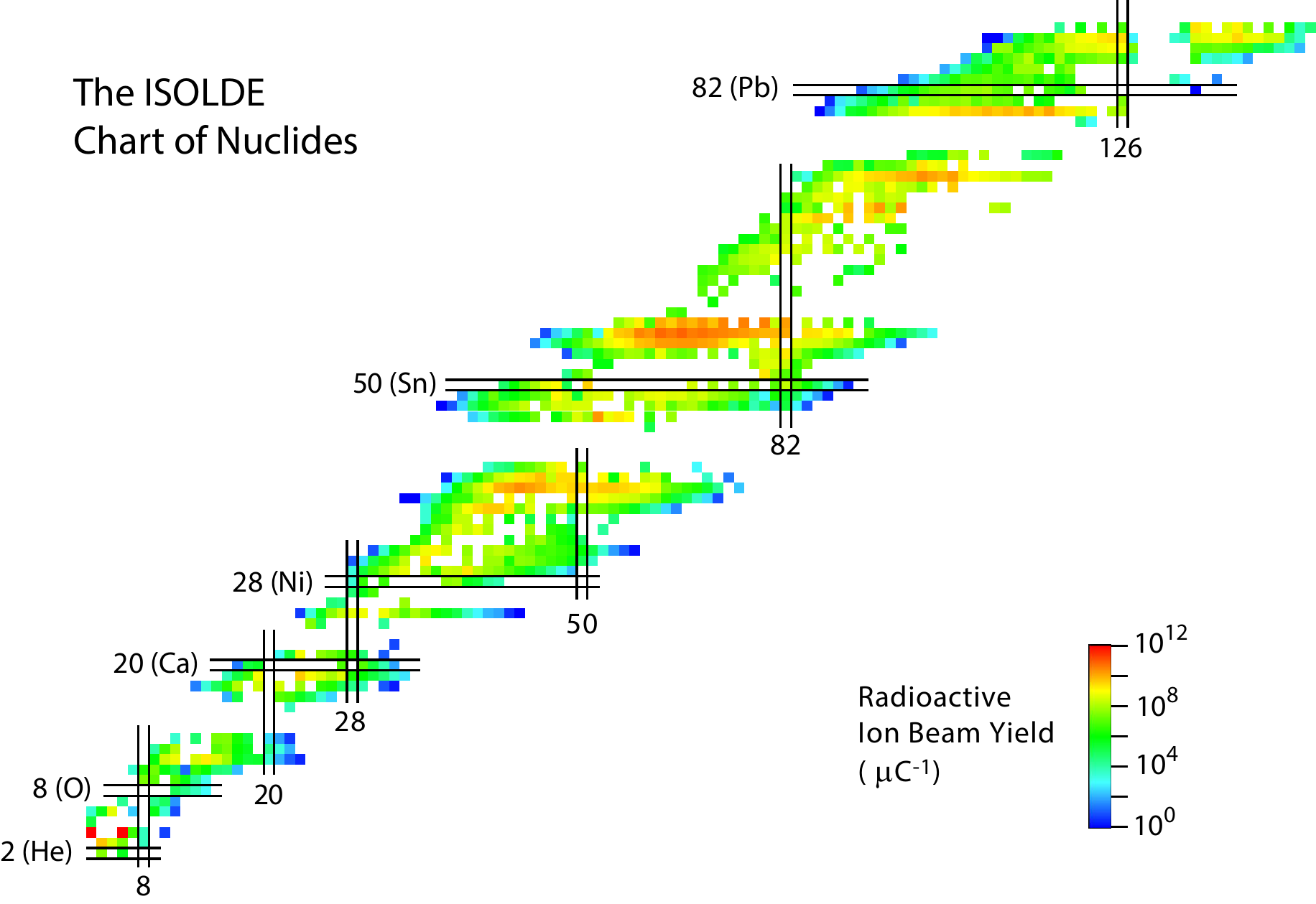}
\caption{RIB intensities produced at ISOLDE with a 1 $\mu$C proton current on production target. From the ISOLDE data base \cite{ISOLDEDB,isoyielddb2020}. Courtesy J. Ballof, CERN.}
\label{figure_ISOLDEbeams}
\end{figure}

Nuclear physicists from CERN were the pioneers of RI production and RI physics. The first experiments started in 1967 at the 600 MeV Proton Synchrotron.
Since 1992, the ISOLDE production targets have been irradiated with the 1 GeV proton beam from the CERN PS Booster~\cite{kugler1992}. In 2004, the beam energy was increased to 1.4 GeV. The pulsed high-energy and high-intensity proton beams of the PS Booster are unique assets to produce RI beams. An average proton beam intensity of 2 $\mu$A impinges on the ISOLDE production target ($3.3 \times 10^{13}$ protons per pulse), resulting in high production rates as illustrated in Fig.~\ref{figure_ISOLDEbeams}. The combination of target material~\cite{ramos2016} and ion source is optimized for each isotope to produce. More than 1300 different isotopes from 75 different chemical elements can be produced at ISOLDE. The ion sources used at ISOLDE are based on different techniques: the traditional surface and plasma ionisation and, more recently, laser-induced ion source with the RILIS development~\cite{mishin1993,fedoseev2008,fedoseev2012}. Due to the high selectivity achieved and thanks to the development efforts of the CERN ion-source teams, RILIS is requested for over 50\% of ISOLDE experiments. Ionization schemes have been developed for more than 30 chemical elements at ISOLDE. They require several lasers, often a combination of dye, Ti:Sa and Nd:YAG lasers. By triggering lasers on the same pulse, and by use of delay generators, it is possible to synchronize the lasers so that the successive atomic excitations of the ionization scheme are made by specific lasers.

\medskip
\noindent
ISOLDE has two production targets followed by two distinct isotope separators: a rather traditional one allowing to dispatch the beams to three low-energy experimental areas, and the so-called high-resolution separator (HRS) leading to a mass separation up to $\frac{\Delta M}{M}\sim 7000$.
RI beams after the HRS can be re-accelerated up to 10 MeV/nucleon. REX-ISOLDE makes use of a unique combination of the cooling and bunching device REXTRAP \cite{rextrap} and the charge breeder REXEBIS \cite{rexebis} prior to RIs' acceleration. The post-acceleration complex is composed of the REX-ISOLDE (a RFQ and accelerating cavities for energies up to 3 MeV/nucleon) followed by the HIE-ISOLDE (High Intensity and Energy) composed of five additional accelerating cavities allowing a post-acceleration from 3 to 10 MeV/nucleon. These beam energies cover the necessary range to perform Coulomb excitation and nucleon-transfer studies \cite{kadi2017}.
In the coming years, the proton-beam energy of the PS Booster will be increase to 2 GeV, leading to an increase of production rates up to a factor of 10. In addition, the current is expected to be increased to a factor 3, improving in total the RI beam intensities by a factor of 30.

\subsubsection{The RIBF, the first new-generation facility}
\label{ribf}
\begin{figure}[!]
\centering
\includegraphics[trim=4cm 0cm 4cm 0cm,clip,angle=-90,width=14cm]{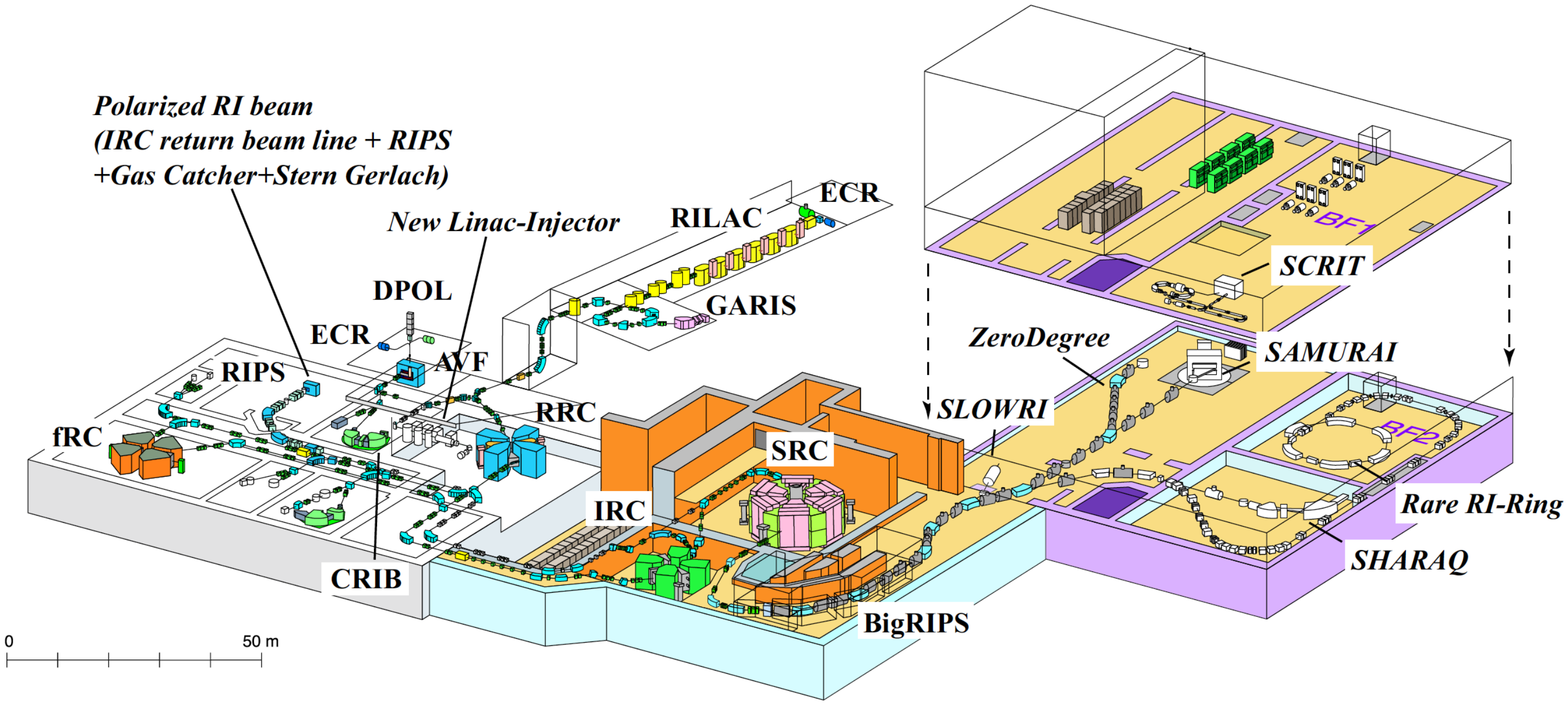}
\caption{Schematic view of the RIBF layout. The primary beam is accelerated in series, the last acceleration stage being the SRC cyclotron. Radioactive isotopes are produced from the reaction of the primary beam with a production target. The isotopes of interest are separated and identified event by event by use of the BigRIPS spectrometer. Secondary beam studies can be carried out at different experimental areas depending on the measurement: the high-resolution Zero-Degree ans SHARAQ spectrometers, the large-acceptance SAMURAI spectrometer or the RI-RING storage ring for mass measurements. The $\gamma$ spectroscopy program, mentioned in this section, was carried at both the Zero-Degree and SAMURAI spectrometers.}
\label{figure_RIBF_layout}
\end{figure}

The RIBF has been in operation since 2007. A layout of the facility is shown in Fig.~\ref{figure_RIBF_layout}. The RIBF delivers intense heavy-ion primary beams accelerated by a series of coupled cyclotrons up to energies of 345 MeV/nucleon. Its concept is based on an existing accelerator complex composed of a LINAC and several cyclotrons used as an injector to the new facility~\cite{yano2005,sakurai2008}. The last acceleration stage is composed of the SRC, one of the largest research cyclotron worldwide~\cite{okuno2007,yamada2008}. After in-flight production from fission, fragmentation, or Coulomb dissociation, reaction products are separated from the beam and analyzed by the BigRIPS fragment separator~\cite{kubo2007,fukuda2013} and delivered to experimental areas.

\begin{figure}[!]
\centering
\includegraphics[trim=0cm 0cm 0cm 0cm,angle=-90,clip,width=12cm]{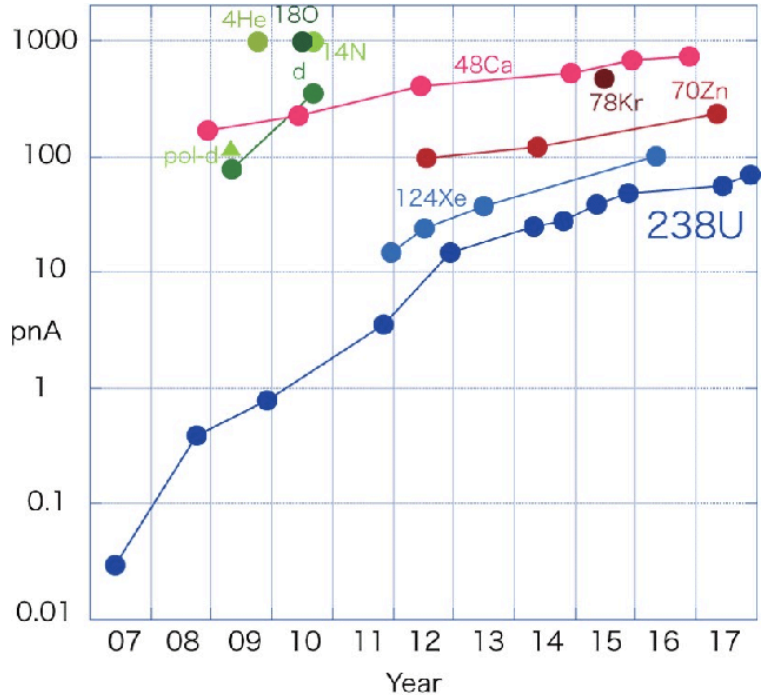}
\caption{Evolution of primary-beam intensities reached at the RIBF after the last acceleration stage (SRC) as a function of time since 2007. Figure from \cite{okuno2020}.}
\label{figure_RIBF_intensities}
\end{figure}

\begin{figure}[!]
\centering
\includegraphics[trim=0cm 6cm 0cm 6cm,clip,width=12cm]{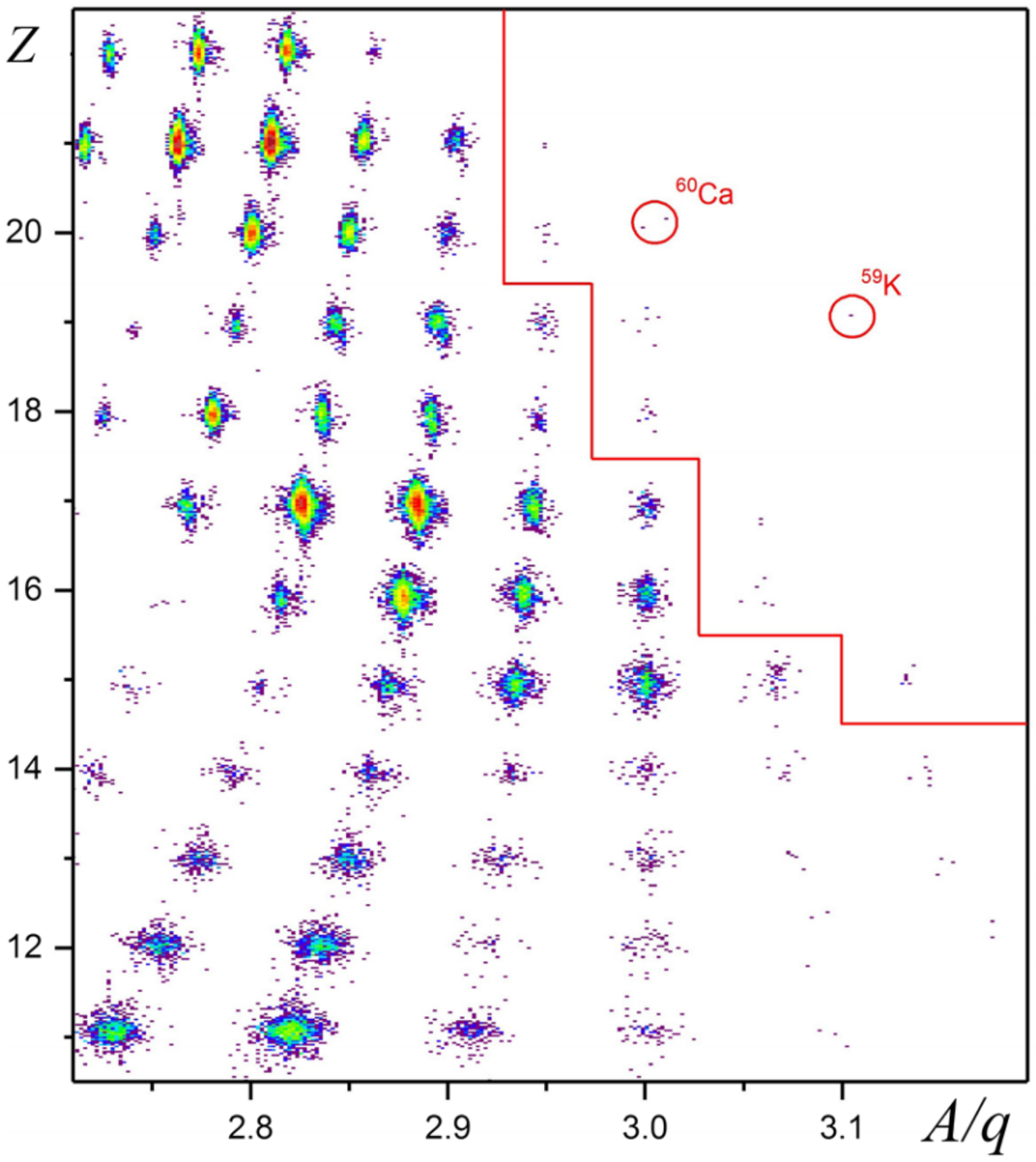}
\caption{Identification of fragments with the BigRIPS separator after the interaction of a $^{70}$Zn beam at 345 MeV/nucleon and a $^9$Be target. In this measurement, $^{60}$Ca has been identified for the first time~\cite{tarasov2018}. Reprinted figure with permission from ~\cite{tarasov2018}. Copyright 2020 by the
American Physical Society.}
\label{figure_60Ca}
\end{figure}

\medskip
\noindent
The primary beam intensities delivered at the RIBF are the highest reached so far for intermediate-energy heavy ions. The nominal beam currents of 1 p$\mu$A have been reached for light ions. Ca beams are routinely delivered at 500 pnA. U beams have reached an intensity of 72 pnA\footnote{Value given on the RIBF website~\cite{RIBFinfo} as of April 2020.}. These achievements are the product of a long-term commitment of the RIKEN accelerator teams, as illustrated in Fig.~\ref{figure_RIBF_intensities}, while the intensities are continuously improved~\cite{okuno2020}. An upgrade of the facility is foreseen to reach 1 p$\mu$A of intensity for accelerated $^{238}$U ions.
The RIBF was designed to be a discovery machine and it is naturally the place where most of the new isotopes, over the past 10 years, were discovered. Two recent highlights demonstrate the performances of the RIBF: the discovery of the very neutron-rich Ca isotope $^{60}$Ca~\cite{tarasov2018}, as illustrated in Fig.~\ref{figure_60Ca}, and the experimental determination of the neutron dripline at $^{31}$F and $^{34}$Ne~\cite{ahn2019} for the F and Ne isotopic chains, respectively.

\subsubsection{Upcoming new-generation facilities}
\label{UpcomingFacilities}
\begin{figure}[!]
\centering
\includegraphics[trim=0cm 6cm 0cm 6cm,clip,width=12cm]{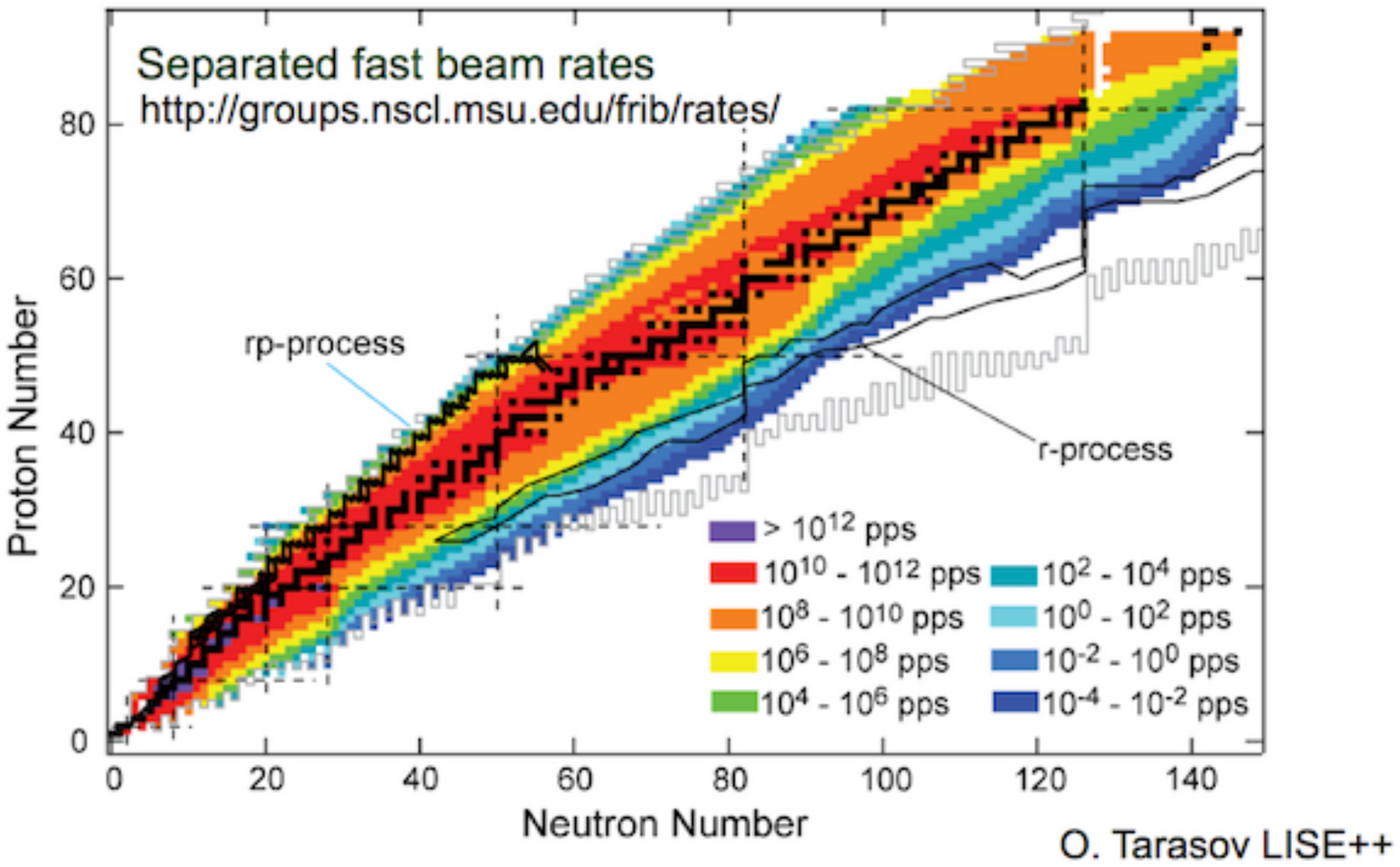}
\caption{Projected secondary beam intensities at FRIB with full power estimated with the program LISE++~\cite{lise++}. Figure from the FRIB User Organization website~\cite{FRIBref}.}
\label{FRIBrates}
\end{figure}

The future for RI physics is exciting: new facilities are emerging with the promises of the access to a broad range of nuclei and orders of magnitude of beam intensities compared to the first generation. 
FRIB will replace the NSCL as the USA leading nuclear physics facility in 2021. Built on the same site as the NSCL, the accelerator complex will be brand new. The primary beams will be accelerated with a LINAC up to 200 MeV/nucleon for $^{238}$U (higher for lighter ions) at a first stage, to be upgrade to 400 MeV/nucleon in the future. The accelerator is design for a maximum primary beam power on target of 400 kW (for energies of 200 MeV/nucleon). The foreseen FRIB secondary-beam intensities (first stage) with maximum beam power on production target are given in Fig. \ref{FRIBrates}. RI beams will be produced via fragmentation and in-flight fission. As an example, $^{78}$Ni, after fragmentation of $^{86}$Kr at 233 MeV/u, is predicted to be produced at 194 MeV/u at an intensity of 15 pps. The experimental area of FRIB and devices allow experiments at intermediate energy or low energy. The concept of the low energy part relies on the following steps: (i) intermediate-energy RI beams are stopped in a gas cell, (ii) charge bred, and (iii) possibly post-accelerated to energies up to 12 MeV/nucleon (ReA12 post-acceleration complex). In terms of beam energy domain, intensities and production method, FRIB and RIBF will have a similar profile.

\medskip
\noindent
The new international FAIR facility will be the largest facility dedicated to nuclear physics in Europe. It aims at ground-breaking discovering not only in low-energy nuclear physics, but in hadronic physics, heavy ion collisions, and atomic and plasma physics \cite{durante2019}. The facility is located at the site of the existing GSI facility. FAIR will use the GSI primary beams (the last acceleration stage of GSI is the SIS18 synchrotron, accelerating stripped ions to 18 Tm) as an injector to the new SIS100 synchrotron. RI beams will be produced from fragmentation or in-flight fission and separated by a new large-acceptance separator called Super-FRS \cite{geissel2004}. The secondary beam energy will be limited by the Super-FRS maximum rigidity of 20 Tm, typically several hundreds to 2000 MeV/nucleon. The primary beam intensities of FAIR compared to those of the previous GSI facility will be increased by a factor 1000 to 10000 \cite{FAIRbaseline}. This factor originates from an upgrade of the existing facility leading to a factor 10 more source intensities, a factor 10 from the acceleration of ions with a smaller charge state (allowed by the SIS100) and a factor 10 to 100 coming from the larger acceptance of the Super-FRS compared to the existing FRS spectrometer. The first phase of FAIR is expected to deliver its first RI beams from the SIS100 and the Super-FRS spectrometer in 2025. FAIR will be unique in terms of beam energies and the associated detection.

\medskip
\noindent
Based on its pioneering and strong expertise in ISOL technique, the low-energy community in Europe developed the concept of a large scale ISOL facility beyond today's ISOLDE capabilities~\cite{EURISOL}. This third-generation facility is out of reach in one step and several ISOL facilities, in parallel to ISOLDE, are being developed. The SPIRAL2 facility on the site of GANIL, France, was thought to be the next new-generation ISOL facility for some time with high-intensity fission fragment beams produced from the interaction of light ions or light-ion-induced neutron beams onto a uranium-carbide (UCx) target at fission rates up to 10$^{13}$ s$^{-1}$. The first stage of the SPIRAL2 facility, composed of a RFQ (A/Q=3) followed by a LINAC has been commissioned in 2019. The first phase of SPIRAL2 focuses on heavy elements and physics with neutron beams. The status of the RI part (second phase) has been suspended and its future is today under debate. In parallel, the SPES facility at LNL, Italy, is being built~\cite{SPES} and the first post-accelerated fission-fragment beams are expected in 2022. The RI production of SPES is based on the interaction of light-ions accelerated by a cyclotron (35-70 MeV) with a uranium-carbide target. Fission fragments will then be post-accelerated with the existing ALPI superconducting LINAC.  

\medskip
\noindent
The HIAF, High Intensity heavy-ion Accelerator facility, is a multifunction ring complex planned in China \cite{hiaf}. Ions will be accelerated through a LINAC and a 45 Tm accumulation and booster ring (ABR-45) followed by a multipurpose ring ensemble. After the ABR-45 booster, as an example of primary beam, U$^{76+}$ will be accelerated up to 3.4 GeV/u at intensities up to 2.5 10$^{10}$ particles / spill. Radioactive beams will be produced via in-flight method.

\medskip
\noindent
The RAON facility is a heavy-ion accelerator under construction in Korea \cite{raon} with the objective of producing radioactive beams from in-flight and ISOL methods. Primary beams for the in-flight production will be accelerated by a LINAC up to 200 MeV/nucleon with a power on target of 400 kW. The ISOL production will be driven by 70 MeV, 1 mA proton cyclotron. ISOL fragments will be re-accelerated by a LINAC. 

\subsubsection{Proposed new concepts to produce more neutron-rich RIs}
The above-mentioned upgrades or new facilities based on the traditional ISOL or fragmentation of stable beams are believed to reach their limitations in terms of beam intensities: space-charge effects limit the maximum intensities of primary beams, the maximum power that can withstand stripper foils are also a limitation for in-flight techniques. Unfortunately, realizing the r-process in the laboratory is so far not possible until a revolutionary concept to produce RI beams is proposed. Nevertheless, new concepts based on two-step reactions, such as fragmentation of neutron-rich beams produced from fission, have been suggested to give access to unexplored regions of the nuclear landscape and are on the way to be implemented.

\begin{figure}[!]
\centering
\includegraphics[trim=0cm 9cm 0cm 6cm,clip,width=16cm]{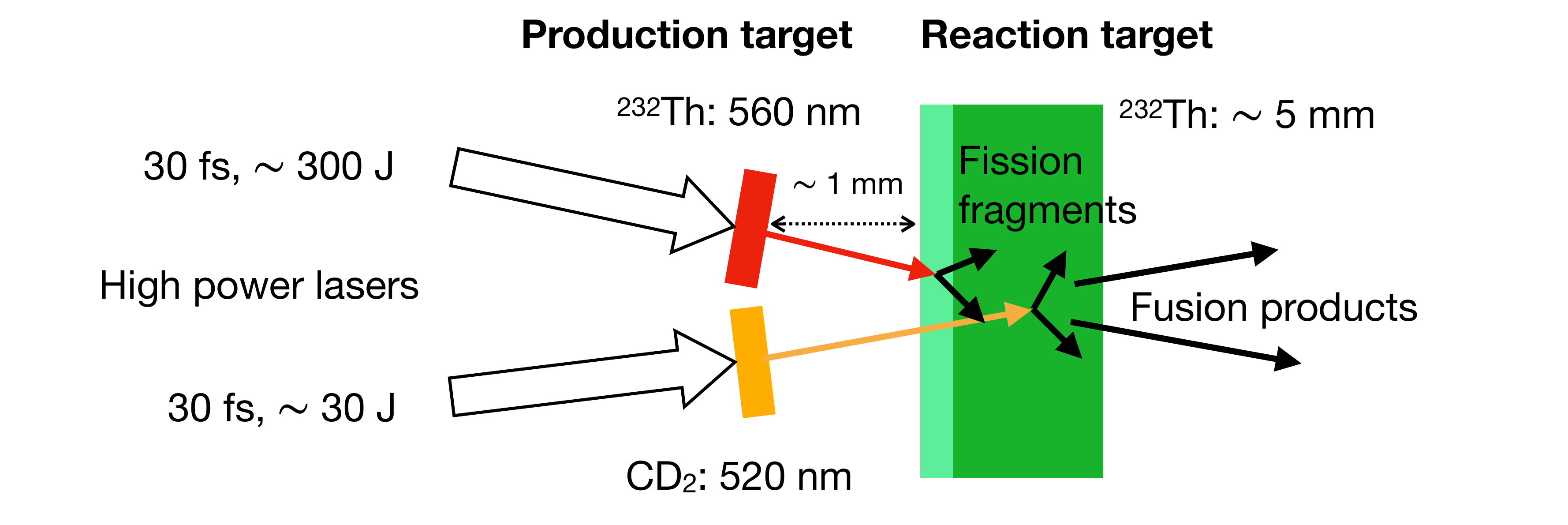}
\caption{Sketch of the RI production scheme based on the fission-fusion reaction process after laser ion acceleration. The first layer of the secondary Thorium target is used to decelerate the fission fragments from about 7 MeV/nucleon to about 3 MeV/nucleon, suitable for efficient fusion of in-flight fission fragments with target-like fission fragments. Modified from~\cite{habs2011}.}
\label{figure_LaserProduction}
\end{figure}

\medskip
\noindent
The fast development of laser-induced acceleration led to a new concept to produce heavy neutron-rich isotopes from the fusion of fission fragments. The proposed technique implies the laser-acceleration of Th stable isotopes and the laser acceleration of light ions, for example carbon and deuteron from a $CD_2$ target. The laser acceleration is foreseen to be performed in the Radiative Pressure Acceleration (RPA) regime~\cite{esirkepov2004,macchi2005,henig2009,kars2012}. The production targets are followed by a secondary Th target. Accelerated Th ions can fission by Coulomb excitation from the secondary target Th nuclei, while the accelerated light ions might induce fission of Th isotopes in the secondary target. There is a probability that target-like and projectile like fission fragments fuse together to produce so-far-unreachable r-process nuclei. The production scheme as proposed by D. Habs, P. G. Thirolf and collaborators is illustrated in Fig.~\ref{figure_LaserProduction}. A first milestone has been reached with the recent laser acceleration of Au ions with an energy above 5 MeV/nucleon~\cite{lindner2019}.
As mentioned above, projects of future facilities, EURISOL \cite{EURISOL} and Beijing ISOL (BISOL) \cite{BeijingISOL}, are based on the secondary fragmentation of intense post-accelerated  fission-induced neutron-rich ISOL beams. In the case of BISOL, rates of $10^{15}$ fissions per second are expected from the interaction of a $8\times10^{14}$ cm$^{-2}$s$^{-1}$ neutron flux onto a Uranium target. Taking into account transmission from the target, charge breeding and post-acceleration efficiency, high intensities of post-accelerated very neutron-rich nuclei are expected: $7\times10^{10}$ s$^{-1}$ of $^{132}$Sn, for example.

\subsection{Experimental techniques and methods}
The description of atomic nuclei in terms of energy shells is at the origin of our representation of nuclear structure, built upon few key observables: masses, charge and matter radii, $\beta$-decay half-lives, electric and magnetic moments, spectroscopy and exclusive cross sections from so-called direct reaction cross sections. Progress in understanding the properties of rare isotopes was also made possible by news ideas in experimental methods and improvements in the sensitivity of measurement tools to measure theses properties in inverse kinematics with low-intensity secondary beams. 

\subsubsection{Ground-state properties: masses, charge radius and moments} 
The mass of a nucleus reflects the sum of all interactions at play among its nucleons. Following the Einstein's relation $E=mc^2$ in the center of mass frame, the mass $m(Z,N)$ of a nucleus is 
\begin{equation}
m(N,Z) = N \, m_n + Z \, m_p - B(N,Z),
\end{equation}
where $B(N,Z)$, the total binding energy of the nucleus, can be split in a mean field part (the largest one, including the Coulomb repulsion among the protons) plus contributions coming from
the irreducible two body correlations dominated by the pairing and the quadrupole-quadrupole interactions. 
Different mass measurements techniques have been developed in the past century with a major improvement in precision from traps and storage rings over the past decades \cite{blaum2006,dilling2018}. The current capabilities in terms of sensitivity is beautifully illustrated by the ISOLTRAP measurement for $^{52,53,54}$Ca \cite{wienholtz2013} and storage-ring measurements at the Institute of Modern Physics (IMP) of Lanzhou, China, for $^{52-54}$Sc \cite{xu2019}, that we shall discuss in section 5. 

\medskip
\noindent
The rather-recent phase-imaging ion-cyclotron-resonance (PI-ICR) technique allows to increase the sensitivity of mass measurements. It is based on the determination of the cyclotron frequency via the projection of the ion motion in the trap onto a high-resolution position-sensitive micro-channel plate detector~\cite{eliseev2013}. The Ramsey method can also be applied to the excitation of the cyclotron motion of short-lived ions to improve the precision of the mass measurement \cite{george2007}.

\begin{figure}[!]
\centering
\includegraphics[trim=0cm 4cm 0cm 5cm,clip,width=16cm]{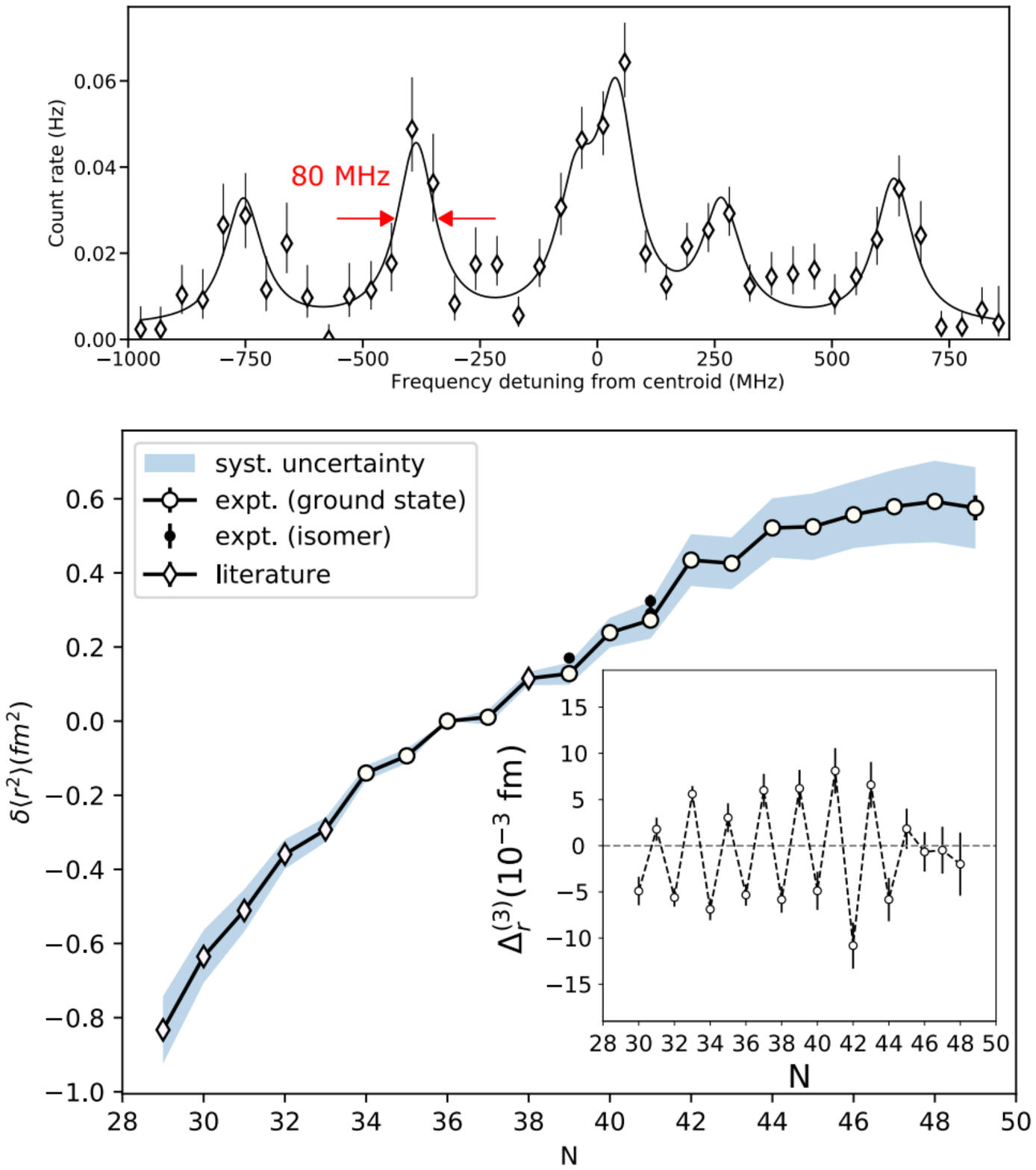}
\caption{(Top) Hyperfine spectrum of the ground state of $^{78}$Cu obtained at CRIS. (Bottom) Charge radii difference relative to $^{65}$Cu. The inset shows the odd-even staggering of charge radii given by $\Delta^{(3)}_r = \frac{1}{2}(r_{A+1} -2 r_A + r_{A-1})$. Figure from~\cite{groote2020}.}
\label{figure_CRIS}
\end{figure}

\medskip
\noindent
The fine and hyperfine structure of atoms gives access with high precision to the relative charge radius and quadrupole and magnetic moments, respectively, of the measured isotope.  In particular, the colinear resonance ionization spectroscopy (CRIS) method allows to perform background free measurement and therefore to access the above information for nuclei produced at very low rates. In practice, neutralized atoms interact with two pulsed laser beams. The first of these laser systems is tuned to the optical transition to
resonantly excite the atoms, while the second laser excites
them to an auto-ionizing state. The required coincidence of the laser pulse and the detection of ions creates the quasi-background-free measurement. The achieved precisions are illustrated in Fig. ~\ref{figure_CRIS}.
Nuclear moments provide unique information on the nucleonic distribution inside the nucleus: magnetic moments give information on the angular momentum and spin distribution in the nucleus, related to the single-particle distribution of valence nucleons, while electric quadrupole moments are the key observable for nuclear deformation. Although the several experimental techniques to measure these moments are beyond the scope of this review, we emphasize here the potential of the two-step projectile fragmentation to produce RI with a significant spin alignment (measured at 30(5)\% in the case of $^{75}$Cu produced from an intermediate $^{76}$Zn) for very neutron-rich nuclei as described in~\cite{ichikawa2012,ichikawa2019}. This production method combined with the time-differential perturbed angular distribution (TDPAD) technique makes it possible to determine the magnetic moment of an isomer from the angular distribution of its gamma decay for rather exotic species. The presence of low-energy isomers together with the value of their magnetic moment can sign shape coexistence in some regions of the nuclear landscape, as we will discuss later.

\subsubsection{Spectroscopy} 
\begin{figure}[!]
\centering
\includegraphics[trim=0cm 0cm 0cm 0cm,clip,width=14cm]{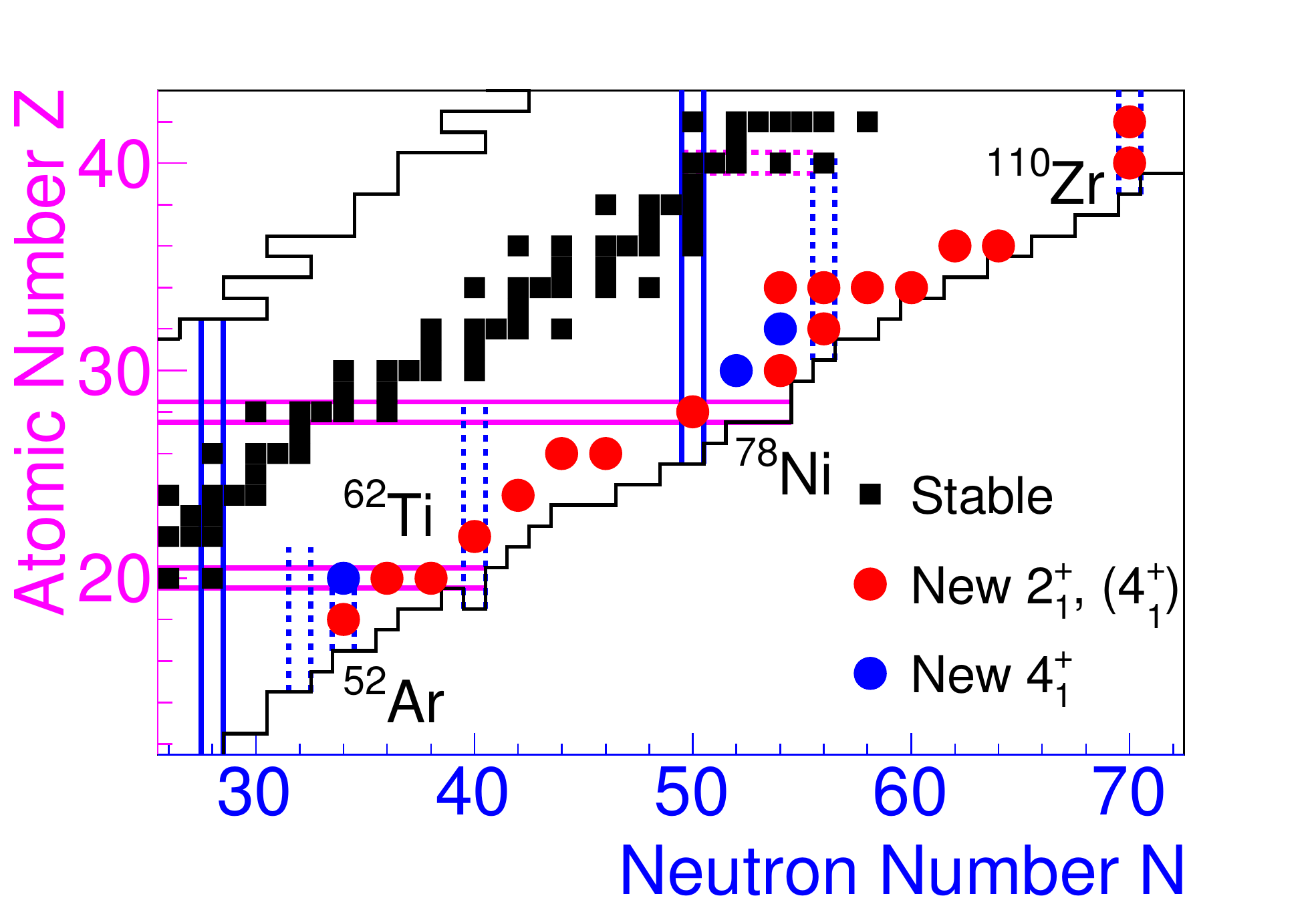}
\caption{Spectroscopy of low-lying states in even-even neutron-rich nuclei performed within the experimental program with the MINOS target-tracker system and the DALI2 array at the RIBF over 2014-2017. Courtesy of P. Doornenbal, RIKEN Nishina Center.}
\label{figure_SEASTAR}
\end{figure}

\begin{figure}[!]
\centering
\includegraphics[trim=0cm 6cm 0cm 6cm,clip,width=12cm]{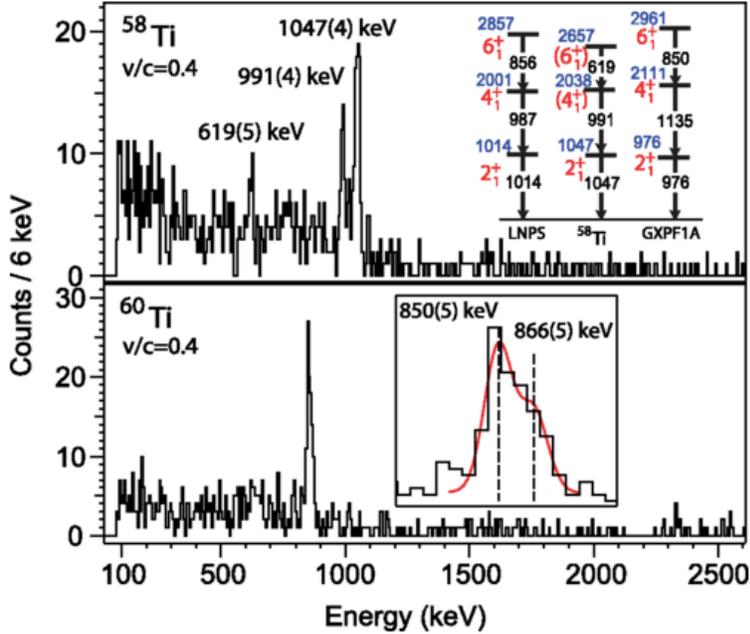}
\caption{Doppler-corrected gamma-ray spectra measured with GRETINA at the NSCL in coincidence with $^{58,60}$Ti reaction residues produced by nucleon-removal from $^{61}$V projectiles at 90 MeV/nucleon (velocity $\beta \sim 0.4$). The indication of a transition doublet in $^{60}$Ti, made possible by the high intrinsic resolution  of high-purity germanium and the tracking capabilities of GRETINA, is shown as an inset in the lower panel. Figure from \cite{gade2014}.}
\label{60TiGRETINA}
\end{figure}

The spectroscopy of neutron-rich nuclei is essential to study the nuclear shell evolution. In particular, in-beam gamma spectroscopy with fast beams allows the best luminosity and gives access to rare isotopes from beam-particles produced around one particle per second and even below in some limit cases. The technical challenge of these studies is to achieve the best resolution while maximizing the luminosity.
In addition to the beam intensity, the target thickness has often been a limiting factor to preserve the energy resolution in these measurements. Since 2014, the MINOS device developed for experiments at the RIBF allows to use thick liquid hydrogen target, up to twenty centimeters (1.3 10$^{23}$ cm$^{-2}$ ($i.e.$ close to the limit authorized by the reaction cross section beyond which all beam particles have reacted in the target) thanks to the use of a vertex tracker which allows the determination of the reaction vertex position inside the target for $(p,2p)$-like reactions where charged particles are scattered with large momentum transfer. Three successful experimental campaigns in 2014, 2015 and 2017 were dedicated to measure the first spectroscopy of eighteen new 2$^{+}$ states in neutron-rich even-even isotopes ~\cite{santamaria2015,shand2017,flavigny2017,paul2017,taniuchi2019,liu2019,cortes2020}.  A summary of the achieved measurements is shown in Fig.~\ref{figure_SEASTAR}.  A highlight of these recent campaigns is the first spectroscopy of $^{78}$Ni ~\cite{taniuchi2019}: its first 2$^{+}$ state was measured at 2.6 MeV while a state at 2.9 MeV was tentatively assigned to a second deformed 2$^+$ state, signing the competition of spherical and deformed configuration in the vicinity of $^{78}$Ni~\cite{santamaria2015,nowacki2016}. Nuclear structure in neutron-rich nuclei with N=50 will be discussed further in section 7. 
Gamma spectroscopy with scintillating-material detectors shows strong limitations to high level densities \cite{olivier2017} and calling for detectors with better intrinsic resolution. In-beam experiments with fast beams require a sub-centimetric resolution on the first interaction point of the photon with the detector so that the uncertainty on the scattering angle of the photon does not spoil the energy resolution after Doppler correction. In that perspective, the new generation tracking array detectors GRET(IN)A~\cite{gretina} and AGATA~\cite{agata} are perceived as the ultimate tool for the high-resolution spectroscopy of RI. Several experiments were already performed with GRETINA at the NSCL. Fig.~\ref{60TiGRETINA} illustrates the excellent resolution obtained with GRETINA at the NSCL at a beam velocity of $\beta\sim$0.4. In addition, new developments along the MINOS concept but based on in-vacuum high-granularity semiconductor trackers should allow for combined gamma and missing-mass spectroscopy with fast beams.
 
\begin{figure}[!]
\centering
\includegraphics[trim=2cm 9cm 2cm 9cm,clip,width=12cm]{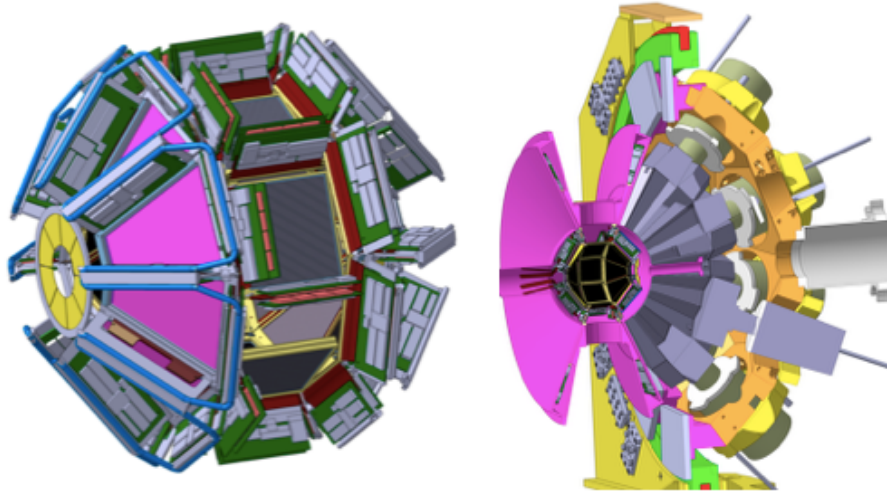}
\caption{(Left) Drawing of the GRIT silicon array (annular detector at small forward angles (yellow), trapezoidal detectors at forward angles (pink) and square detectors at $\sim$90 degrees (dark grey)) and its front-end electronics (light grey). The design and concept of the array is such that it allows particle detection and identification over a large range of angles and energies, while in a very compact geometry to be inserted in a $\gamma$-detection shell. (Right) The GRIT array in its vacuum chamber surrounded by the 1-$\pi$ configuration of the AGATA germanium array. Courtesy of F. Flavigny, LPC Caen.}
\label{GRIT}
\end{figure}

\medskip
\noindent
In general, the development of combined particle and gamma spectroscopy is indeed the doorway to access the absolute excitation energy of populated states with important information of transferred angular momentum to build up the level scheme while conserving the advantages of the excellent energy resolution provided by the gamma-ray detection. The combination of high resolution gamma spectroscopy and charged-particle spectroscopy for experiments at energies around 10 MeV/nucleon have been made possible by several compact semiconductor arrays combined with HPGe gamma detectors (see for example Refs. ~\cite{trex,sharc}). The under-development GRIT device should become the state of the art in the coming years: it is designed as a high granularity 4$\pi$-acceptance silicon array, with digital electronics allowing pulse-shape discrimination (PSD)~\cite{grit1,grit2,grit3} technique and seamless integration inside the Advanced Gamma Tracking array AGATA, as well as in the PARIS scintillator detector~\cite{paris}. The design of the GRIT array is presented in Fig.~\ref{GRIT}. It is based on a conical-shaped set of 8 trapezoidal telescopes in both the forward and backward hemisphere assembled with a ring of squared-shape silicon telescopes around 90$^{\circ}$. The integration of special targets such as a pure and windowless hydrogen target~\cite{chymene}, as well as $^{3,4}$He cryogenic targets, is planned. An intermediate semi-compact configuration was recently commissioned and operated at GANIL with AGATA in its 1$\pi$ configuration. 

\medskip
\noindent
In the case of ISOL secondary beams available at relatively high intensities ($\geq 10^5$ pps), missing-mass spectroscopy in inverse kinematics can be performed with superb excitation energy resolution following the concept of HELIOS~\cite{wuosmaa2007} developed in ANL, USA. HELIOS is based on the measurement of direct transfer reactions in a constant magnetic field along the beam axis. Based on the same concept, the new ISS device has been successfully developed and installed at HIE-ISOLDE. The first published results show the promising perspectives of the method \cite{tang2020}.
The most exotic species or reactions with very-low-energy recoils strongly benefit from recent advances in active targets based on the time-projection chamber (TPC) concept. The MAYA development~\cite{demonchy2007} at GANIL following IKAR \cite{ikar1,ikar2}, the first TPC for low-energy nuclear physics developed at the Petersburg Nuclear Physics Institute, Russia, triggered a new instrumentation trend at RI facilities and TPCs have flourished meanwhile. ACTAR~\cite{actar}, AT-TPC~\cite{at-tpc}, CAT~\cite{cat} are visible examples of such active targets dedicated to rare isotopes.

\medskip
\noindent
In this review, we shall discuss and interpret, within the shell-model framework (mostly), data measured over the past decades at RIB facilities to address the question of shell evolution across the nuclear chart. Our objective is to give a comprehensive view of shell structure with an emphasis on the competition between shell migration and onset of correlations.  Thanks to the fantastic experimental efforts sketched above, new regions where structural changes in atomic nuclei occur have been recently identified and are a focus of this work. They also raise new questions which should be addressed at the upcoming RIB facilities. 


\section{The unified shell-model view in a nutshell.}  
 The platonic goal of the nuclear theory  is to predict (explain) the properties of all the nuclei spanning the Segr\'e chart, starting from first
 principles. The observables include:  binding energies, radii, spins and parities, electromagnetic moments, excitation energies,
 decay modes and their lifetimes, reaction cross sections, etc. A first epistemological dilemma is which elementary constituent should one choose. 
 It is fair to say that nucleons are the common choice, although recently certain aspects of Quantum Chromodynamics (QCD) are taken into account in the built-up of
 the nucleon nucleon interaction (more on that later). From the very beginning,  the nuclear theory is forcefully an effective theory of QCD
  with quarks and gluons as elementary constituents.
 Given the difficulty of solving the many body problem, approximations are inevitable, which, in some sense produce new effective theories
 at a lower resolution scale. Historically, two main roads have been explored to attack the many body problem: the mean field approach 
 with energy density functionals as effective interactions (MF), and the shell model with configuration mixing
 in a basis provided by the (spherical) independent particle model (SM). In both cases efforts were made to keep the contact with the underlying
 bare nucleon nucleon interaction. The local density approximation \cite{negele1971} gave a microscopic basis to  the density dependent
 functionals of Skyrme \cite{vautherin1972} and Gogny \cite{gogny1975} type. However, in order to comply with the experimental data, it was
 necessary to fit the parameters of the force to a set of (cleverly chosen) experimental data, a first example of educated phenomenology. In the Shell Model (SM)
 context
 the effective interaction can be cast in terms of its k-body matrix elements among the orbits of the valence space. For a long time k=1 and k=2
 seemed to be enough, now we have realized that k=3 can play a role as well.  Solving a many body problem is computationally intensive, therefore
 advances in computation go hand with hand with the possibility of doing realistic calculations for nuclei of larger masses, or, better, for nuclei
 whose dynamics involve a larger number of valence nucleons. The p-shell is the smaller valence space which  makes physical sense, the first to
 be used for SM calculations, and the first where purely phenomenological effective interactions were extracted by fitting all the one-body and two-body
 matrix elements (2 and 13) to the available experimental excitation energies \cite{cohen1965}. In parallel, huge efforts were devoted to obtain
 these SM effective interactions "ab initio", {\it i. e.}, rigorously from a nucleon nucleon interaction able to reproduce the nucleon-nucleon scattering   data and the properties of the deuteron. The main problem was to regularize the short range repulsion of the bare interaction, solved by  Brueckner  theory  \cite{brueckner1954}, and to renormalize once again the resulting  G-matrices to take into account the finiteness of the valence spaces. This task was achieved fifty years ago for the $sd$ and $pf$ shells by Kuo and Brown \cite{kuo1966, kuo1968}.
 
\subsection{The first "ab initio" campaign and its failures}
The Kuo-Brown (KB) effective interaction for the $pf$-shell was released in 1968~\cite{kuo1966}. 
Had they had a powerful enough computer and shell-model code, they could have calculated the spectrum of $^{48}$Cr, 
getting the results shown in Table~\ref{tab:KB} compared with the experimental results (not available yet in 1968). 
An excellent spherical shell-model description of the yrast band of a well deformed nucleus indeed!!
Somehow, the "ab initio" description of nuclear collectivity dates 50 years back in time.
And why not to try doubly magic $^{56}$Ni? In this case the results would have puzzled them quite a lot!
Because the KB interaction makes  it a perfect oblate rotor, with $\beta$= 0.43, and  E(4$^+$)/E(2$^+$)= 3.2.
The doubly magic N=Z=28 configuration is completely absent in the yrast states' wave functions, which are, on average,
of  8p-8h nature instead. A  4p-4h prolate band of  similar deformation shows up about 2 MeV higher.
In addition, KB does not produce a doubly magic $^{48}$Ca, but it does an excellent job for $^{52}$Ca and makes
$^{54}$Ca the strongest doubly magic nuclei ever.  All in all, with lights and shadows,  a rather impressive first attempt.
But we are being counterfactual; there were neither the codes nor the computers to do the calculations, and  the
neutron-rich isotopes of calcium had not been synthesised  yet. What is true is that agreement of the "ab initio" SM  results
with the experimental data deteriorated as the number of valence particles increased in the relatively few cases that were 
computationally accessible. Because of that and because of deep formal problems in the "ab initio" theory,
the purely phenomenological approach took the lead in the SM calculations for many years, reaching its climax
with the USD interaction \cite{wildenthal1984}, which
was a fit of the three one-body and sixty three two-body matrix elements of the $sd$-shell to a database of hundreds of experimental
excitation energies, and which has enjoyed an immense popularity since then.

\begin{table}[h]
\begin{center}
\caption{Full pf-shell calculations for yrast states in  $^{48}$Cr and $^{56}$Ni with the original Kuo-Brown effective interaction.} 
\label{tab:KB}
\begin{tabular*}{\linewidth}{@{\extracolsep{\fill}}|l|llll|lll|}
   \hline 
   & \multicolumn{4}{c|}{E$^*$ in MeV} & \multicolumn{3}{c|}{B(E2)(J$\rightarrow$J-2) in e$^2$fm$^4$} \\
      & 0$^+$ & 2$^+$  & 4$^+$ &  6$^+$  & 2$^+$ & 4$^+$ &6$^+$ \\ 
\hline
  $^{48}$Cr &&&&&&& \\
  KB     & 0.0 & 0.65  & 1.66    & 3.14 & 268 & 376  &  389   \\
   Exp.  &  0.0 &  0.75 & 1.86 &  3.45 & 320(40) &  330(50) &  300(80)   \\
\hline 
\hline
  $^{56}$Ni &&&&&&& \\ 
   KB     & 0.0 &  0.52 & 1.65 & 3.35 & 546 & 771  & 818     \\
   EXP  &  0.0 &  2.70 & 3.92 & 5.32  & 98(25) &  - &  -   \\
\hline 
\end{tabular*}
\end{center}
\end{table}   

\subsection{The monopole Hamiltonian $vs$ the nuclear correlations}
To bridge the gap between the "ab initio" effective interactions and the purely fitted ones, the work of Zuker and collaborators
was instrumental. Full details can be found in the review of ref.~\cite{caurier2005} and in ref.~\cite{dufour1995}. The basic asset of the new approach is the separation of the monopole and multipole parts of the effective interaction, the former containing only number operators in the proton and neutron orbits of the valence space,    
 $H = \mathcal{H}_{m} +  \mathcal{H}_{M}$.
 
\begin{equation} 
 \mathcal{H}_m = \sum_{i} n_i  \left[  \epsilon_i +
\sum_{j}   {\displaystyle 1 \over \displaystyle (1+\delta_{ij})}
 \overline{V}_{ij}\, (n_j-\delta_{ij}) \right]
\end{equation} 

\noindent
where the coefficients $\overline{V}$ are angular averages of the two body matrix elements, or
centro\"{\i}ds of the two-body interaction:

\begin{equation} 
	\overline{V}_{ij}={\sum_J V_{ijij}^{J}[J] \over \sum_J [J]}
\end{equation} 
 
\noindent
and the sums run over Pauli allowed values. It can be written as well as:

\begin{equation} 
 \mathcal{H}_m = \sum_{i} n_i  \left[  \epsilon_i +
\sum_{j}   {\displaystyle 1 \over \displaystyle (1+\delta_{ij})}
 \overline{V}_{ij}\, (n_j-\delta_{ij}) \right]
\end{equation}  

\noindent
thus,
\begin{equation} 
 \mathcal{H}_m = \sum_{i} n_i  \; \;  \hat{\epsilon}_i (\left[ n_j \right]) 
 \end{equation}  

\begin{figure}
\begin{center}
\vspace{2cm}
\hspace{-2cm}\includegraphics[width=0.65\textwidth]{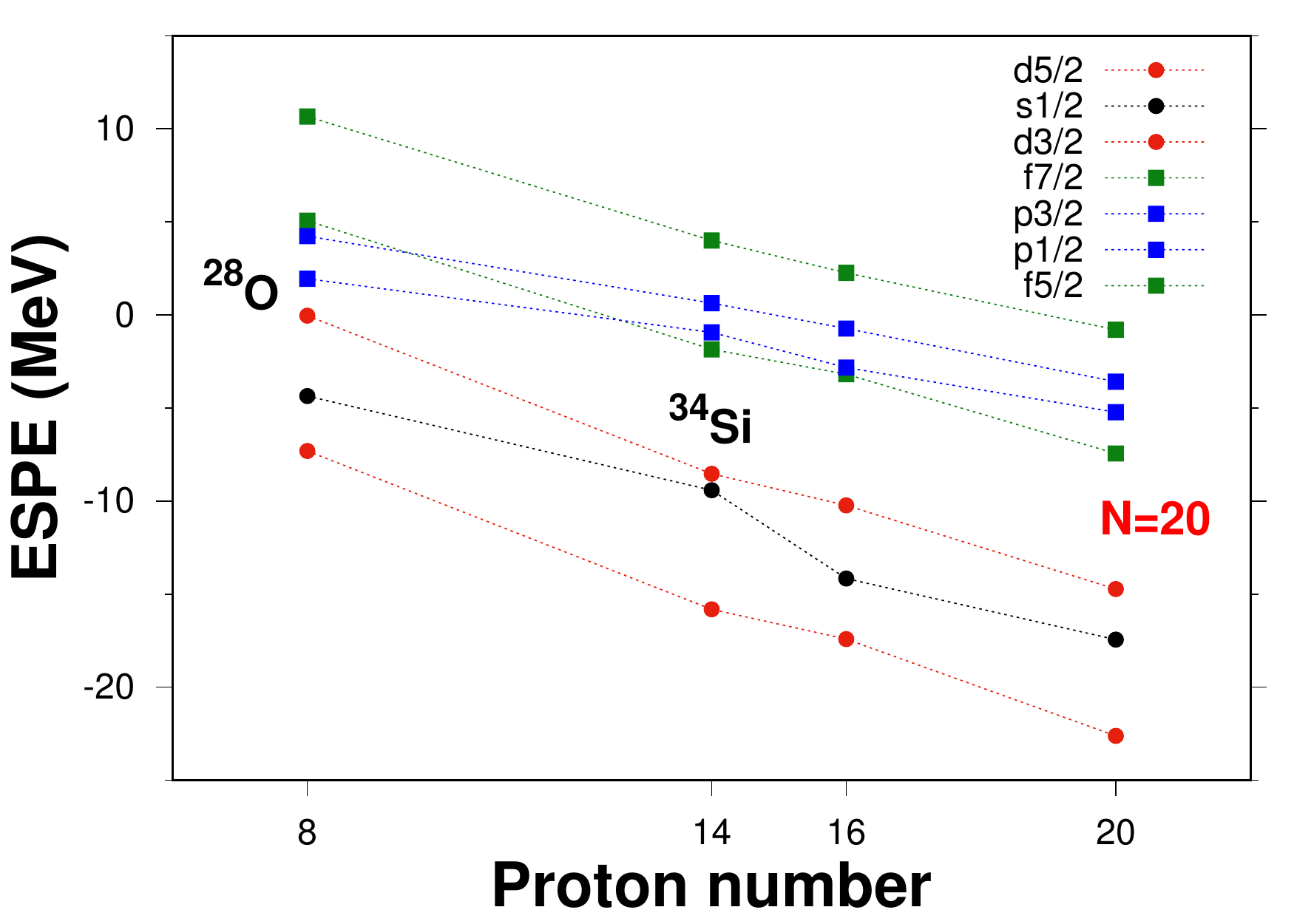}
\caption{Shell Evolution: ESPE's at N=20, SDPFU-MIX interaction.\label{ESPE}}
\end{center}
\end{figure}

\noindent
 These $\hat{\epsilon}_i (\left[ n_j \right])$ which are called effective single-particle energies (ESPE), are just
 the spherical Hartree-Fock energies corresponding to the effective interaction.
 It is seen that the monopole Hamiltonian determines  the evolution of the underlying (non observable) spherical mean field (aka, shell evolution)
 as we add particles in the valence space. Figure \ref{ESPE} gives an example for the sd and pf-shell neutron orbits at N=20 as a function of Z.

\medskip
\noindent
 The monopole Hamiltonian is diagonal in the spherical Hartree-Fock
 basis, therefore in this formalism there is a clean separation between the mean field and the correlations. The multipole Hamiltonian is fully responsible for
 the configuration mixing and hence for the coherent phenomena in the nucleus.
 Zuker's surmise was that the defects of the realistic interactions, which made them spectroscopically invalid when the number 
 of particles in the valence space increased, were limited to its monopole part, the only one which is "extensive". Indeed, as the monopole 
 Hamiltonian contains quadratic terms in the number operators, small imperfections in the centro\"{\i}ds of the two body interaction can result
 in spectroscopic catastrophes, as we have discussed for the Kuo Brown interaction, which does not reproduce the doubly magic character of $^{48}$Ca and 
 $^{56}$Ni. The modification of just two  centro\"{\i}ds of the Kuo Brown interaction was seen to be enough to restore the N=28 and Z=28 magicity and led
 to a very successful description of the $pf$-shell nuclei \cite{poves1981,caurier1994}, which became soon the new frontier of the shell model with large scale
 Configuration Interaction (mixing) (SM-CI)  approaches. Similar monopole defects plague all the realistic two body interactions irrespective of the adopted bare nucleon nucleon force and of the details of the renormalization procedure, as seen in the ones calculated with the techniques of the Oslo group \cite{hjort1995,machleidt2001}. The monopole problems were identified as due to the absence of three-body forces in ref.~\cite{zuker2003} (see also ref.~\cite{caurier2005}). 
 The monopole Hamiltonian can be trivially extended to incorporate three body terms, without relevant increases in the computational 
 cost of the SM-CI calculations.
 
\begin{figure}
\begin{center}
\includegraphics[width=0.7\textwidth]{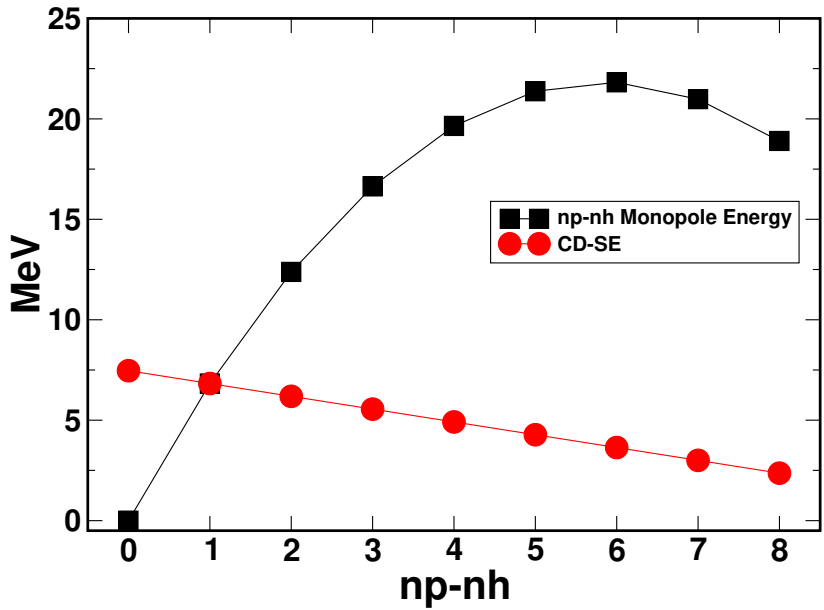}
\caption{Monopole energy of the $(0d_{3/2})^{8-n}$ $(1p_{3/2})^{n}$ configurations with respect to  doubly magic $^{40}$Ca, and the effective splitting of the single-particle energies of both orbits in the different np-nh configurations,
CD-SE.\label{cdse}} 
\end{center}
\end{figure}

 \medskip
\noindent
 Let's stress that the monopole Hamiltonian, let alone, attaches a certain binding energy to each and every configuration, characterized by the neutron and proton
  occupancies of the spherical shell-model orbits, labeled by the quantum numbers (n, l, j). And, obviously, the effective single-particle energies are 
  configuration dependent, hence they vary along an isotopic or an isotonic chain. In the literature this effect is  called shell evolution (see Fig. \ref{ESPE}).
  In the same nucleus, different configurations give rise to different ESPE's, a property which we call  CD-SE (configuration driven shell evolution).
  The Tokyo  group has put forward the term Type II shell evolution \cite{tsunoda2014}, which is less telling in our view. We offer an example of CD-SE in Fig. \ref{cdse} where it is shown how the splitting between the $0d_{3/2}$ and the $1p_{3/2}$ orbits in $^{40}$Ca changes depending on the configuration chosen. 
  It is worth noticing that, compared to the naive linear shell evolution, the 8p-8h configuration is more bound by about 30 MeV, and it 
  is located at the same excitation energy than the 4p-4h one. This effect can be of overwhelming importance to explain the occurrence of
  low lying intruder states in doubly magic \cite{caurier2009} or semi-magic nuclei \cite{caurier2014}. 
  
\medskip
\noindent  
 The multipole Hamiltonian turns out to be universal and  correctly given by the realistic effective interactions. The analysis of ref.~\cite{dufour1995} confirms that only a few channels are coherent and attractive: isovector and isoscalar pairing, plus quadrupole-quadrupole and octupole-octupole interactions. All of them are of very simple nature; BCS-like in the former case and $(r^{\lambda} Y^{\lambda} \times r^{\lambda} Y^{\lambda})^0$ for the latter. We display in Table \ref{DZ-sd} the results of the Dufour-Zuker analysis of several $sd$ shell interactions of very different nature.  KB is the original renormalized G-matrix from Kuo and Brown~\cite{kuo1966}. USD-A is a fully fitted interaction from ref.~\cite{brown2006}, CCEI \cite{jansen2014} and  NN+NNN-MBPT \cite{bogner2014} are two-body effective interactions obtained from chiral perturbation theory, which include the restriction of their three-body part to one and two body terms in the valence space.
 The similarities are astonishing and bear out the idea of multipole universality. A previous study of another set of effective interactions in the $pf$-shell led to the same conclusion \cite{caurier2005} (see Table \ref{DZ-pf}). A very interesting property stems out of the comparison of the last two rows of the table. NN-MBPT does not contains real three-body forces whereas NN+NNN-MBPT does. However the values in the table are almost identical (with a little excess of isoscalar pairing in the later, not a big issue). One is naturally driven to submit that the effects of the real three-body forces are just of monopole type, fully consistent with Zuker's surmise.
 
\begin{table}[tbh]
\caption{Dominant terms of several effective interactions for the $sd$-shell in MeV (see text). To the left particle-particle channels, isovector, isoscalar and quadrupole pairing, to the right particle-hole channels,  quadrupole, hexadecapole, spin, and spin-isospin.}
\label{DZ-sd}
\begin{tabular*}{\linewidth}{@{\extracolsep{\fill}}|c|ccc|cccc|}
\hline
 & \multicolumn{3}{c|}{pp(JT)} & \multicolumn{4}{c|}{ph($\lambda\tau$)} \\
     & 10  & 01  & 21 &    20  & 40 & 10 & 11\\ 
\hline
KB  &  -5.83 &  -4.96 & -3.21 &  -3.53 & -1.38 &  +1.61 &  +3.00   \\
USD-A      & -5.62 & -5.50  & -3.17    & -3.24  & -1.60  & +1.56  &  +2.99    \\
CCEI     & -6.79  & -4.68  & -2.93    &  -3.40  &  -1.39  &  +1.21 & +2.83     \\
NN+NNN-MBPT    & -6.40  &  -4.36  &  -2.91    & -3.28 & -1.23 & +1.10 &  +2.43    \\
NN-MBPT   &  -6.06 &  -4.38  &  -2.92  & -3.35 & -1.31 &  +1.03 &  +2.49    \\
\hline
\end{tabular*}
\end{table}

\begin{table}[tbh]
\begin{center}
   \caption{Strengths of the coherent multipole components of different interactions for the $pf$-shell. \label{DZ-pf}}
   \begin{tabular*}{\linewidth}{@{\extracolsep{\fill}}|c|cc|ccc|}\hline
  Interaction & \multicolumn{2}{c}{particle-particle} &
                \multicolumn{3}{c}{particle-hole} \\  
                & JT=01  & JT=10 & $\lambda \tau$=20 &
                                 $\lambda \tau$=40 &
                                 $\lambda \tau$=11   \\ \hline
 KB3 & -4.75 & -4.46 & -2.79 & -1.39 & +2.46 \\
 FPD6 & -5.06 & -5.08 & -3.11 & -1.67 & +3.17 \\
 GOGNY & -4.07 & -5.74 & -3.23 & -1.77 & +2.46 \\
 GXPF1 & -4.18 & -5.07 & -2.92 & -1.39 & +2.47 \\
 BONNC & -4.20 & -5.60 & -3.33 & -1.29 & +2.70 \\
 \hline
    \end{tabular*}
 \end{center}
\end{table}

\medskip
\noindent
Let's emphasize that the utility of the above analysis is mainly heuristic, because to do precision shell-model spectroscopy, all the channels must be taken into account. Nonetheless, it is a precious tool for the design of physically sound valence spaces and for the interpretation in simple terms of the huge-dimensional wave functions produced by the SM-CI diagonalizations.
 
\medskip
\noindent 
 We have seen that the realistic effective two-body interactions of Kuo and Brown have monopole pathologies. These are common to all the G-matrices regardless of the details of the many body calculation  and of choice of the bare nucleon-nucleon interaction. Chiral perturbation theory produces naturally three body forces which seem to cure these problems. We will discuss this issue in section 3.4 when we deal with the VS-IMRG \cite{stroberg2017} approach. As of now, we assume that the effective interaction has a good monopole behaviour either by fitting the necessary centro\"{\i}ds of the two-body interaction to selected experimental data or modifying them by three-body monopole terms.
Once the effective interaction is under control, the problem is to find the adequate valence space, which must be physically sound and computationally tractable. The dialectical relationship between the spherical mean field configurations (monopole Hamiltonian) and the nuclear correlations (the multipole Hamiltonian) should be able to explain the full panoply of nuclear manifestations from the extreme single particle behaviours to the fully coherent ones, usually known as nuclear collective modes.

\medskip
\noindent
  A simple example, using a schematic monopole plus isovector pairing 
  interaction, can shed light on the demeanour's of this competition. Imagine the case of two identical nucleons interacting via schematic pairing in a valence space of N orbits with total angular momentum $j_i$ and mean field energies $\epsilon_i$. The quantity of interest is the
  value of the gap, $\Delta$. The many body problem can be cast in matrix form and the solutions
  given by its eigenvalues and eigenvectors. Assuming that the pairing strength is -G, the matrix of H has the following form:
  
\begin{equation}  \left(  \begin{array}{rrrr}
2\epsilon_1 - G\Omega_1 & - G \sqrt{\Omega_1 \Omega_2}&- G \sqrt{\Omega_1 \Omega_3}& \ldots \\
- G \sqrt{\Omega_2 \Omega_1} &2\epsilon_2 - G\Omega_2&- G \sqrt{\Omega_2 \Omega_3}& \ldots \\
- G \sqrt{\Omega_3 \Omega_1} & - G \sqrt{\Omega_3 \Omega_2} & 2\epsilon_3 - G\Omega_3 & \ldots \\
. & . & . & \ldots \\
. & . & . & \ldots \\
. & . & . & \ldots \\
\end{array} \right)
\end{equation}

\noindent
There is a limit in which maximum coherence is achieved; when the orbits have the same $\Omega$ and they are degenerate at zero energy. Then the coherent pair is evenly distributed among all the orbits and its energy  is $ \Delta = -G \; N \; \Omega$. All the other solutions remain at their unperturbed (zero) energy. This result is independent of the value of $G$. Therefore degeneracies at the monopole level favour the built-up of collectivity. In fact the control parameter in this problem is:
\begin{equation}
\overline{\eta} = \frac{1}{N} \sum \frac{G}{(\epsilon_{i+1} - \epsilon_i)} =
\frac{1}{N} \sum \eta_i 
\end{equation}

\noindent
and the superfluid limit is approached when  $\eta \rightarrow \infty$. In the limit $\eta \rightarrow 0$ the spectrum in given by $ E_i = 2 \epsilon_i - G \; \Omega_i$, the ground state being i=1. The coexistence of a normal ground state with a superfluid phase occurs when 
$\eta_1 \rightarrow 0$ and $\eta_{i \ne 1} \rightarrow \infty$, and it is a proxy for the cases of shape coexistence to be explored farther on in this review.

\subsection{Heuristic of the quadrupole-quadrupole interaction: SPQR, Elliott's realm}
The fact that the spherical nuclear mean field is close to the Harmonic Oscillator (HO) has profound consequences, because the dynamical symmetry of the HO, responsible for the "accidental" degeneracies of its spectrum, is SU(3), among whose generators stand the five components of the quadrupole operator.
When valence protons and neutrons occupy the degenerate orbits of a major oscillator shell, and for an attractive Q$\cdot$Q interaction (Q being the
quadrupole operator $r^2 \; Y^2(\theta, \phi)$), 
the many body problem has an analytical solution in which  the ground state of the nucleus is maximally deformed and exhibits  rotational spectrum (Elliott's model) \cite{elliott1956}.
The basic simplification of the model is threefold; i) the valence space is limited to one major HO shell;
ii) the monopole Hamiltonian makes the orbits of this shell degenerate and iii) the multipole Hamiltonian only contains the quadrupole-quadrupole 
interaction. This  implies (mainly) that the spin orbit splitting and the pairing interaction are put to zero.
Let's then start with the isotropic HO which in units m=1 $\omega$=1  can be written as:

\begin{equation}
H_0 = \frac{1}{2} (p^2 + r^2) = \frac{1}{2} (\vec{p} + i \vec{r})  (\vec{p} - i \vec{r}) +  \frac{3}{2} \hbar = \hbar  (\vec{A}^{\dagger} \vec{A} +  \frac{3}{2} )
\end{equation} with

\begin{equation}
\vec{A}^{\dagger} = \frac{1}{\sqrt{2\hbar}}(\vec{p} + i \vec{r})  \; \;  \vec{A} = \frac{1}{\sqrt{2\hbar}}(\vec{p} - i \vec{r})
\end{equation}
which have bosonic commutation relations. $H_0$ is invariant under all the transformations which leave invariant
the scalar product  $\vec{A}^{\dagger}\vec{A}$. As the vectors are three dimensional and complex, the symmetry group is U(3).
We can built the generators of U(3) as  bi-linear operators  in the A's. The anti-symmetric combinations
produce the three components of the orbital angular momentum L$_x$,  L$_y$ and  L$_z$, which are in turn the generators 
of the rotation group O(3). From the six symmetric bi-linears we can retire the trace that is a constant; the mean field energy.
Taking it out we move into the group SU(3). The five remaining generators are the five components of the quadrupole operator:

\begin{equation}
q^{(2)}_{\mu} = \frac{\sqrt{6}}{2\hbar} (\vec{r} \wedge  \vec{r})^{(2)}_{\mu}  + \frac{\sqrt{6}}{2\hbar} (\vec{p} \wedge \vec{p})^{(2)}_{\mu}
 \end{equation} 
 The generators of SU(3) transform single nucleon wavefunctions of a given {\bf p} (principal quantum number)
 into themselves. In a single nucleon state there are  {\bf p}  oscillator quanta which behave as l=1 bosons. When we have several
  particles we need to construct the {\it irreps} of SU(3) which are characterized by the Young's tableaux  (n$_1$, n$_2$, n$_3$) 
  with  n$_1$$\ge$n$_2 $$\ge$n$_3$ and  n$_1$+n$_2$+ n$_3$=N{\bf p} (N being the number of particles in the open shell). 
   The states of one particle in the  {\bf p} shell correspond
  to the representation (p,0,0). Given the constancy of  N{\bf p} the {\it irreps} can be labeled with only two numbers. Elliott's choice 
  was $\lambda$=n$_1$-n$_3$ and $\mu$=n$_2$-n$_3$. In the cartesian basis we have;  n$_x$=a+$\mu$,   n$_y$=a, 
   and n$_z$=a+$\lambda$+$\mu$, with 3a+$\lambda$+2$\mu$=N{\bf p}. 
   
\bigskip
\noindent
The quadratic Casimir operator of SU(3) is built from the generators

\begin{equation}
\vec{L}=\sum_{i=1}^N \vec{l}(i) \; \;   \; \; \; \;  Q^{(2)}_{\alpha}=\sum_{i=1}^N q^{(2)}_{\alpha}(i)
\end{equation}as:

\begin{equation}
C_{SU(3)}=\frac{3}{4} (\vec{L} \cdot  \vec{L})  +  \frac{1}{4} (Q^{(2)} \cdot Q^{(2)})
\end{equation} 
and commutes with them. With the usual group theoretical techniques, it can be shown that the 
eigenvalues of the Casimir operator in a given representation ($\lambda,\mu$) are:

\begin{equation}
C(\lambda,\mu)=\lambda^2 + \lambda \mu + \mu^2 + 3(\lambda + \mu)
\end{equation}

\bigskip
\noindent
Once these tools ready we come back to the physics problem as posed by Elliott's Hamiltonian

\begin{equation}
H=H_0 + \chi (Q^{(2)} \cdot Q^{(2)})
\end{equation} which can be rewritten as:

\begin{equation}
H=H_0 + 4 \chi C_{SU(3)} - 3 \chi  (\vec{L} \cdot  \vec{L})
\end{equation} 
The eigenvectors of this problem are thus characterized by the quantum numbers  $\lambda$, $\mu$, and L.
We can choose to label our states with these quantum numbers because O(3) is a subgroup of SU(3) and therefore the
problem has an analytical solution: 

\begin{equation}
E(\lambda, \mu, L) = N \hbar \omega (p + \frac{3}{2}) + 4 \chi   (\lambda^2 + \lambda \mu + \mu^2 + 3(\lambda + \mu)) - 3 \chi  L (L + 1) 
\end{equation} 
This important result can be interpreted as follows: For an attractive quadrupole-quadrupole interaction ($\chi < 0 $) the 
ground state of the problem pertains to the representation which maximizes the value of the Casimir operator, and this corresponds
to maximizing $\lambda$ or $\mu$ (the choice is arbitrary). If we look at  that in the Cartesian basis, this state is the one which has 
the maximum number of oscillator quanta in the Z-direction, thus breaking the symmetry at the intrinsic level. We can then speak of 
a deformed solution even if  its wave function conserves the good quantum numbers of the rotation group, i.e. L and L$_z$. For that one
(and every) ($\lambda, \mu$) representation, there are different values of L which are permitted, for instance for the representation 
($\lambda, 0$) L=0,2,4$\ldots\lambda$. And their energies satisfy the L(L+1) law, thus giving the spectrum of a rigid rotor. The problem 
of the description of the deformed nuclear rotors in the laboratory frame is thus formally solved.

\bigskip
\noindent
We can describe the intrinsic states and its relationship with the physical ones using another chain of subgroups of SU(3).  The one
we have used until now is; SU(3)$\supset$O(3)$\supset$U(1) which corresponds to labeling the states as $\Psi([\tilde{f}] (\lambda \mu) LM )$.
$[\tilde{f}]$ is the representation of U($\Omega$) conjugate of the U(4) spin-isospin representation which guarantees the antisymmetry of the total wave function.
For instance, in the case of $^{20}$Ne, the fundamental representation  (8,0) (four particles in p=2) is fully symmetric, $[\tilde{f}]$=$[4]$,  and its  conjugate
representation in the U(4) of Wigner  $[1,1,1,1]$, fully antisymmetric. The other chain of subgroups, SU(3)$\supset$SU(2)$\supset$U(1), does not
contain O(3) and therefore the total orbital angular momentum is not a good quantum number anymore. Instead we can label the wave functions as;
$\Phi([\tilde{f}] (\lambda \mu) q_0 \Lambda   K)$, where $q_0$ can have a maximum value $q_0 = 2\lambda + \mu$. The intrinsic quadrupole moment 
Q$_0$=(q$_0$+3) b$^2$, where b is the length parameter of the HO. 
K is the projection of the angular momentum on the Z-axis and $\Lambda$ is an angular momentum without physical meaning.  Both
representations provide a complete basis, therefore it is possible to write the physical states in the basis of the intrinsic ones. Actually, the
physical states can be projected out of the intrinsic states with maximum quadrupole moment as:

\begin{equation}
\Psi([\tilde{f}] (\lambda \mu) LM )= \frac{2L+1}{a(\lambda \mu KL)} \int D^L_{MK}(\omega) \Phi_{\omega}([\tilde{f}] (\lambda \mu) (q_0)_{max} \Lambda   K) d\omega
\end{equation}Remarkably, this is the same kind of expression used in the unified model; the Wigner functions $D$ being the eigenfunctions of the rigid rotor
and the intrinsic functions the solutions of the Nilsson model.

\medskip
\noindent
 To compare the intrinsic quadrupole moment of a given SU(3) representation  with the experimental values we need to know the 
transformation rules from intrinsic
to laboratory frame quantities and vice versa. In the Bohr Mottelson model these are:

\begin{equation}
\label{bmq}
 Q_0(s)=\frac{(J+1)\,(2J+3)}{3K^2-J(J+1)}\,Q_{spec}(J), \quad K\ne 1
\end{equation} 
\begin{equation}
 B(E2,J\;\rightarrow\;J-2)=
\frac{5}{16\pi}\,e^2|\langle JK20|
 J-2,K\rangle |^2 \, Q_0(t)^2\quad K\ne 1/2,\, 1;
\label{bme2} 
\end{equation}

\noindent
The expression for the quadrupole moments is also valid in Elliott's model. However the one for the B(E2)'s is only approximately
valid for low spins. 

\bigskip
\noindent
Besides Elliott's SU(3) there are other approximate symmetries related to the quadrupole-quadrupole interaction which are
of great interest. Pseudo-SU3 applies when the valence space consists of a quasi-degenerate harmonic oscillator shell except 
for the orbit with maximum $j$, this space will be denoted by r$_p$. Its quadrupole properties are the SU(3) ones of the shell
with (p-1) \cite{arima1969,hecht1969} except for the radial integrals of r$^2$. Quasi-SU3 \cite{caurier2005,zuker1995} applies 
in a regime of large spin orbit splitting,  when the valence space contains the intruder orbit and the
$\Delta$j=2, $\Delta$l=2; $\Delta$j=4, $\Delta$l=4; etc,
orbits obtained from it. 
In addition, single orbits of large $j$ can give non negligible contributions to the quadrupole moment of a given configuration, the more so
in the case of holes which add prolate coherence. The final scheme which incorporates the latter three realizations of the quadrupole quadrupole  interaction has been dubbed SPQR, where R stands for renormalized, in ref.~\cite{zuker2015}, where all the technical details are thoroughly  discussed.

\medskip
\noindent
 There are plenty of uses of the SPQR scheme in conjunction with the monopole Hamiltonian. (i) It makes it possible to anticipate in which nucleus or region of nuclei can the quadrupole correlations thrive. It suffices to examine the structure of the spherical mean field above and below the Fermi level. If there are
 quasi-degenerated orbits which abide by the Quasi-SU(3) or Pseudo-SU(3) rules they can. Single shell quadrupole collectivity alone does not drive the nucleus into the deformed phase, but can act coherently with the other two coupling schemes. (ii) It can serve as a guide for the choice of the minimal valence spaces which can encompass the quadrupole collectivity in full, describing deformed rotors in the laboratory frame.
 (iii) If the quadrupole operator is diagonalized in the SU(3), Quasi-SU(3), or
 Pseudo-SU(3) closed set of orbits, one obtains the Nilsson-SU(3) intrinsic levels which can be characterized by their intrinsic quadrupole moments and K-values and which were baptized as ZRP diagrams in \cite{zuker2015}. This makes it possible to calculate the maximum quadrupole value of the different nuclei and configurations, and, in particular to signal the possibility of shape coexistence and shape transitions before performing the full fledged SM-CI calculations. (iv) The self-consistent SU(3)-Nilsson method, developed also in ref.~\cite{zuker2015}, permits more quantitative applications of the SPQR scheme, which reinforce
 as well its heuristic value. The crucial finding of \cite{zuker2015} is that the  SPQR scheme, which in principle requires the degeneracy of the relevant single particle orbits, is,  to a large extent,  resilient to the presence of single particle splittings, which is the more realistic and usual situation. As the SPQR results
 are barely eroded by reasonable modifications of the spherical mean field, they can be taken as bare face predictions.
 We will give abundant examples of the heuristic use of the SPQR scheme along with our explorations of the neutron-rich shores in Section 4, which indeed, when compared with the SM-CI provide an extra support to the validity of the
 SPQR approach in the quadrupole dominated regime. When N$>>$Z, the regime of quadrupole dominance will hold or not depending mainly on the product of the total quadrupole moments attainable by the protons and by the neutrons, because the quadrupole interaction is much stronger in the isoscalar than in the isovector channel. However, it is important to
 keep in mind that the latter channel is not negligible at all.
 We shall offer a fully worked example of this situation when studying the case of $^{60}$Ca in Section 6.
 
\begin{figure}[h]
\begin{center}
\includegraphics[width=0.49\textwidth]{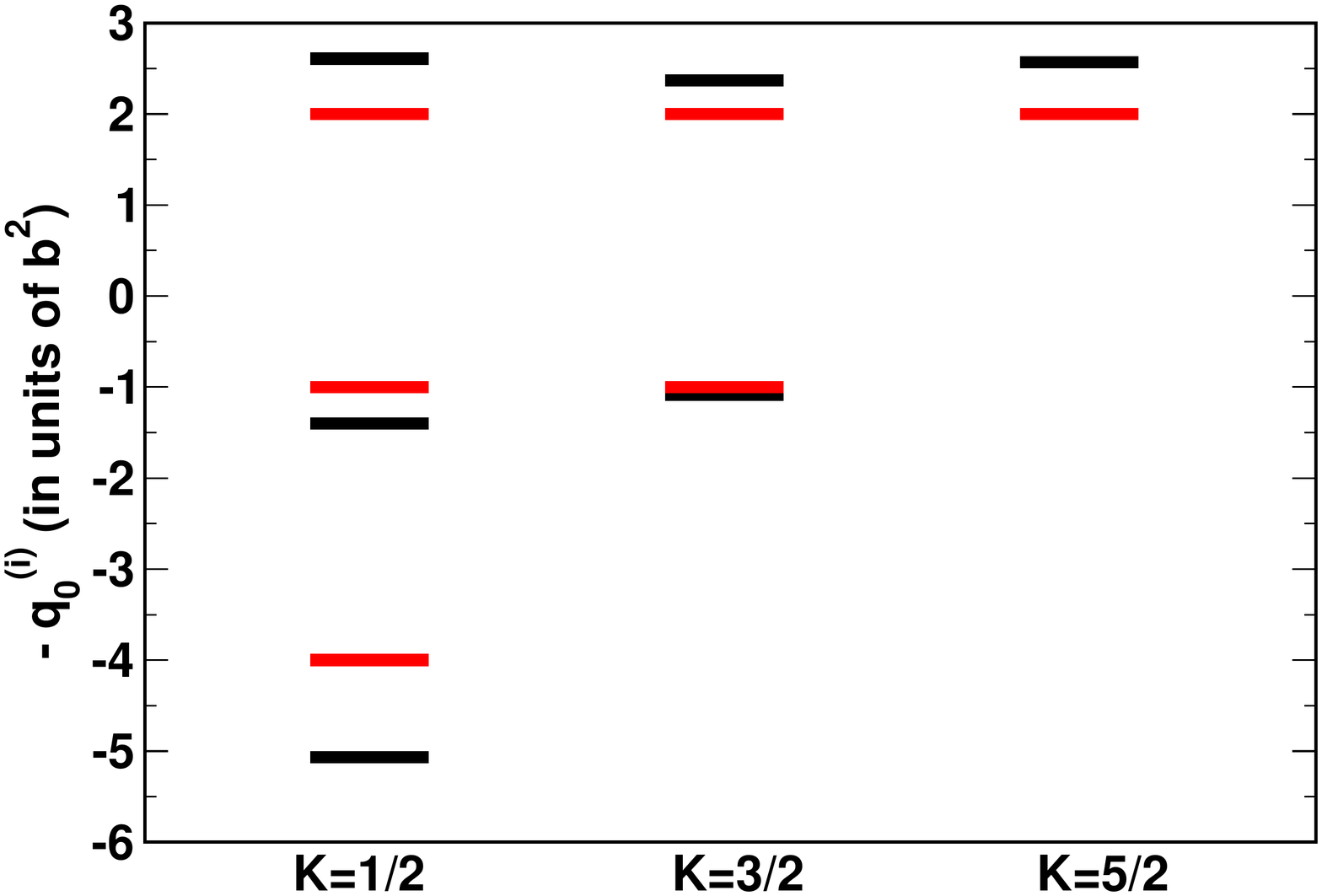}
\includegraphics[width=0.49\textwidth]{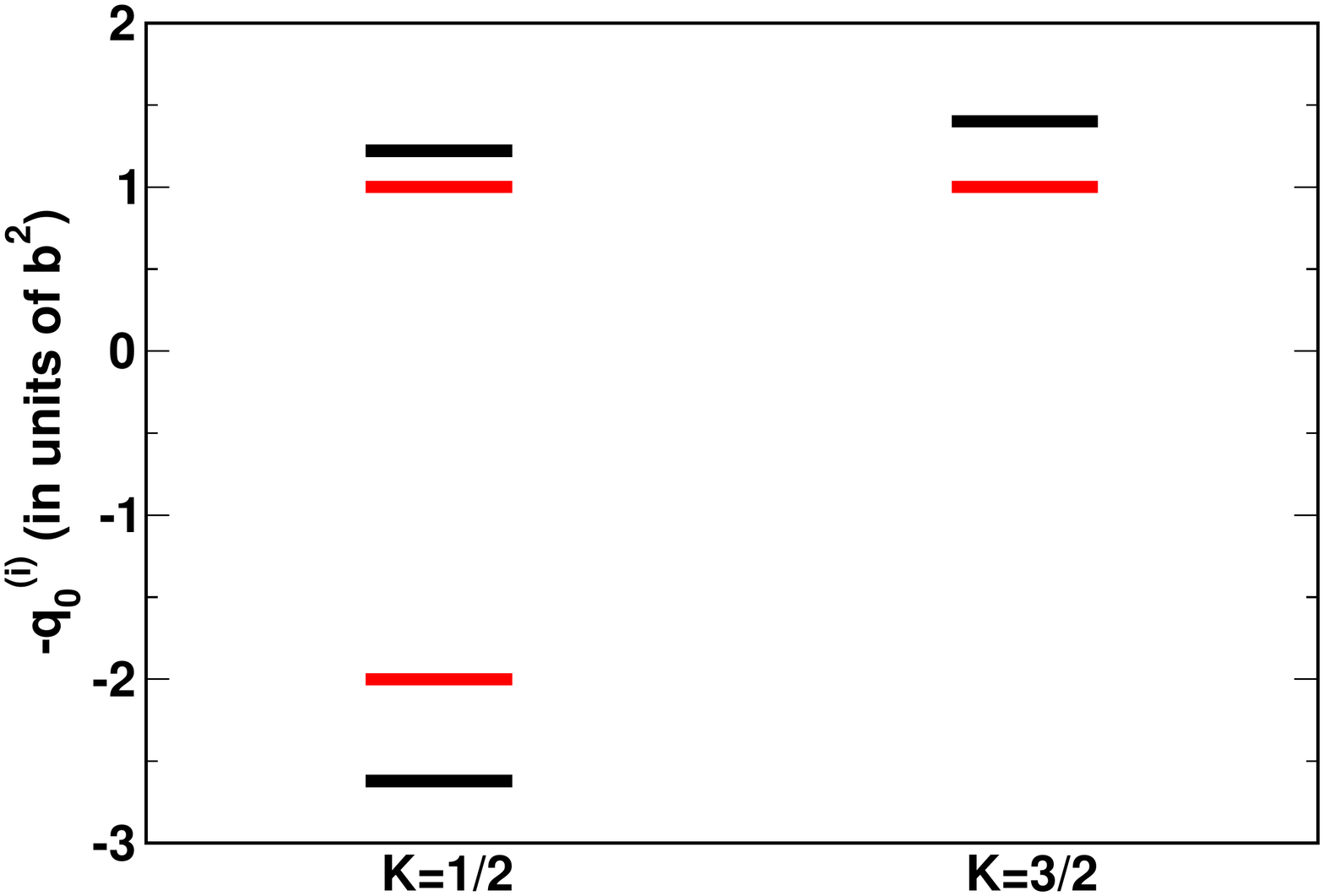}
\caption{ZRP diagrams; (Left) Intrinsic states of the Elliott's model interaction in the $sd$-shell labeled by their SU(3) quadrupole moments q$_0^{(i)}$ and K's (red bars). (black bars) Id for the case of  Pseudo SU(3) in the $pf$-shell. (Right) SU(3) in the $p$-shell (red lines) and Pseudo-SU(3) in the $sd$-shell (black lines). b$^2$ is the length parameter of the HO,  b$^2$=$\displaystyle{\frac{\hbar}{m \omega}}$.  \label{su3}} 
\end{center}
\end{figure}

\medskip
\noindent
  Before moving on, let's discuss briefly the simplest of the ZRP diagrams, the one corresponding to strict SU(3)
  in the $sd$-shell depicted in Fig. \ref{su3}.
  The diagram shows six intrinsic states (in reality twelve because the positive and negative values of K are degenerated. Therefore each of them can accommodate two neutrons and two protons. The prolate solutions are found filling the platforms in the diagram from below and the oblate ones filling them  from above. The value of
  the intrinsic SU(3) quadrupole moment q$_0$ is obtained as $\sum_i q_0^{(i)}$ and equals $2\lambda + \mu$. It is trivial to check it for the paradigms of SU(3) nuclei, $^{20}$Ne, $^{24}$Mg, and $^{28}$Si, corresponding to the representations ($\lambda,\mu$) (8,0), (8,4) and (12,0) or (0,12) respectively. Notice that for $\mu > \lambda$, 
  q$_0$= -($2\mu + \lambda$). For simplicity we shall use electric effective charges 1.5e for the protons and 0.5e
  for the neutrons. In dealing with mass quadrupole moments we take 2.0 for both protons and neutrons. Our choice of  $\hbar \omega$ is  45 A$^{-1/3}$ - 25 A$^{-2/3}$. 
   The black platforms in Fig. \ref{su3} correspond to the Pseudo-SU(3) intrinsic states in the $pf$-shell, i.e.
  in the space of the orbits 1p$_{3/2}$, 1p$_{1/2}$, and 0f$_{5/2}$. Were it not because of the difference in the 
  values of $<r^2>$, the platforms would be identical to those of SU(3) in the $sd$-shell. It is seen that they are not, although they follow the same  pattern. The largest deviation corresponds to the more prolate K=1/2 state, and, because of that, the prolate quadrupole moments for the four and eight particle cases are noticeably enhanced. The operating rules are always the same. 
  
 \medskip
\noindent
  To verify the resilience of the SU(3) predictions let's work out the case of $^{20}$Ne. The theoretical values K=0 and  Q$_0$=19~b$^2$
  give through equations (17) and (18) Q$_{spec}$=-17~efm$^2$ and B(E2)($2^+ \rightarrow 0^+$)=72~e$^2$fm$^4$, which compare nicely with the experimental ones Q$_{spec}$=-23(3)~efm$^2$ and B(E2)($2^+ \rightarrow 0^+$)=66(3)~e$^2$fm$^4$, irrespective of the fact that the three orbits of the $sd$ are far from being degenerated. When the filling of the platforms is not unique, several K values correspond to the same quadrupole moments
  and their energies are the same. This is the case for $^{24}$Mg which can have K=0 and K=2.
  In strict SU(3) the K=0 and K=2 bands do not mix, but in reality they  do, making it triaxial and breaking the degeneracy  of the two lowest  2$^+$ states in accord with the experimental situation.

\subsection{How to solve the nuclear  many-body problem; the configuration interaction approach (SM-CI) and some variants}

       The solution of the nuclear many body problem in the basis of the spherical independent-particle model (SM-CI) is conceptually straightforward but computationally very demanding, because the dimensions of the matrices to be built and 
       diagonalized grow rapidly with the number of states in the valence space (the orbits that 
       are explicitly considered in the calculation, usually a few above and below the Fermi level,
       the orbits which are assumed to be always complete make the core, and those which will be never occupied the external space)
       and the number of valence particles. If the number of available states is N$_{\pi}$ for the protons
       and  N$_{\nu}$ for the neutrons and the number of valence protons and neutrons is n$_{\pi}$ and  n$_{\nu}$, the dimension of the many body basis is:

\begin{equation} D =  \left( \begin{array}{c}
  N_{\pi}\\
   n_{\pi}
\end{array} \right)
\cdot
\left( \begin{array}{c}
  N_{\nu}\\
   n_{\nu}
\end{array} \right)
\end{equation}

\noindent
This gives D=853776 as the maximum value for the $sd$-shell valence space with 6 protons and 6 neutrons. In reality, this number includes a lot of redundant Slater determinants, if one takes just those with M=0 a large reduction ensues and D(M=0)=93710. In addition the basis can be partitioned in subspaces with well defined  J or well defined
J and T, to further reduce the dimensions (for instance, in this case D(J=0)=3372). However,  the computation of the many-body matrix elements in the uncoupled basis (m-scheme)
is much simpler than in the coupled schemes, and most often  compensates for the dimensional issues.  The above quoted  dimensions were the largest attainable until the end of the 80's using shell-model codes like the Oak-Ridge Rochester Multishell
\cite{french1969}, the Glasgow code \cite{whitehead1977} and  Oxbash \cite{brown1985,brown1988}, among others.
More details can be found in ref.~\cite{caurier2005}, where the new generation of codes pioneered by
Antoine in  m-scheme  are discussed as well. At present, direct diagonalizations taking advantage of the Lanczos method make it possible to reach D(M=0) $\sim$ O(10$^{11}$).
Another avenue is provided by the Monte Carlo Shell Model (MCSM) (or quantum Monte Carlo diagonalization method), \cite{otsuka1998,otsuka2001}, which is based on  deformed Hartree-Fock solutions in the valence space, and stochastically chosen Slater determinants using the Hubbard-Stratonovich auxiliary field techniques. The selected states are projected to good angular momentum providing a basis in which the Hamiltonian is diagonalized. The size of the basis is increased till convergence is achieved. Another path, more in the mean field tradition relies in the use of self-consistent symmetry breaking Hartree-Fock solutions, using energy  
density functionals of contact type (Skyrme) or finite range (Gogny). To reach spectroscopic accuracy
the symmetries must be restored, and states remixed via the generator coordinate method (SCCM). For a recent review see reference \cite{robledo2019}.

\medskip
\noindent
 There has been a recent revival of the "ab initio" methods which try to connect the SM-CI calculations with QCD using Chiral perturbation theory. In this approach three-body forces appear naturally \cite{machleidt2011}. There are many different procedures to regularize the interaction and  to adapt it to the valence space. All of them  share problems, first with the selection of the
 starting interaction and the fit of its free parameters to the nucleon-nucleon data. Then with the choice of the values of the different cut-offs and of the order 
 at which the Chiral series is  truncated. This gives rise to a plethora of different interactions that all carry the "ab initio" label. This fact somehow 
 blurs the very many merits of the present search for the final connection between QCD and the low energy structure of the atomic nucleus.

\medskip
\noindent 
  For very light nuclei the No Core Shell Model (NCSM) has been successfully applied,\cite{barrett2013,hergert2013}. Other methods include the Coupled Cluster (CC),
 \cite{hagen2014}, the In Medium Similarity Renormalization Group (IMSRG), and many body perturbation theory (MBPT) \cite{coraggio2009}. In the cases of the NCSM, CC, and IMSRG, there
 has been efforts to generate effective one-body  and two-body interactions to be used in SM-CI calculations in restricted valence spaces. These where first proposed for the $sd$-shell in refs. \cite{jansen2014}
 (CC) and \cite{bogner2014} (IMSRG) as previously discussed. The idea was to make summations of the three body
 interactions involving one or two particles in the core, to obtain their contributions to the one-body  and two-body interactions in the valence space, disregarding 
 the irreducible three-body terms in 
 the valence space. It was soon noticed that the results deteriorated rapidly as the number of active 
 particle was increased. The good news, as already mentioned, was that their multipole Hamiltonians were extremely similar to the traditional renormalized G-matrices. Very recently effective interactions for the $sd$-shell have been obtained by projecting the NCSM results in the valence space \cite{smirnova2019}. The effect of disregarding explicit three-body forces in the valence space
 can be minimized in the VS-IMSRG approach (VS stands for Valence Space). The clever trick is
 \cite{stroberg2017,morris2018} to take as the reference state for the reduction of the interaction to one-body and two-body terms in the valence space, the Hartree-Fock solution (in the equal filling approximation) for each nucleus. Notice than in the equal filling approximation only the monopole-like terms have non null expectation values.  In this way the problem of the evolution of the interaction with the number of particle is  approximately solved. New improved  effective interactions based on the CC method (CCSM)  have been
 calculated in ref.~\cite{sun2018} as well. An excellent review of the present situation of the "ab initio"
 approaches to the SM-CI  can be found in ref.~\cite{stroberg2019}. Seemingly, for technical reasons the "ab initio" program is not yet able to cope properly with problems which demand valence spaces encompassing more than one major oscillator shell  for the protons and/or the neutrons. This severely hampers its applicability to the regions of interest in this review, notably to the Islands of Inversion and where the phenomena of shape coexistence show up. 
 Another recent result, worth to mention, comes from the MBPT approach taking into account again the modifications of the two body Hamiltonian in the valence space due to real three-body forces. In ref.~\cite{ma2019}, it is shown that the three body effects are of monopole character, restoring (in their $pf$-shell calculation) the doubly magic nature of $^{48}$Ca and  $^{56}$Ni. The SM-CI framework requires the use of effective operators as well. In particular, the Gamow Teller $\beta$ decay operator $\vec{\sigma} \cdot \vec{\tau}$ needs to be quenched by a factor $\sim$ 0.7 to reproduce the experimental data. This value has been recently deduced theoretically by several of the "ab initio" methods sketched above \cite{menendez2011,gysbers2019}.

\subsection{The meaning of the nuclear shape: Kumar invariants.}
The description of collectivity in nuclear structure has its roots in the unified model of Bohr and Mottelson~\cite{bohr1953}. The treatment of the dominant quadrupole correlations and the description of nuclear deformation have often been carried out in the intrinsic frame, and 
quadrupole shapes have been characterized by the intrinsic deformation parameters $\beta$ and $\gamma$. We commonly characterize a nucleus as 
prolate when $\gamma$=0$^\circ$, oblate when $\gamma$=60$^\circ$, fully triaxial when $\gamma$=30$^\circ$, prolate triaxial when $0^\circ < \gamma < 30^\circ$, and oblate triaxial when $30^\circ < \gamma < 60^\circ$. There has been a longstanding debate as to whether nuclei with rigid triaxial deformation exist, or instead, can only be $\gamma$-soft. Often the same nucleus exhibits signatures of intrinsic structures with different values of $\beta$ and $\gamma$, a phenomenon known as shape coexistence~\cite{heyde2011}. The intrinsic deformation parameters can change along isotopic or isotonic chains, referred to as shape evolution or shape transition.

\medskip
\noindent
   The intrinsic shape parameters are usually inferred from laboratory frame
  experimental values of observables such as excitation energies, $E2$ transitions, and spectroscopic quadrupole and magnetic moments. 
  They can also be extracted from calculations carried out in the laboratory frame. In both cases it is necessary to agree on
  a set of rules of transforming between the laboratory frame and the intrinsic frame. These existed since long ago, e.g., in the work of Davidov and Filipov~\cite{davidov1958}
  and many others.  However, the only rigorous method to relate the intrinsic parameters to laboratory-frame observables is provided
  by the so-called quadrupole invariants $\hat Q^n$ of the second-rank quadrupole operator $\hat Q$  introduced by  Kumar~\cite{kumar1972} (see also ref.~\cite{cline1986}). The calculation of $\beta$ and $ \gamma$  requires knowledge of the expectation values of the second- and third-order invariants defined, respectively, by $\hat Q^2=\hat Q\cdot \hat Q$ and $\hat Q^3=  (\hat Q\times \hat Q)\cdot \hat Q$ (where $\hat Q \times \hat Q$ is the coupling of $\hat Q$ with itself to a second-rank operator).  These invariants were recently applied to describe the evolution of
 collectivity in cadmium isotopes~\cite{schmidt2017}.  However, it is not meaningful to assign effective values to $\beta$ and $\gamma$ without also studying their fluctuations, and this task requires to calculate all the moments up to the sixth. The fourth order invariant was used in Refs.~\cite{alhassid2014,gilbreth2017} to estimate the fluctuations in $\beta$ in heavy rare-earth nuclei and  in Refs.~\cite{hadynska2016,hadynska2018} for the ground state and the deformed excited band of $^{42}$Ca. Only very recently the exact calculation of all the moments became possible in the work of ref.~\cite{poves2020}. The new results imply a change of paradigm which is of general relevance when speaking in general about nuclear shapes and shape coexistence and in particular for the discussion of these phenomena at the neutron-rich Islands of Inversion (IoI) which follows.
 In brief, what the fluctuation in $\beta$ and $\gamma$ tell us is that the concept of shape is very much blurred in the laboratory frame. The conclusions of \cite{poves2020} that we need to keep in mind when talking about shapes are:
 \begin{itemize}
 \item (I) Even in well deformed nuclei $\beta$ has a non-negligible degree of softness.
 \item(II) The $\gamma$ degree of freedom is so soft that it is no longer meaningful to assign to it an effective value.
 \item (III) In some cases the full spread of  $\gamma$ at one sigma  occurs between 0$^{\circ}$ and 
 30$^{\circ}$ or 30$^{\circ}$ and 60$^{\circ}$, thus we can loosely speak of a prolate-like
  or oblate-like nucleus.
 \item(III) The very concept of shape is meaningless for doubly magic nuclei or in general for states which are not dominated by the quadrupole correlations, because of the huge fluctuations both in
 $\beta$ and $\gamma$.
 \item(IV) This implies that the concept of a spherical nucleus is a quantal oxymoron\footnote{In spite of this bold assertion, and to comply with
 the established semantics in the nuclear physics community, we shall continue using the tag spherical nuclei loosely, attaching it to those
 nuclei which do not have any intrinsic shape.}.
 \item (V) In the SU(3) limit the fluctuations in $\beta$ and $\gamma$ are zero and  they are small
  when a nucleus is close to this limit, for instance in $^{20}$Ne. Heavy, well deformed, nuclei
  may be close to this limit as well, and therefore the shape concept bears a revival in the semi-classical limit.

\end{itemize}
\noindent

\section{ Islands of Inversion (IoI) at the neutron-rich shores.}

The different variants of the SU(3) symmetry turn
out to be at the root of the appearance of islands of inversion far from stability. They are more prominent at the neutron-rich side
and occur when the configurations which correspond to the neutron shell closures at N=8, 20, 28,  40  and 50 are less bound than the
intruder ones (more often deformed) built by promoting neutrons across the Fermi level gap. The reason of the inversion is that
the intruder configurations maximize the quadrupole correlations and thus their energy gains. This is only possible when the
orbits around the Fermi level can develop the symmetries of the quadrupole interaction. There are however cases in which nuclei with only active 
neutrons in the valence space may belong to the IoI, provided the gap at the  Fermi level is small enough.
In addition to the present mechanism, far from stability, the monopole drift makes it possible that other regimes may develop.

\subsection{The IoI's below doubly magic $^{34}$Si and $^{48}$Ca, at N=20 and N=28} 

The exploration of this region began with the mass measurements made at CERN by what would later be the Orsay group of the ISOLDE Collaboration \cite{thibault1975}. They found an excess of binding in $^{31}$Na and $^{32}$Na which could not be explained by the available theoretical models. Simultaneously, a group of theorist at Orsay tried to reproduce this effect with Projected Hartree-Fock, PHF,  calculations using density dependent interactions \cite{campi1975}. They got a normal, spherical, solution for $^{31}$Na with a small shoulder in the prolate side of the energy versus deformation curve. 
Nevertheless they  argued that applying a rotational correction this shoulder might become a minimum and explained the anomaly as due to the gain in energy due to deformation. Even with this proviso, their model was unable to explain the $^{31}$Na data. Three decades and huge developments in beyond mean field (BMF) techniques were necessary to produce this minimum, in the easier to treat nucleus  $^{32}$Mg. All in all, their idea was
 correct even if the calculations were not conclusive. Already in the paper of Thibault  {\it et al.}  \cite{thibault1975}, the possibility of having a new region of deformation was contemplated.

\medskip
\noindent   
   Further work by the same group \cite{huber1978,detraz1979} extended the mass measurements 
   to the Magnesium isotopes finding the same kind of anomaly in  $^{32}$Mg. In addition, studying the beta decay
   of   $^{32}$Na, they populated the first 2$^+$ state in   $^{32}$Mg at 881~keV \cite{guillemaud1984}. Indeed this is a very low
   excitation energy compared with the corresponding one in $^{30}$Mg  (1.54~MeV). The more so, because
   common knowledge  would indicate  that the  2$^+$ excitation energy should increase at the magic neutron number N=20
   and not to decrease almost by a factor two.
   The idea of the existence  of an unexpected new region of deformation, precisely where the
   current models would have predicted enhanced sphericity, was reinforced.  

\medskip
\noindent        
    New shell-model studies of these nuclei were triggered by a paper by Wildenthal and Chung \cite{wildenthal1980}
    in 1980 with the provocative title "Collapse of the conventional shell-model ordering in the very-neutron-rich isotopes of Na and Mg".
    In it, the authors, using their newly fitted high quality interaction for the upper part of the $sd$-shell, concluded that it was impossible
    to explain the behaviour of the neutron-rich N=20 isotones in this valence space. The question that remained unanswered and which
    has continued to produce a lot of confusion until now is the meaning of the sentence "Collapse of the conventional shell-model ordering".
    For many, this had to be interpreted as if, at the mean field level, some  single particle orbits of the $pf$-shell  would lay below
    the upper orbits of the $sd$-shell. Work in this direction was carried out by the Glasgow group \cite{watt1981,storm1983} by including
    the 0f$_{7/2}$ orbit in the valence space, essentially degenerated with the  0d$_{3/2}$. With this prescription they indeed 
    got  an increase in the binding energy of the N=20 isotones, but the proposed solution was far from being physically sound, and did 
    not address at all  the issue about deformation risen by the PHF calculations on ref.~\cite{campi1975}.

\medskip
\noindent
         The work of ref.~\cite{poves1987} interpreted the collapse  of the shell-model ordering in a different way: they were not the spherical
     single particle orbits of the two adjacent major harmonic oscillator shells  which collapsed far from stability, instead, what happened was that 
     the intruder configurations of  (neutron) 2p-2h nature, somehow became more bound than the normal, 0p-0h, ones. Therefore, the
     collapse must be due to the nuclear correlations. This interpretation makes a natural link with the surmise of ref.~\cite{campi1975} provided
     the correlations are of quadrupole-quadrupole type and the intruder configurations become deformed. The problem was how to get
     this behaviour out of a realistic shell-model calculation. And the solution was to include the  1p$_{3/2}$ orbit --the quadrupole partner
     of the  0f$_{7/2}$-- in the valence space. This calculation produced deformed  solutions for $^{32}$Mg,  $^{31}$Na,  $^{30}$Ne, keeping
     a substantial single particle gap between the $sd$ and the $pf$-shells.
     Shortly after, Warburton, Becker, and Brown  \cite{warburton1990},  made extensive shell-model calculations in the same model space
     reaching the same conclusions. And they coined the term "Island of Inversion", IoI, which has enjoyed a great success since then. Further work in the same context was carried out in  refs.~\cite{heyde1991,fukunishi1992}. Another important result was the discovery of the 
     doubly magic character of $^{34}$Si at ISOLDE \cite{baumann1989} which established the border in Z of the IoI.

\medskip
\noindent
      Since then, there has been a plethora of new experimental results. A very important milestone was the measurement 
      during a Coulomb-excitation experiment at RIKEN of the B(E2)  connecting the  2$^+$ with the ground state of $^{32}$Mg, which confirmed its well
      deformed nature corresponding to $\beta$=0.45 \cite{motobayashi1995}. Other Coulomb-excitation experiments followed soon \cite{pritychenko2000,iwasacki2001,yanagisawa2003,
      chiste2001,church2005,li2015}. A lot of work was made at ISOLDE. including mass measurements, radii isotope shifts, magnetic
      and quadrupole moments, spectroscopic factors etc., \cite{neyens2005,yordanov2007,himpe2008,neyens2011}
       and in other laboratories worldwide 
       \cite{mach1989,ibbotson1998,hofmann1990,nummela2001,mittig2002,iwasa2003,grevy2005,gade2007,terry2008,kanungo2010,hinohara2011}. The existence of an IoI
      at N=20 is now well established, even if  its boundaries are not fully delineated yet.
      Crucial experiments to establish the relationship between shape/phase coexistence and the
      onset of the IoI, where the discovery of the long sought excited 0$^+$ states in $^{32}$Mg
      \cite{wimmer2010} and $^{34}$Si \cite{rotaru2012}.

\medskip
\noindent      
      Indeed, the activity
      in the theory sector continued, both in the shell-model framework 
      \cite{caurier1998,utsuno1999,wood1999,caurier2001,brown2001,caurier2002}
      and with BMF methods \cite{terasaki1997,reinhard1999,lalazissis1999,peru2000}. The latter
      only got quantitative agreement with experiment after including the triaxial degrees of freedom, breaking time reversal invariance
      and allowing configuration mixing via the Generator Coordinate approximation \cite{rodriguez2002}.

\subsection{Valence space; the SPQR model in action}
\begin{figure}[H]
\begin{center}
\includegraphics[width=0.49\textwidth]{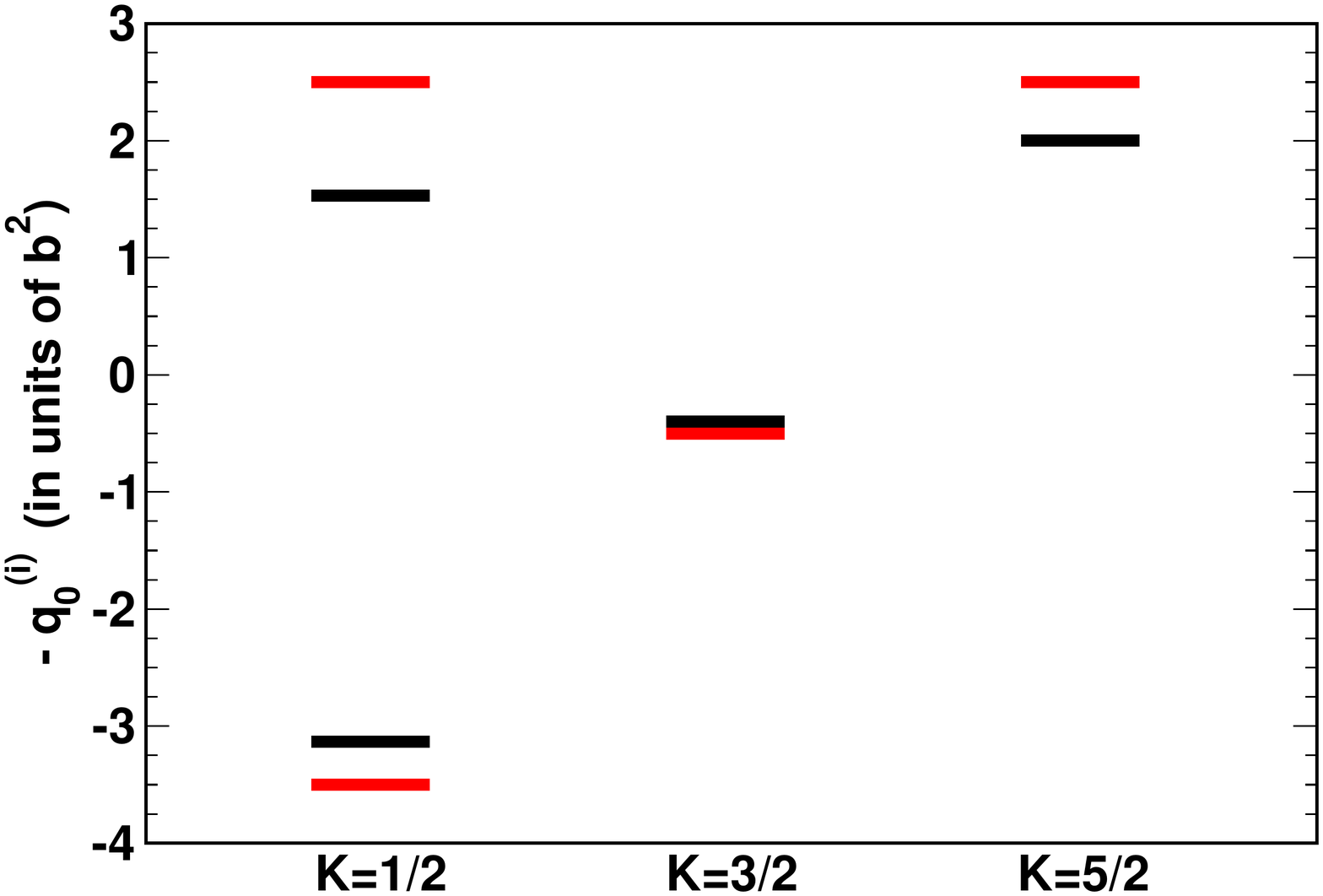}
\includegraphics[width=0.49\textwidth]{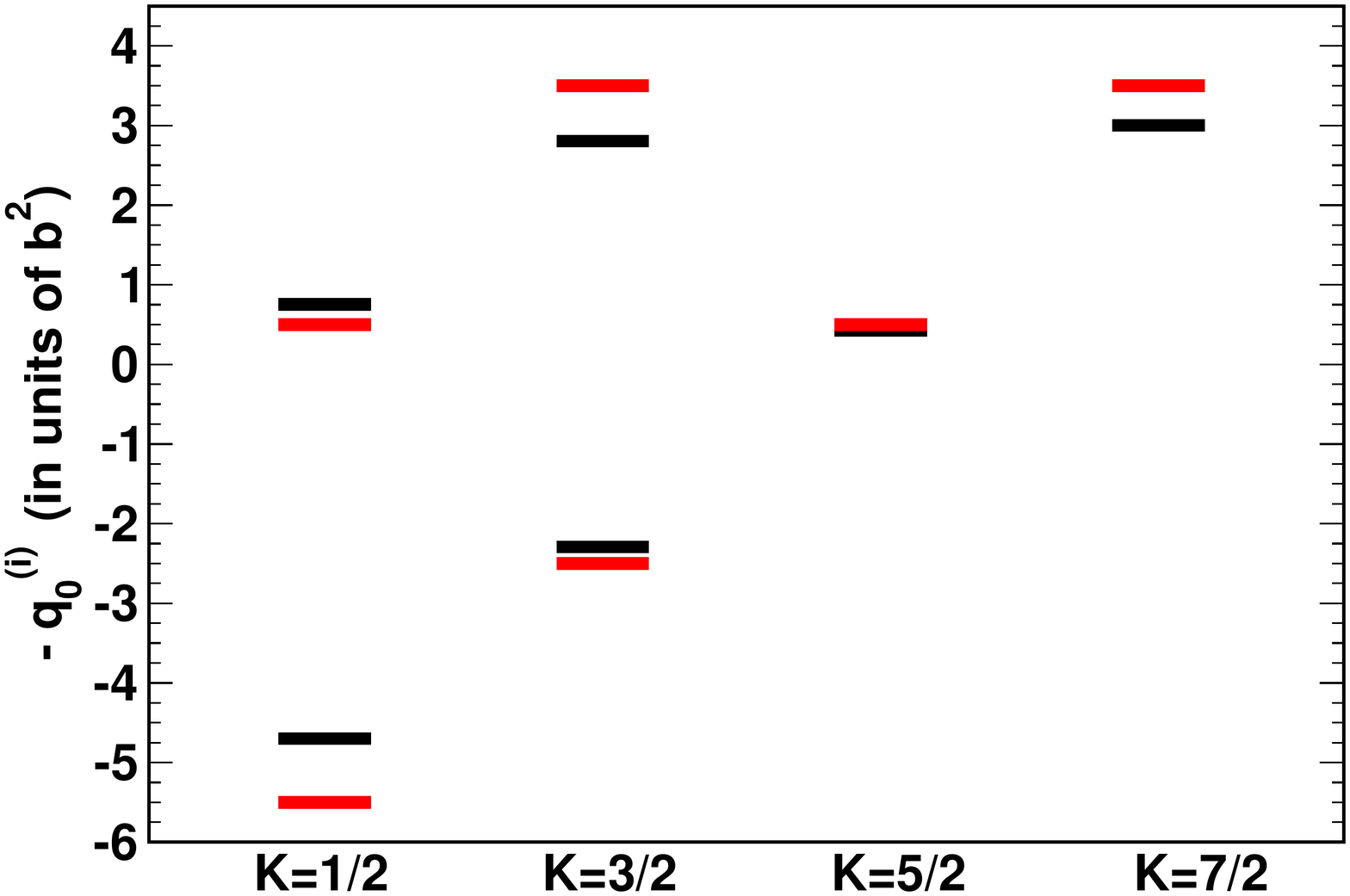}
\caption{(Left) ZRP diagrams for Quasi-SU(3) in the $sd$-shell. Red lines, schematic Quasi-SU3, black lines, 
 exact Quasi-SU3. (Right) The same for the $pf$-shell. \label{quasi-2}} 
\end{center}
\end{figure}

  The natural valence space for this region comprises the $sd$-shell for the neutrons and protons and in addition the
  $pf$-shell for the neutrons.
  Around N=20 the np-nh (neutron) intruder states in  $^{34}$Si and its lighter isotones 
 can take advantage of the presence two Quasi-SU3 sets of orbits, for the protons
 in the $sd$-shell and for the neutrons in the $pf$-shell (Fig. \ref{quasi-2}),
 as well as the Pseudo-SU3 set for the neutron holes in the $sd$ shell (Fig. \ref{su3}).
 This coupling scheme makes it possible to get large quadrupole moments and leads to  huge gains of correlation energy 
 (which go roughly as the product of the mass quadrupole moments of neutrons and protons, see  ref \cite{zuker2015}) which, in some cases,
 suffice to turn the intruders into ground states. Lets work out explicitly the case of the 0p-h, 2p-2h and 4p-4h
 neutron configurations in $^{32}$Mg. From figures  \ref{su3}  and \ref{quasi-2}, 
 it obtains  Q$_0$(0p-0h) = 10 $\chi_m$ b$^2$,  Q$_0$(0p-0h) = 24 $\chi_m$ b$^2$, and \mbox{Q$_0$(0p-0h) = 31 $\chi_m$ b$^2$}, where
 b is the HO length parameter and $\chi_m$ the quadrupole "effective" mass.
 This means that the maximum quadrupole energy gains of the three types of configurations are roughly in the ratio 1/6/10. With the
 value of the quadrupole coupling constant for his mass region, this translates into 2.5~MeV, 15~MeV and 25~MeV respectively. Indeed,
 the extra energy gains of the 2p-2h and 4p-4h maximally deformed configurations values are commensurable with the energy cost of promoting two or four neutrons across the N~=~20 closure. Therefore all the necessary conditions for the inversion of configurations are already
 disposable; that they become sufficient depends on subtle balances between the spherical mean field and the correlations.
 The SM-CI calculations utilise the effective interaction SDPFU-MIX described in ref.~\cite{caurier2014}.

\begin{figure}[h]
\begin{center}
\includegraphics[width=0.5\textwidth]{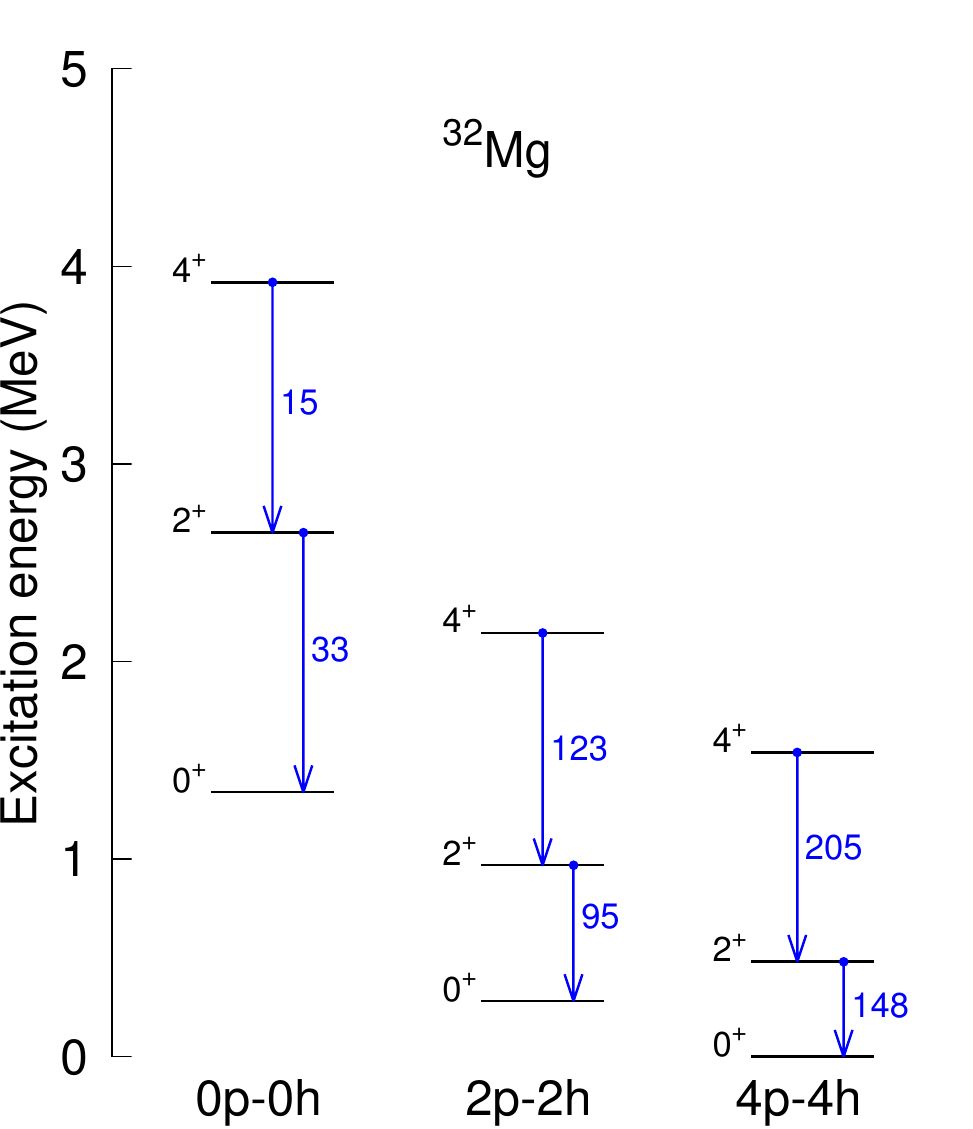}
\caption{Low energy spectra of $^{32}$Mg at fixed np-nh configuration with the SDPFU-MIX interaction. The numbers in the plot are the B(E2)(J $\rightarrow$ J-2) in e$^2$fm$^4$. \label{mg32tf}}
\end{center}
\end{figure}

\medskip
\noindent
  In Fig. \ref{mg32tf} we present the results of the diagonalizations at fixed number of particle-hole neutron excitations.
  From now on, effective electric charges q$_{\pi}$=1.31e and q$_{\nu}$=0.46e, microscopically derived in ref.~\cite{dufour1995}, will be used.
  Consistently with these values $\chi_m$=1.77. It is seen that, not so surprisingly in view  of the previous discussion, the lowest energy corresponds to the 4p-4h configuration, which, given the in-band B(E2) values could be called super-deformed. Next the 2p-2h deformed band shows up and finally the 0p-0h corresponding to the normal filling, N=20 closure. This is a textbook example of configuration inversion. 


\begin{figure}[H]
\begin{center}
\includegraphics[width=0.49\textwidth]{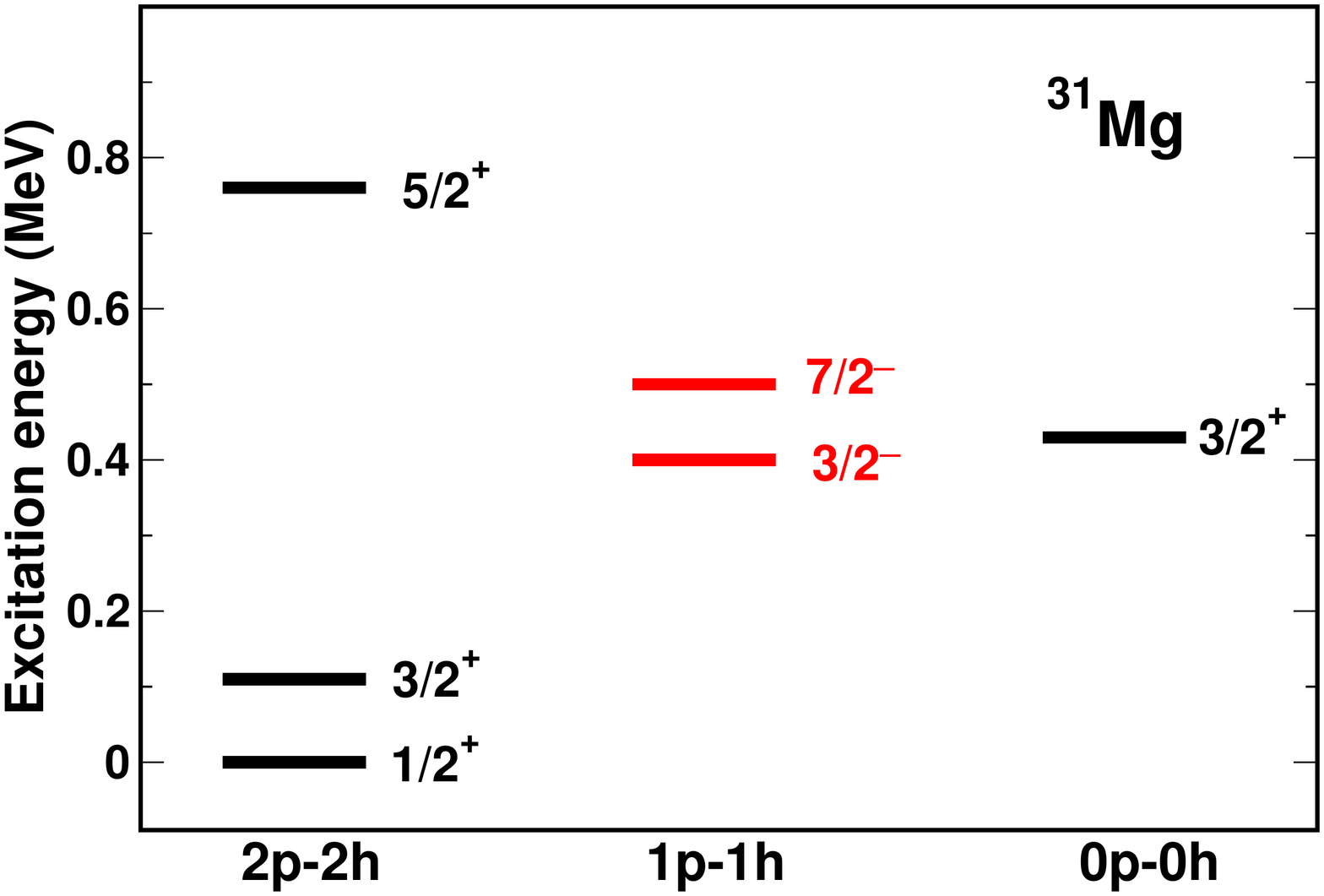}
\includegraphics[width=0.49\textwidth]{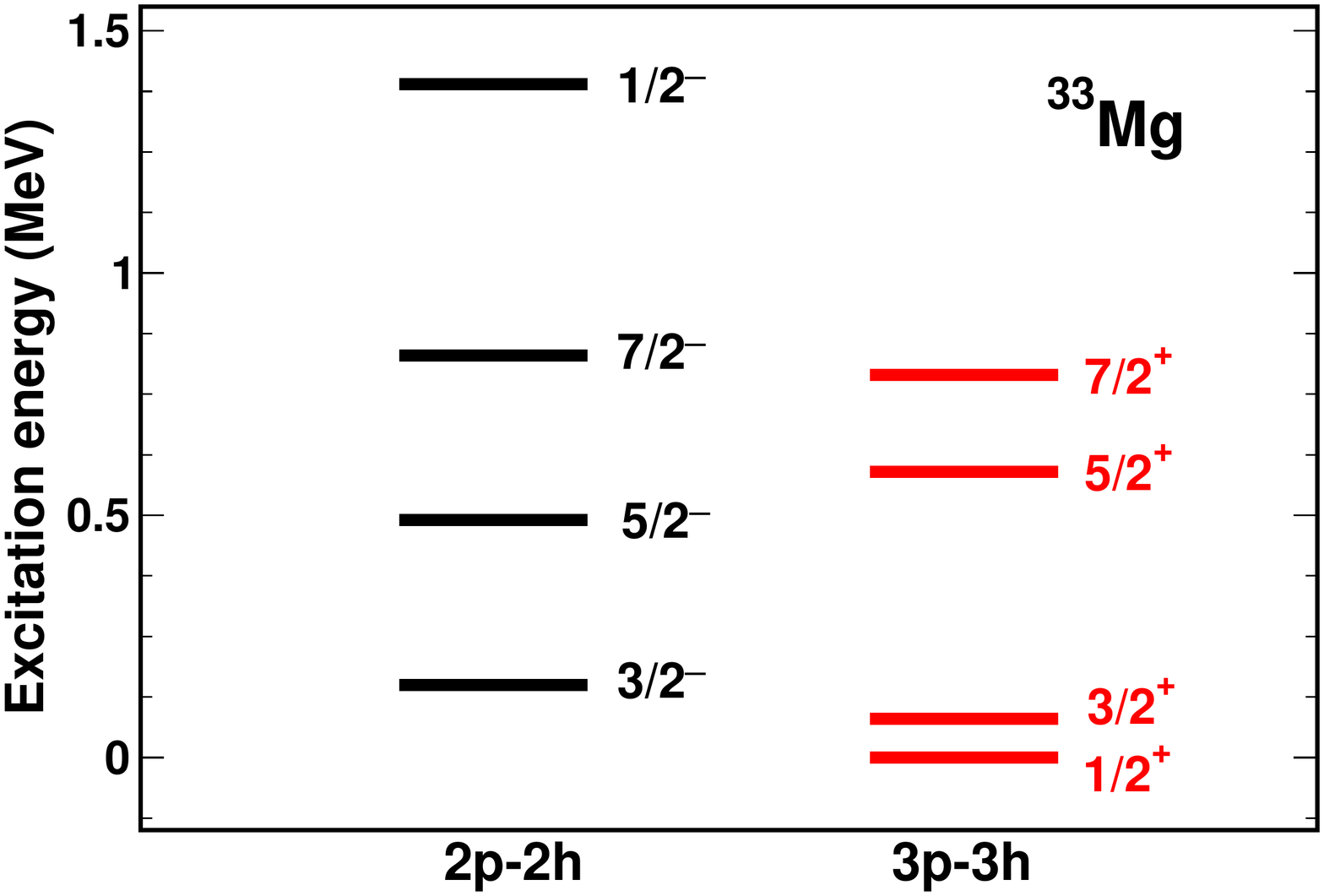}
\caption{Low energy spectra of $^{31}$Mg (left) and  $^{33}$Mg (right) at fixed np-nh configuration with the SDPFU-MIX interaction. \label{mg31tf}} 
\end{center}
\end{figure}

\medskip
\noindent
In Fig. \ref{mg31tf} we have plotted the results at fixed np-nh for $^{31}$Mg and $^{33}$Mg.
In the former the lowest configuration is 2p-2h giving rise to an yrast band of  K=1/2$^+$
as can be easily inferred from Fig. \ref{su3} (three neutrons in the Pseudo-SU3 of the $sd$-shell).
Experimentally the ground state is indeed a 1/2$^+$ \cite{neyens2005}.
The situation in $^{33}$Mg is different. In this case the competition is between
the 2p-2h negative parity and the 3p-3h positive parity configurations which are almost degenerated before mixing. The final result gives a 3/2$^-$ as a ground state in accord with the experimental result of ref. \cite{yordanov2007}
(the assignment of the experimental value went through a heated debate, see also refs. \cite{neyens2011,kanungo2010}).

\subsection{Flagship experimental and theoretical results at the N=20 IoI} 
 As the experimental and theoretical work in this region is too extensive to be fully covered in this review we have selected a few flagship cases to develop in detail the experimental challenge which were duly  overcome to unveil their structure, and the theoretical predictions available in the SM-CI framework.
 
\begin{figure}[h]
\begin{center}
\includegraphics[width=0.7\textwidth]{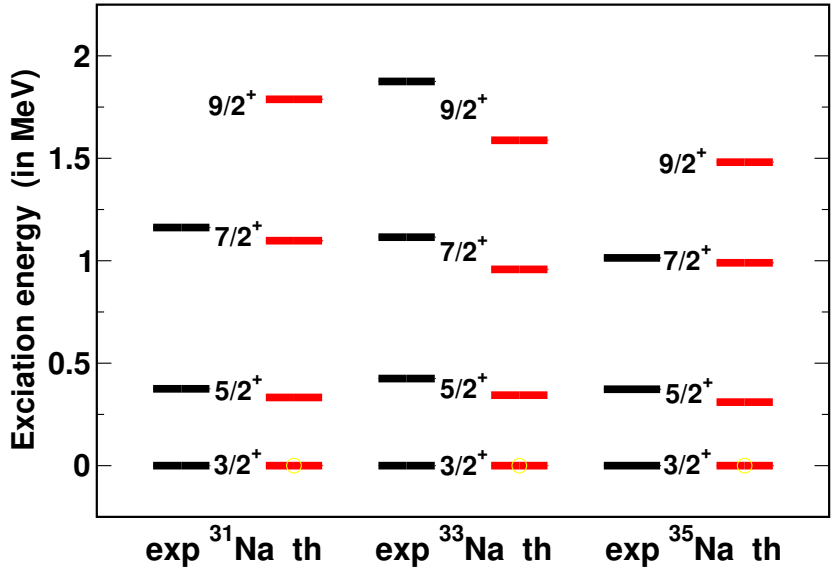}\\
\caption{Low energy spectra of  $^{31}$Na,  $^{33}$Na and  $^{35}$Na. Full calculation with the SDPFU-MIX interaction (red bars) compared with the experimental data (black bars). \label{na31}} 
\end{center}
\end{figure}

 \bigskip
 \noindent
 {\bf $^{\bf 31}$Na}. We start paying homage to the nucleus that started all this
 \cite{thibault1975}. There is a large amount of experimental results available at present, including the magnetic moment of the ground state $\mu$=+2.298~$\mu_N$, and its rms radius, 3.171 fm. In Fig. \ref{na31}, we show the coulomb excitation results of refs. \cite{doornenbal2010,gade2011}, compared with the predictions of the SDPFU-MIX interaction. The theoretical prediction $\mu$=+2.26~$\mu_N$ compares very well with the datum as do the excitation energies and quadrupole properties. That the yrast band is K=3/2$^+$ is straightforward from Fig. \ref{quasi-2}. The experimental B(E2) from the ground state to the excited 5/2$^+$ \cite{pritychenko2000} of 310(+170/-130)e$^2$fm$^4$,  is consistent with the calculated one, which corresponds to an intrinsic quadrupole moment Q$_0$=~65~efm$^2$. The agreement with the experiment extends to {\bf $^{\bf 33}$Na} and  {\bf $^{\bf 35}$Na} \cite{doornenbal2014} whose yrast bands have K=3/2$^+$ as well, and Q$_0$=~75~efm$^2$ and 73~efm$^2$ respectively.
Notice that the SPQR limits for the Q$_0$'s are 64~efm$^2$ for the 2p-2h configuration of {\bf $^{\bf 31}$Na},  65~efm$^2$ and 92~efm$^2$ for the
0p-0h and 2p-2h configurations of {\bf $^{\bf 33}$Na}, and 93~efm$^2$ and 87~efm$^2$ for the 0p0h and 2p-2h configurations of {\bf $^{\bf 35}$Na}.


\bigskip
\noindent
{\bf $^{\bf 32}$Mg}. This nucleus is probably the better studied and more fascinating of the N=20 region. After the pioneer experiments 
which measured its mass \cite{detraz1979}, the excitation energy of the 2$^+$ \cite{guillemaud1984}, it took some time to converge on an accepted experimental value for the reduced transition probability B(E2)($0^+ \rightarrow 2^+)$. The first measurement performed by T. Motobayashi and collaborators was obtained from intermediate-energy Coulomb excitation \cite{motobayashi1995}. A B(E2$\uparrow$) of 454(78) e$^2$fm$^4$ was extracted without considering a contribution from unobserved feeding. Several intermediate-energy Coulomb excitation measurements confirmed this result \cite{pritychenko2000,iwasacki2001,church2005} except one from GANIL \cite{chiste2001}. The correction to take into account unobserved feeding from high-lying states can be estimated, for example from Distorted Wave Born Approximation (DWBA) calculations. It leads to typical reductions of 10-25\% of the extracted value of the B(E2$\uparrow$). After nuclear excitation and feeding subtractions, the latest independent measurements give a value of 328(48) \cite{church2005} and 432(51) e$^2$fm$^4$ \cite{li2015}. The difference between the two values could come from a different estimate of the feeding correction: it is considered 25\% in the NSCL measurement of ref.~\cite{church2005} and 14\% in the RIKEN measurement at 195 MeV/nucleon of ref.~\cite{li2015}. As one would expect the feeding to increase with incident energy, these two corrections are in tension with each other.
It took then a while and some debates to find the yrast 4$^+$ state  
\cite{klotz1993,gelin2007,mattoon2007,tripathi2008,takeuchi2009}.
The band was recently continued up to the 6$^+$ in ref.~\cite{crawford2016}.

\begin{figure}[h]
\begin{center}
\includegraphics[width=0.7\textwidth]{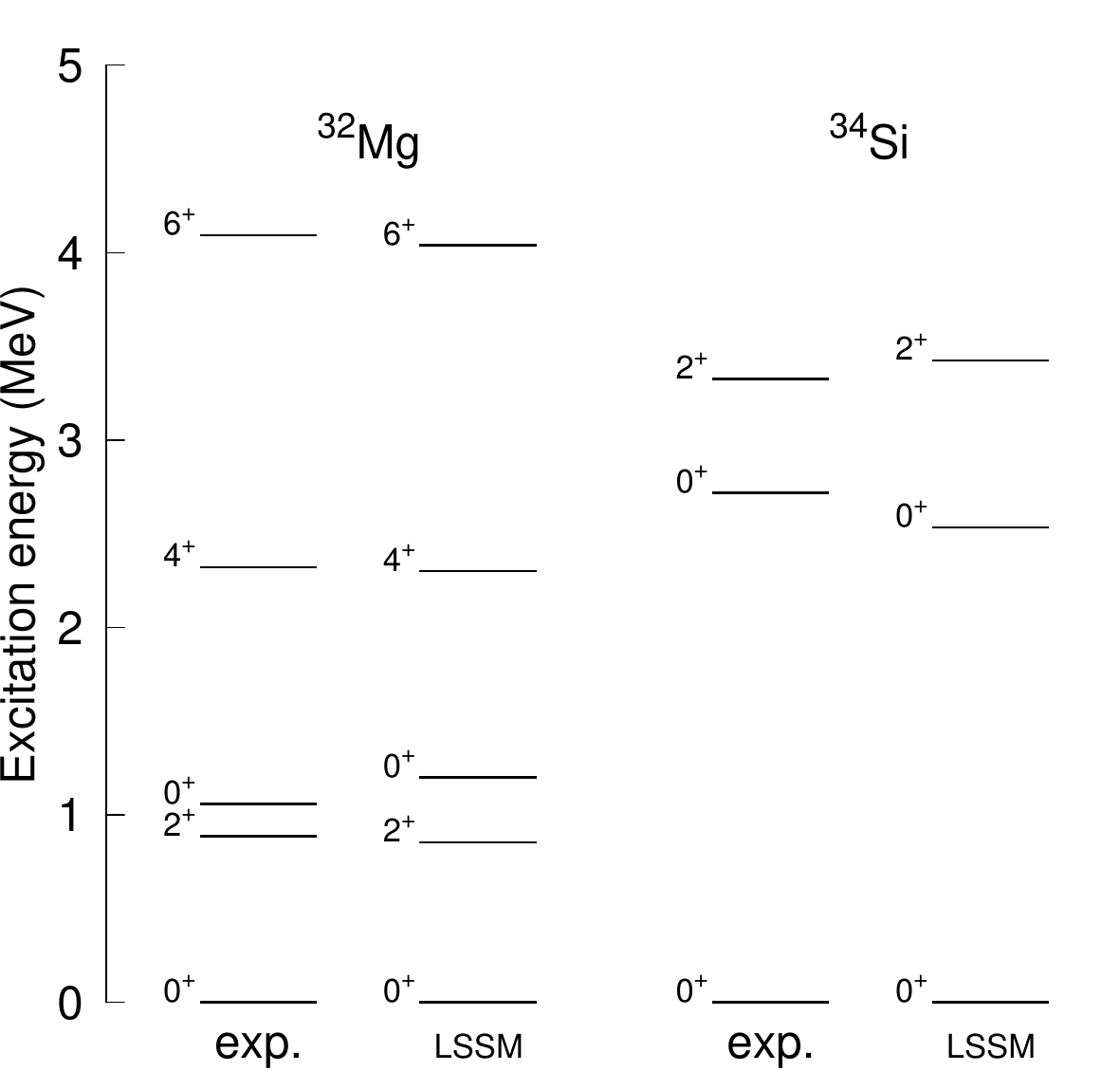}
\caption{Low energy spectra of  $^{32}$Mg and  $^{34}$Si. Full calculation with the SDPFU-MIX interaction compared with the experimental data \label{spec32-34}} 
\end{center}
\end{figure}

\medskip
\noindent
 All these experimental data are well reproduced by the SM-CI calculations with the SDPF-U-MIX interaction (see ref.~\cite{caurier2014}). The experimental and theoretical level schemes are compared in Fig. \ref{spec32-34}. It is seen that the agreement is excellent. In fact, no other calculation in the literature has been able to locate the excited 0$^+$ state close to its experimental energy. As we mentioned  apropos of Fig. \ref{mg32tf}, in addition to the 2p-2h intruders, the 4p-4h ones should play a crucial role in the structure of the 0$^+$ states. And, indeed. this is what the SM-CI results indicate. The ground state is 9\% 0p-0h, 54\% 2p-2h, 35\% 4p-4h and 1\% 6p-6h, thus, it is a mixture of deformed and superdeformed prolate shapes. However, the excited 0$^+$ has 33\% 0p-0h, 12\% 2p-2h, 54\% 4p-4h and 1\% 6p-6h, hence it is a surprising hybrid of a semi-magic configuration and  a superdeformed band-head, whose direct mixing matrix element is strictly zero. The first excited 0$^+$ state in $^{32}$Mg was measured at REX-ISOLDE by use of the two-neutron transfer $^{30}$Mg$(t,p)$ at 1.8 MeV/u \cite{wimmer2010}. The experiment combined gamma and particle spectroscopy, allowing the unambiguous assignment a  0$^+$ nature to the state measured at 1058 keV from recoil proton angular distribution, and a precise determination of the excitation energy of this 0$^+$ state from its partial decay branch $0^+_2 \rightarrow 2^+_1$. The setup was composed of the Miniball Ge array and the T-REX transfer setup. A feature of the measurement was the use of a tritium-loaded metallic Ti target with an atomic ratio $^3$H/Ti of 1.5 equivalent to a 40 $\mu$g/cm$^2$ tritium target. The target activity was 10 GBq, which is one of the limiting factors for the use of such targets. At a given moment there was some tension between the cross sections to the two 0$^+$ states and the very intruder character of the ground state of 0$^+$, when analysed in a two state model 0p-0h plus 2p-2h \cite{fortune2011}. Indeed the tension is fully relaxed when the superdeformed 4p-4h configuration is taken into account as explicitly demonstrated in ref.~\cite{macchiavelli2016} (see also ref.~\cite{elder2019}). 

\bigskip
\noindent
{\bf $^{\bf 34}$Si}. The doubly magic character of this nucleus was recognized
very soon, because of the the large excitation energy of its first 2$^+$ state at 3.326 MeV
measured by the team of ref.~\cite{baumann1989} in the $\beta^-$ decay of {\bf $^{\bf 34}$Al}
in an experiment carried out at ISOLDE. The abrupt change from doubly magic Silicon  to well deformed 
Magnesium, only two protons away, is a spectacular example of structural phase transition.
Two questions soon arose, where is the deformed structure in  {\bf $^{\bf 34}$Si}, and where is the
semi-magic 0$^+$ state in {\bf $^{\bf 32}$Mg}? We have answered the latter in the previous
paragraph; it is not the first excited 0$^+$ at 1.06 MeV. In fact according to the calculations 
there should be a third 0$^+$ at about 2~MeV dominated by the closed N=20 neutron configuration.
The former was searched for many years with contradictory or non reproducible claims, see refs.
\cite{nummela2001,mittig2002,iwasa2003,grevy2005}. Finally it was found by F. Rotaru and collaborators in a clever experiment performed at GANIL \cite{rotaru2012} at 2.719 MeV of excitation energy. Since the $0^+_2$ state had a large chance to be located below the 2$^+_1$ state at 3326 keV, which was confirmed by the experiment, it should have decayed to the ground state via an E0 transition through $e^+e^-$ internal pair creation if its excitation energy were larger than 1022 keV. The state was then looked after by use of electron spectroscopy combined to $\beta$ decay. Indeed, $^{34}$Al being at the edge of the island of inversion, it should exhibit both normal and intruder configurations at low excitation energy. At the time of the measurement, shell-model calculations predicted a 4$^-$ ground state and a 1$^+$ excited state. The excited state was expected to be a  $\beta$-decay isomer with a strong $2p-1h$ configuration favorably decaying to the intruder 0$^+$ state in $^{34}$Si. The experiment was extremely successful and a clear E0 transition was observed from a coincidence measurement of a $e^+e^-$ pair, revealing the intruder-configuration 0$^+$ state in $^{34}$Si at 2719(3) keV. The branching ratio for the decay of the 2$^+_1$ state to the newly-measured first excited 0$^+$ was also determined. Combined to the know B(E2;$2^+_1 \rightarrow 0^+_1$) \cite{ibbotson1998}, a very low value B(E2;$2^+_1 \rightarrow 0^+_2$) = 61(40) e$^2$fm$^4$ could be determined, confirming that the two states 2$^+_1$ and 0$^+_2$ are of different nature. The SM-CI calculations in the sd-pf valence space reproduce fairly well this level scheme (see Fig. \ref{spec32-34}). The calculated excited 0$^+$ is a deformed
(soft-oblate) 2p-2h plus 4p-4h state. Candidates to the 2$^+$ member of its band have been
 found in a recent experiment at ISOLDE \cite{lica2019}. Let's close with a few words on the
 abruptness of the transition. We have seen that in {\bf $^{\bf 32}$Mg} already the deformed band-heads of the 2p-2h and 4p-4h
 configurations were 1 MeV below the semi-magic state, and after mixing,
 this state is pushed-up by about  another MeV. In {\bf $^{\bf 34}$Si} instead, the doubly magic 0p-0h
 state is 1~MeV below the deformed 2p-2h  band-head, and the mixing pushes the deformed state upwards by about another MeV. All in all, a small reduction of the ESPE gap of about 1~MeV between Z=14 and Z=12 produces  the phase  transition.
 
 \medskip
 \noindent
  Besides the mixing of intruder configurations,  detailed recent studies have addressed the underlying shell structure and occupancies in the vicinity of {\bf $^{\bf 34}$Si}. The non-occupation of the $2s_\frac{1}{2}$ proton orbital leads to the conclusion that its  proton density has a 
  "bubble" structure and that this density-depletion has a clear effect on the neutron spin-orbit splitting in this nucleus \cite{Burgunder2014,Mutschler2016,Mutschler-Nature2017,Jongile2020,Sorlin2020}.
  
\begin{figure}
\hspace*{-1.5cm}
\begin{subfigure}[b]{0.49\textwidth}
\includegraphics[width=\textwidth]{ESPE-N20mix.pdf}
\vspace*{0cm}
\end{subfigure}
\begin{subfigure}[b]{0.5\textwidth}
\vspace*{4cm}
\includegraphics[width=\textwidth]{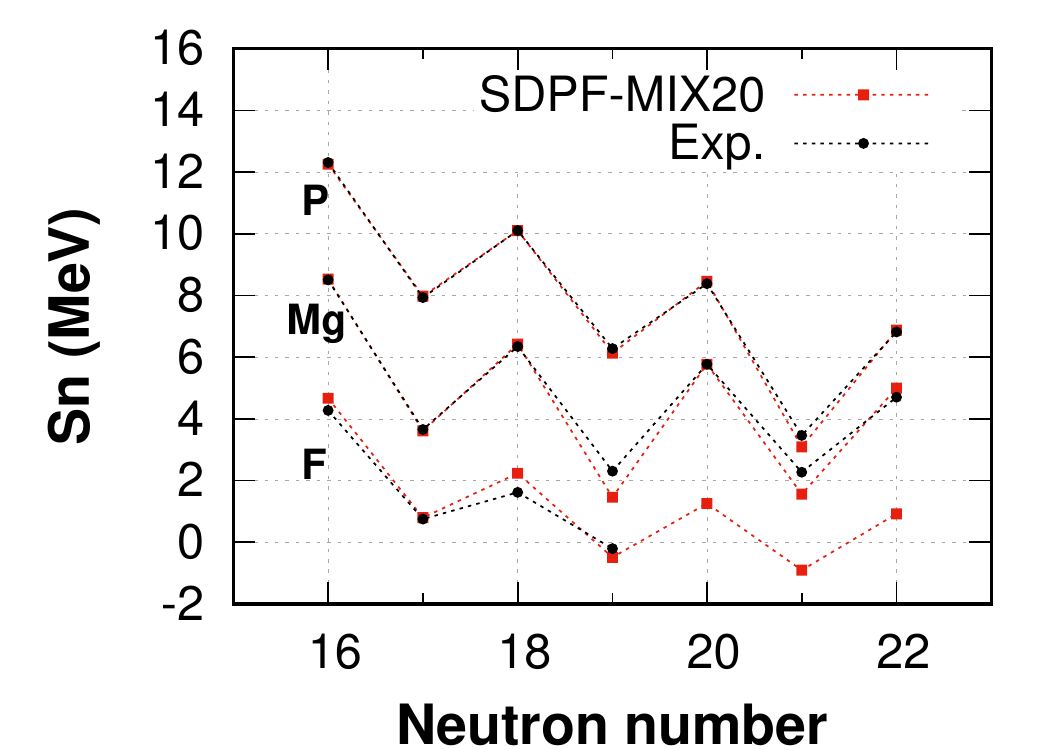}
\end{subfigure}
\caption{(Left)  Effective single particle energies with the SDPFU-MIX interaction. (Right) One neutron separation energies along the F, Mg  and P isotopic chains, SM-CI results compared with the experimental data \label{espe-20}} 
\end{figure}

 \medskip
\noindent
  The remaining N=20 members of the IoI are {\bf $^{\bf 30}$Ne} and (most probably) {\bf $^{\bf 29}$F}. We disregard for the moment {\bf $^{\bf 28}$O} which is unbound, although we will come back to it when discussing the valence proton-less members of the IoI's. The Neon isotopes behave very much as their corresponding Magnesium isotones, as it is seen in the experiments of refs. \cite{yanagisawa2003,doornenbal2016,doornenbal2009,murray2019,liu2017}, and in the SM-CI results of \cite{caurier2014}.  There is a peculiar aspect however, 
  which is due to the possible change in the ordering of the  0f$_{7/2}$ and  1p$_{3/2}$ orbits, which is supported by 
  the experimental data \cite{nakamura2009,takechi2012} and by the SM-CI calculations both of the Strasbourg-Madrid and of the Tokyo 
  shell-model schools.  With a large occupancy of the 1p$_{3/2}$ neutron orbit in the ground state of {\bf $^{\bf 31}$Ne}, and given its very low binding energy, a neutron halo might develop. A similar situation may hold in {\bf $^{\bf 29}$F}
  and {\bf $^{\bf 31}$F}. In addition, recent investigations show also that  the 1p$_{1/2}$ orbit is located below the  0f$_{7/2}$ orbit. This peculiarity gives a natural explanation to the weakening of the odd-even neutron separation energies at the drip line as shown in Figure~\ref{espe-20}. The latest determination of the  $S_n$ has been recently completed at N=19 for $^{\bf 28}$F \cite{revel2020}. The neutron drip line has been established very recently for the Fluorine  and Neon isotopes at N=32 and N=34 respectively, with their N=31 and N=33 isotopes being unbound \cite{ahn2019}.
In addition recent ISOLTRAP mass measurements at N=20 in the Magnesium chain \cite{ascher2019} show peculiar crossing  of $S_{2n}$ energies between  $^{\bf 33}$Mg and $^{\bf 34}$Al.

\begin{figure}[h]
\begin{center}
\includegraphics[width=0.49\textwidth]{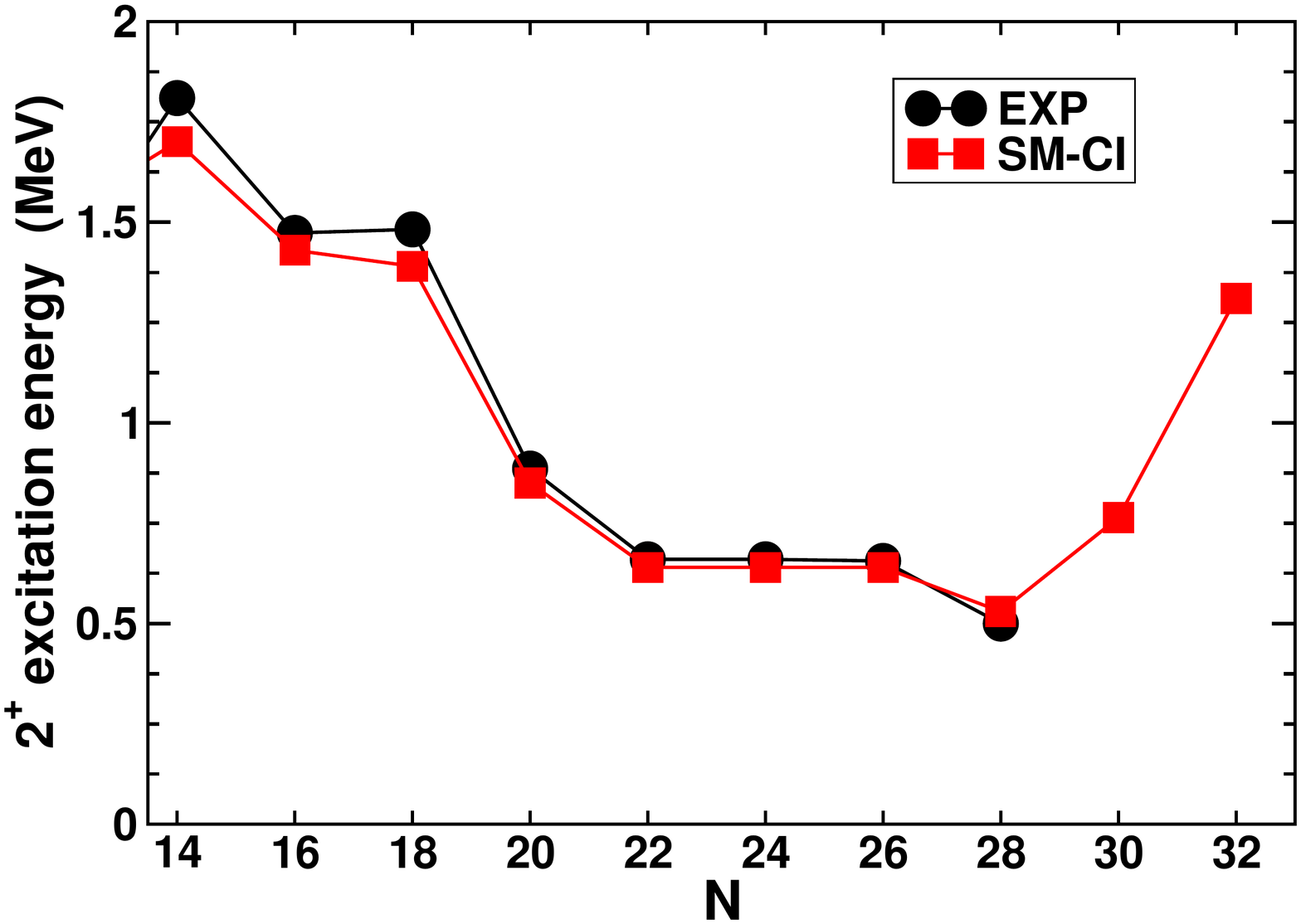}
\includegraphics[width=0.49\textwidth]{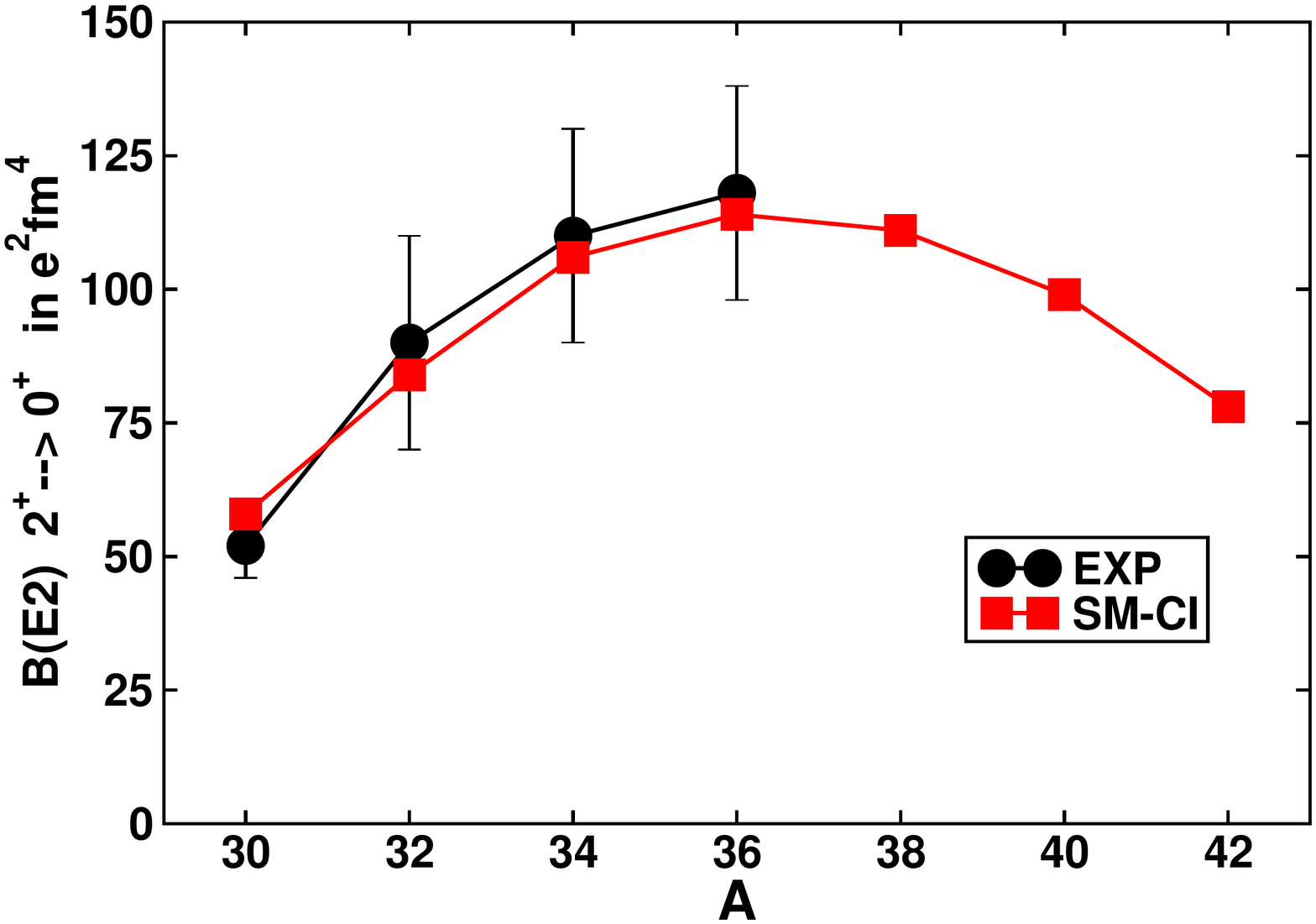}
\caption{(Left) Excitation energies of the first 2$^+$ in the Magnesium isotopes. (Right) B(E2)(2$^+ \rightarrow 0^+$). SM-CI calculations with the SDPFU-MIX interaction {\it vs} experiment.   \label{mg20-44}} 
\end{center}
\end{figure}

\medskip
\noindent  
  The experimental information on the Magnesium isotopes has been extended already until the next neutron shell closure at N=28 
  in experiments carried out at RIKEN (see refs.  \cite{doornenbal2016,doornenbal2013,crawford2014,crawford2019}). The results are gathered in
  Fig. \ref{mg20-44} and compared with the SM-CI predictions with the interaction SDPFU-MIX. Other effective interactions like SDPF-MU
  of the Tokyo group \cite{utsuno2012}produce similar results. Beyond the excellent agreement, what is evident in the figure is that the semi-magic shell closures at N=20 and N=28 are fully washed up, and deformation extends from {\bf $^{\bf 32}$Mg} to {\bf $^{\bf 40}$Mg}. This is borne out by the B(E2) values collected in Fig. \ref{mg20-44} as well, and is the fingerprint of the merging of the two IoI's at N=20 and N=28 in the Magnesium isotopes. Indeed, the models predict that the same merging would occur for the Neon and Sodium chains if they would extend until N=28. This is not the case for the Neon isotopes \cite{ahn2019}. In this reference the authors report one count of {\bf $^{\bf 39}$Na}, but indeed this requires further confirmation.
  {\bf $^{\bf 40}$Mg} is a very intriguing nucleus: it is probably the last Magnesium  isotope which is bound, it is well prolate deformed,
  it has a very low S$_{2N}$ and  a very large occupancy of the 1p$_{3/2}$ orbit which makes it a solid candidate to develop a neutron halo.  In fact  ref.~\cite{crawford2019} suggest that it has a rather exotic behaviour; it reports two  $\gamma$ lines at 500 keV and 670 keV. The first is the one that we have drawn in Fig. \ref{mg20-44}, but the second is difficult to interpret. It seems not to be the yrast $4^+ \rightarrow 2^+$ because the systematics of the Magnesium isotopes and all the calculations give it an energy of about 1.1~MeV. According to the authors, their previous experiment \cite{crawford2014} excludes a 
  $0^+ \rightarrow 2^+$ transition. The interesting  point is that the SM-CI calculation
  does predict such a second $\gamma$ in both {\bf $^{\bf 38}$Mg} and {\bf $^{\bf 42}$Mg} at the right energy as pertaining to the transition between the $2^+_{\gamma}$ (the band-head of the
  $\gamma$ band) and the yrast $2^+$. At N=28   this transition is absent because,
  contrary to his neighbours,  {\bf $^{\bf 40}$Mg} is axial instead of being triaxial. It is tempting to propose a model for {\bf $^{\bf 40}$Mg} as a pair of p-neutrons coupled to J=0,
   weakly attached to {\bf $^{\bf 38}$Mg}, and making a neutron halo. Indeed the halo p-orbit must have a very small overlap with the standard shell-model 1p$_{3/2}$ orbit.

\subsection{The IoI at N=28; {\bf $^{\bf 42}$Si} and its surroundings}

\begin{figure}[h]
\begin{center}
\includegraphics[width=0.49\textwidth]{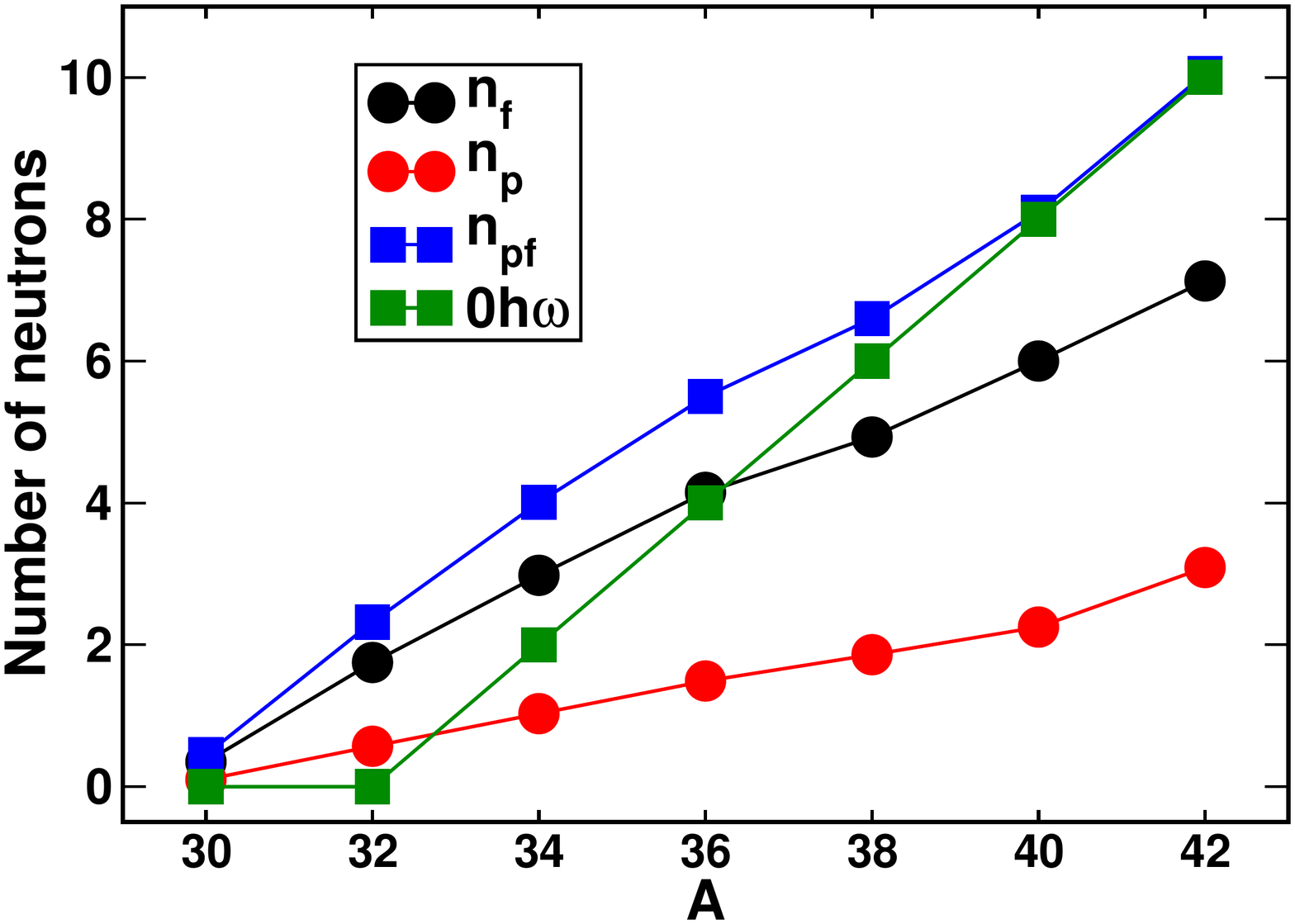}
\includegraphics[width=0.49\textwidth]{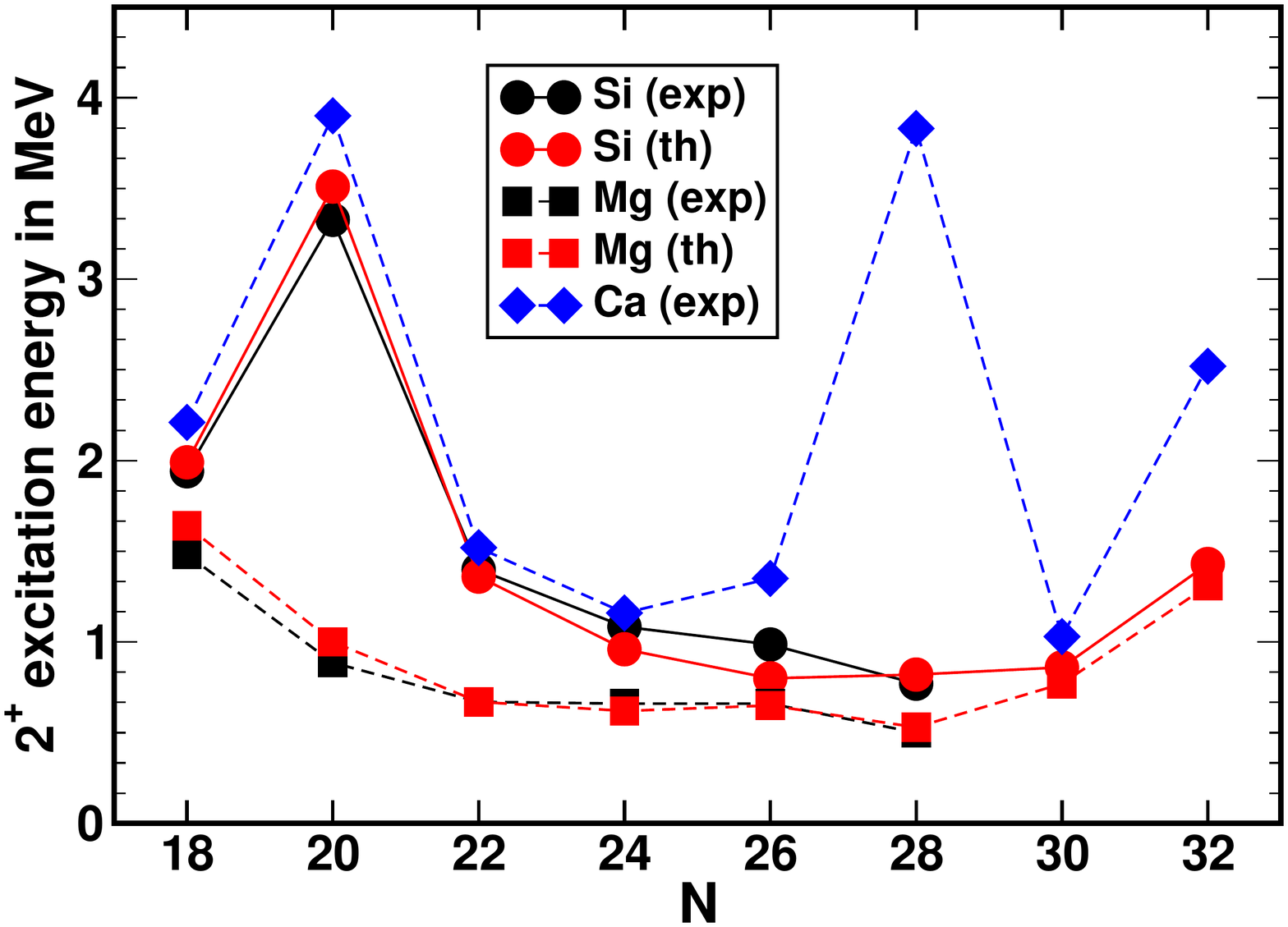}
 \caption{(Left panel). Occupation numbers of the neutron orbits in the chain of Magnesium isotopes. In green the values for closed N=20.
 (Right panel). Excitation energies of the first 2$^+$ in the Magnesium, Silicon, and Calcium isotopes, experiment {\it vs} SM-CI results.
\label{mg_20-28}} 
\end{center}
\end{figure}

The breaking of the N=20 and N=28 neutron shell closure is clearly seen in Fig. \ref{mg_20-28} where the occupation numbers of the neutron orbits in the Magnesium isotopes are plotted. At N=20 more than two neutrons occupy the $pf$ shell, dominantly in the 0f$_{7/2}$
orbit. As the neutron number increases the occupation of the 
1p$_{3/2}$ orbit increase and, at N=28, it contains two neutrons in average. The question now is
twofold; how to interpret this new IoI, and how far it extends. With respect to the SPQR scheme, we can consider now N=20 closed and the natural valence space
turns out to comprise two major HO shells, $sd$ for the protons and $pf$ for the neutrons. Neutron excitation's across N=28 are favored with respect to those across N=20 because the gap is smaller and np-nh excitation's with n odd are not parity forbidden. In the deformed limit neutrons adopt the Quasi-SU3 coupling scheme
depicted in Fig. \ref{quasi-2} and it is seen that for eight neutrons (N=28) the prolate and the oblate fillings of the Nilsson-SU3 orbits lead to quadrupole moments of similar size.
At N=28 the ESPE's of the proton orbits are at variance with the ones at N=20. In particular 
the Z=14 gap is reduced and the 1s$_{1/2}$ and 0d$_{3/2}$ orbits become degenerated. In the case
of {\bf $^{\bf 40}$Mg} the Quasi-SU3 filling of the protons also lead to oblate and prolate
quadrupole moments of similar size. For {\bf $^{\bf 42}$Si}
the oblate option is clearly preferred. For the heavier isotones the SPQR scheme does not give reliable predictions.
The SM-CI calculations of references \cite{nowacki2009} and \cite{utsuno2012} predict a prolate {\bf $^{\bf 40}$Mg}
and oblate {\bf $^{\bf 42}$Si}. Both  reproduce the experimental spectrum of {\bf $^{\bf 42}$Si}. 
The quest for the structure of {\bf $^{\bf 42}$Si} was tortuous. There were indications of a very low energy $\gamma$ in one experiment at GANIL
 \cite{grevy2005}, but simultaneously, the results of a knock out experiment at MSU were interpreted as a proof of its doubly magic nature \cite{fridmann2005}
 (and, by the way, published in Nature). A disclaimer followed soon \cite{fridmann2006}. More and more evidence of the breaking of the N=28 closure
 accumulated (see refs. \cite{campbell2006,campbell2007,jurado2007}). The discussion was settled with the publication of the GANIL data; the first excited 
 2$^+$ state has a very low excitation energy at 743~keV  \cite{bastin2007}, and doubly magicity is definitely excluded. A lot of experimental work
 in the region has ensued (see refs. \cite{gaudefroy2009a,derydt2010,force2010,santiago2011,takeuchi2012,stroberg2014})
 which confirms the existence of the IoI at N=28, connected to the N=20 IoI by the Magnesium's  isthmus. As  of today the yrast states
 2$^+$ at 743~keV and  4$^+$ at 2173~keV of {\bf $^{\bf 42}$Si} seem well understood. However the relative position of the oblate  and 
 prolate 0$^+$ band-heads seems to be interaction dependent, according to the findings of ref.~\cite{gade2019a}.


\noindent
To wrap up this section we have depicted  in the right panel of Fig. \ref{mg_20-28} the evolution of the 2$^+$ excitation energies along the Calcium, Silicon, and Magnesium isotopic chains, where it is seen very pictorially the persistence of the N=20 and N=28 neutron shell closures in the Calcium's, that of the N=20 one in the Silicon's, and the disappearance of both of them in the Magnesium's. It is seen that both in the Magnesium and Silicon isotopes
the orbits 0f$_{7/2}$ and 1p$_{3/2}$ behave as a super-orbit, and therefore the closure effects do not show up until N=32. The situation is completely different in the Calcium isotopes which preserve to a large extent the  purity of the single particle structure of the naive spherical mean field, as we shall discuss in the next section.

\section{The heavy Calcium and Potassium isotopes}
   Initially, the heavy Potassium's, were mainly explored
   by the Strasbourg members of the ISOLDE collaboration. $^{52}$K was produced for the first time  at ISOLDE. It's
   decay produced $^{52}$Ca for the first time as well \cite{huck1985}, whose spectroscopy was very interesting, with a  first 2$^+$
   excited state at 2.56~MeV, a rather high energy. However, the authors didn't dare to claim that it was, as nowadays is accepted, a
    new doubly magic nuclei, which anticipated the appearance of new magic numbers far from stability. 
    Only recently has $^{52}$Ca come back to the scene, fostered by the debate about the doubly magic character of $^{54}$Ca. 
    The results of two new ISOLDE experiments related to it have been published in Nature, one exploring the masses \cite{wienholtz2013}  and another
     the radii isotope shifts \cite{garcia2016} of the heaviest calcium isotopes.  Previously, the  shifts  of radii of the potassium isotopes beyond
     N=28, measured at ISOLDE as well, had been published in ref.~\cite{kreim2014}. Reference~\cite{garcia2016} extends our knowledge of the calcium radii
     up to $^{52}$Ca. Its basic result is that beyond N=28 and up to N=32 the radii increase linearly, with a slope which is far larger that
     all the available theoretical predictions (by the way, the same behaviour holds in the potassium isotopes  \cite{kreim2014,koszorus2021}). That the proton radius
     increases with the neutron excess is well understood as due to the mechanism of equalization of neutron and proton radii; the problem is why
      the slope changes so much when neutrons start filling the orbit 1p$_{3/2}$. A very appealing explanation of this behaviour has been 
      presented in ref.~\cite{bonnard2016}, which surmises that the radius of the $p$ orbits is much larger than expected, rather halo-like, 
      independently of their proximity to the neutron emission threshold. Obviously this mechanism is not directly related to the doubly magic nature
      of  $^{52}$Ca, because it only  depends of the occupancies of the $p$ orbits. This has clearly been overlooked by the authors of ref.~\cite{garcia2016} 
      which write (sic)  {\it The large and unexpected increase of the size of the neutron-rich calcium isotopes beyond N = 28 challenges the
       doubly magic nature of $^{52}$Ca and opens new intriguing questions on the evolution of nuclear sizes away from stability, which are of 
       importance for our understanding of neutron-rich atomic nuclei} (the same arguments are given in ref. \cite{koszorus2021} which extends the 
       Potassium isotope shifts to A=52). In fact,  the linear increase of the radius of  $^{52}$Ca is due to its doubly
       magic nature. At odds with the statement above,  only a very large occupancy of the  0f$_{5/2}$ orbit, i.e. a non magic configuration, can 
       produce a smaller radius of  $^{52}$Ca. Therefore, its (local) doubly magicity is not at stake. The more so in view of the measure of the
       mass of  $^{54}$Ca  in ref.~ \cite{wienholtz2013} which shows a clear change of slope in the S$_{2n}$ curve between N=32 and N=34. The N=32 magicity is
       supported as well  by the recent mass measurements of the heavy Scandium isotopes \cite{xu2019,leistenschneider2021}, that show the persistence of the N=32 shell gap therein. Notice that the later reference reports the disappearance of the N=34 gap already in the Scandium chain.
       Another important experimental result has been recently published in reference \cite{tanaka2020}, a measure of the mass radii isotope shifts of the
       very neutron rich Calcium isotopes. It is seen that the extracted neutron isotope shift radii also exhibit a pronounced kink beyond N=28
       with a large increase of its slope. Whereas this can be consistent with the proposal of reference \cite{bonnard2016}, another explanation has been
       put forward in reference \cite{horiuchi2020} which does not require an abnormal size of the p-orbits. According to this mean field analysis,
       the mechanism of equalization of the proton and neutron radii (anti-neutron-skin effect), in very neutron rich nuclei can be blocked after the Spin-Orbit 
       magic neutron closures, because the the next orbit to be filled (the 1p$_{3/2}$ in the Calciums) has a node and therefore
       contributes more to the core density than the 0f$_{7/2}$ which is peaked at the surface. If the core density is saturated already,
       the anti-neutron-skin solution is not energetically favorable anymore, and neutrons and protons tend to have similar densities and therefore a large neutron skin shows up.
      
\medskip
\noindent
We turn now to $^{54}$Ca itself. How to interpret the experimental value \cite{steppenbeck2013} of the excitation energy of  its first 2$^+$, at 2.04~MeV? Among the several candidates to doubly magic in the calcium isotopes, $^{34}$Ca (3.3~MeV,  taken from its mirror $^{34}$Si), $^{36}$Ca (3.1~MeV), $^{40}$Ca (3.9~MeV), $^{48}$Ca (3.8~MeV), and $^{52}$Ca (2.6~MeV), it has the lowest 2$^+$. In $^{56}$Ca, all the $pf$-shell interactions locate the  2$^+$ at an excitation energy of about 1~MeV (notice however that at this neutron number the effect of the mixing with the 0g$_{9/2}$ orbit can lead to an important increase of this excitation energy). Therefore,  at N~=~34 there will be no peak in the  2$^+$ energy signalling a magic neutron number, only perhaps a shoulder, similar to the one at N~=~16, corresponding to the filling of the orbit 1s$_{1/2}$, or even a plateau. Nonetheless, all the shell-model calculations with different effective interactions predict that the ground state of $^{54}$Ca is dominated to better than 90\% by the neutron configuration      (0f$_{7/2}$)$^8$:(1p$_{3/2}$)$^4$:(1p$_{1/2}$)$^2$ (and that independently of their prediction for the excitation energy of the 2$^+$). Should we give it the doubly magic label, or decide that it only shows a local closure of the 1p$_{1/2}$  orbit favoured by the weak neutron neutron interaction between these orbits?  

\begin{figure}
\begin{center}
\includegraphics[width=0.5\textwidth]{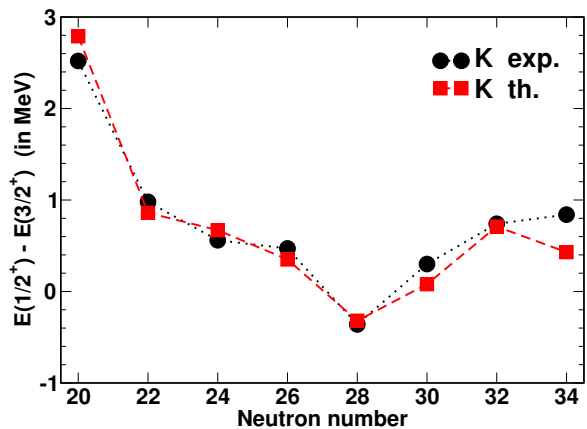}
\end{center}
\caption{\label{k_d3s1} Evolution of the splitting between the lowest 1/2$^+$ and 3/2$^+$ states in the Potassium
                          isotopes. Comparison of the experimental data with the predictions of the {\sc SDPF-U} interaction \cite{nowacki2009}. }
\end{figure}

\medskip          
\noindent         
Concerning the Potassium isotopes, the studies at ISOLDE were fundamental to understand the evolution of the effective single
           particle energies (ESPE) of the proton orbits  1s$_{1/2}$ and 0d$_{3/2}$  when the neutron number increases from N=20 to
           N=28 and above it.  As it is seen in Figure~\ref{k_d3s1} at N=20 the 0d$_{3/2}$ orbit is almost 3~MeV above the  1s$_{1/2}$.
           As neutrons are added in the orbit 0f$_{7/2}$ their splitting  decreases linearly, and in $^{47}$K at N=28 they are
           inverted and 
           quasi degenerated. When neutrons start filling the 1p$_{3/2}$  orbit  the effect is the opposite, and the normal ordering is re-established
           \cite{walter1989}.
           This result predates the discussions of  the "shell evolution"  \cite{sorlin2008}  in terms of the different components of the monopole Hamiltonian 
           (central, tensor and spin-orbit) and, in particular, plays a crucial role in the structure of the very neutron 
           rich N=28 isotopes. Recently the systematics has been extended up to $^{53}$K and the normal ordering of the two orbits persists 
           \cite{sun2020}. The same behaviour shows up at N=28  in $^{45}$Cl  and   $^{43}$P  confirming the
           monopole nature of the effect \cite{gade2006,gaudefroy2006}.  We have gathered all these results in Fig. \ref{k_d3s1}.


\section{From  $^{68}$Ni towards $^{60}$Ca:  The 4$^{th}$ IoI at N=40} 
 The doubly magic status of $^{68}$Ni has been discussed since long \cite{bernas1982}.  In fact, the HO shell closure at Z=40 behaves as such only when a spin-orbit closure for the neutrons reinforces it, as N=50 does in $^{90}$Zr. Could it be also the case of the combination of N=40 and Z=28?
Controversial views can be found in refs. \cite{grawe2001,sorlin2002,bree2008,broda1995,mueller2000,langanke2003}. In addition, evidences were accumulating pointing to an increase of collectivity in the
Iron and Chromium isotopes approaching  N=40. There was an early prediction based in the primitive version of the SPQR heuristics \cite{zuker1995}  of $^{64}$Cr being deformed,
and therefore pertaining to a new IoI \cite{poves1996}. Experimental and theoretical work followed which sustained this claim \cite{caurier2002,hannawald1999,sorlin2003}.

 \subsection{SPQR and the valence space}

\begin{figure}[h]
\begin{center}
\includegraphics[width=0.49\textwidth]{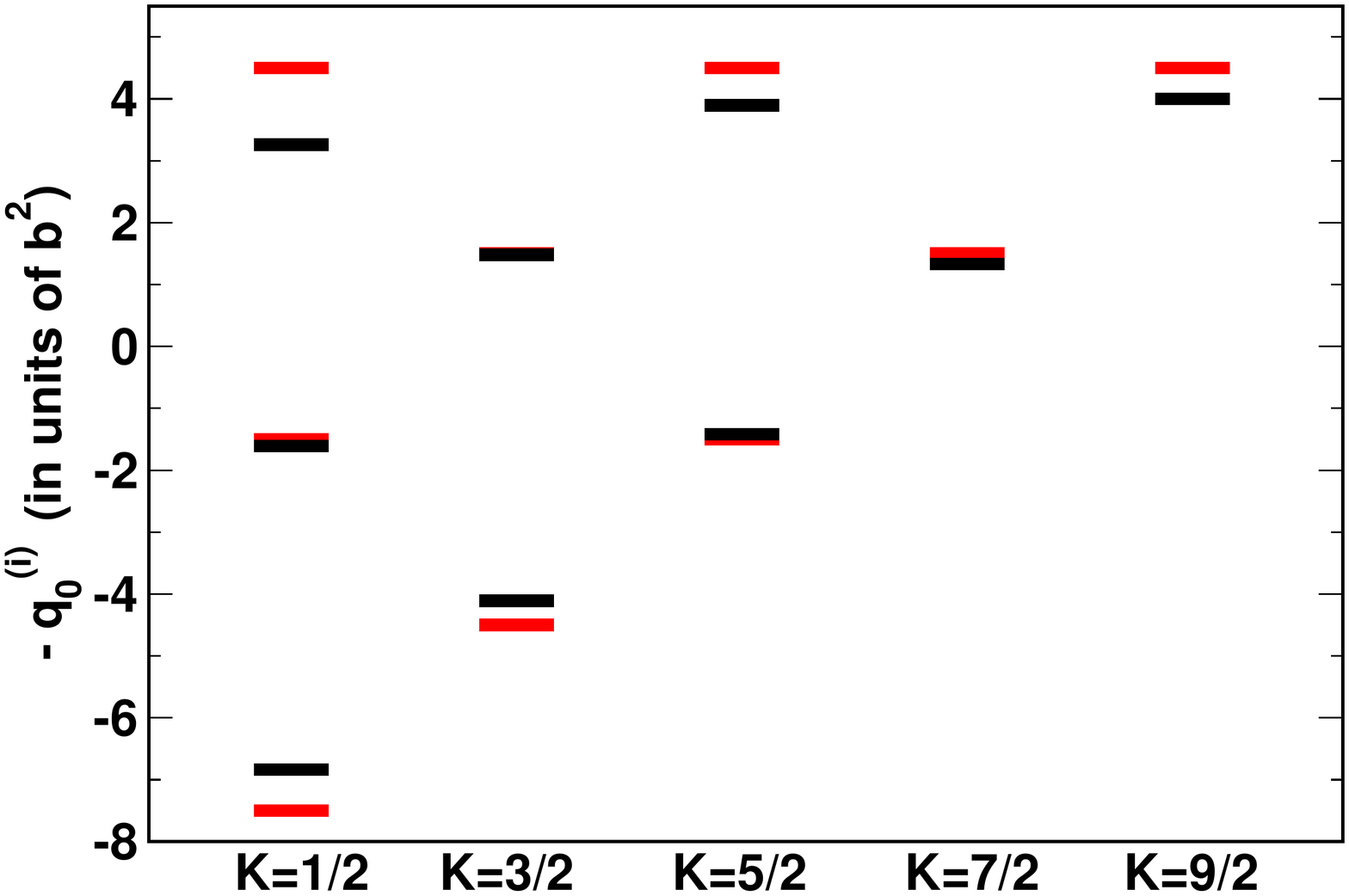}
\includegraphics[width=0.49\textwidth]{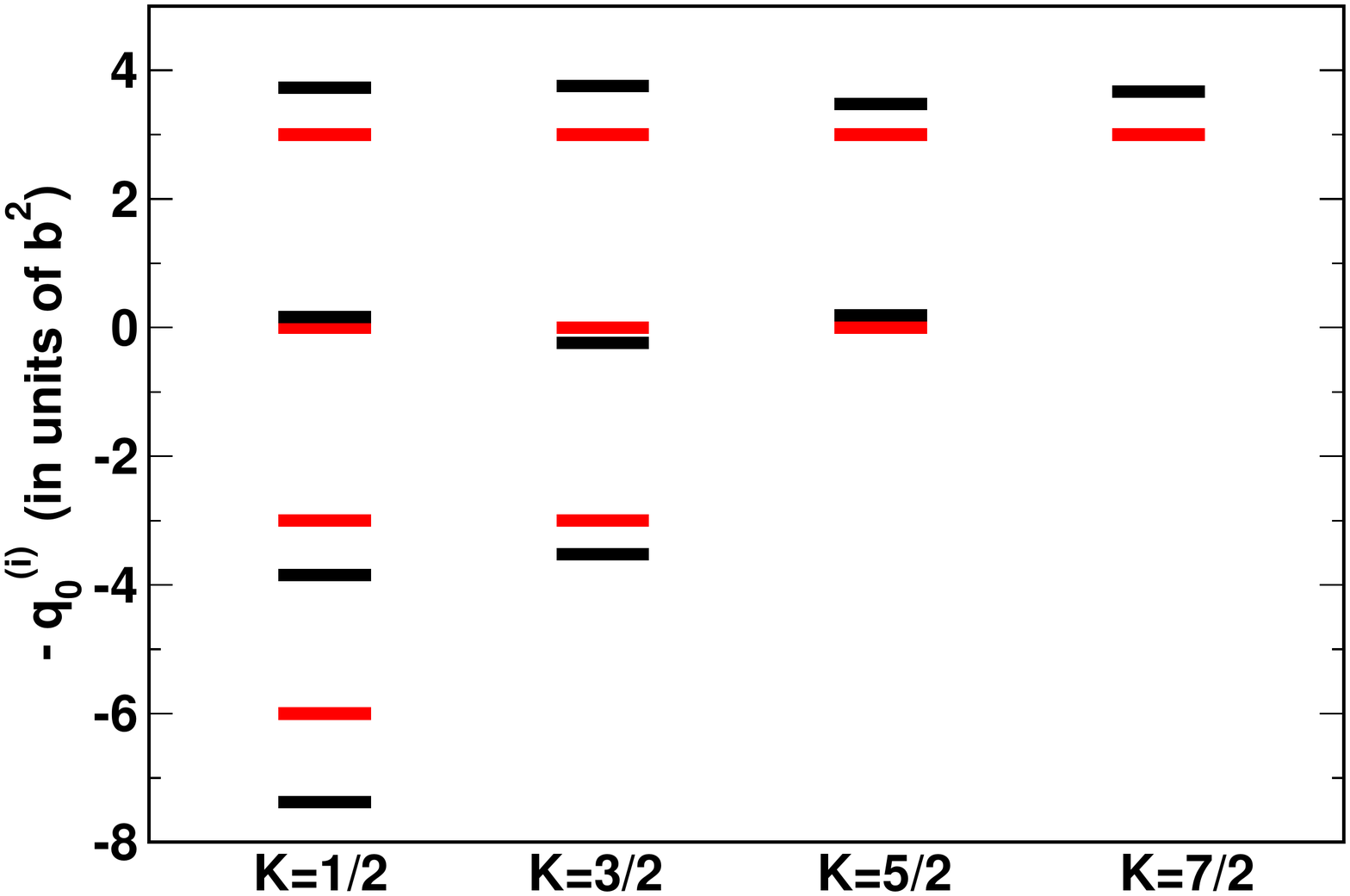}
\caption{(Left) Exact Quasi-SU(3) Nilsson levels for the  $sdg$-shell
(black lines): Schematic Quasi-SU3 (red lines). (Right) Pseudo-SU(3) Nilsson levels for the $sdg$-shell (black lines): SU(3) Nilsson levels for the $pf$-shell (red lines). 
 \label{pseudo-4}} 
\end{center}
\end{figure}

 The patient reader knows already the rules of SPQR, but we dare to remember them again shortly. The valence space relevant to this region  consists of the $pf$-shell for the protons and the quasi-spin triplet 1p$_{3/2}$, 0f$_{5/2}$, and 1p$_{1/2}$ plus
 the $\Delta j$=2, $\Delta l$=2, Quasi-SU3 sequence 0g$_{9/2}$, 1d$_{5/2}$, and 2s$_{1/2}$
 for the neutrons. Correspondingly, np-nh configurations may reach the regime of quadrupole dominance due to the availability of two Quasi-SU3 and one Pseudo-SU3 sets of orbits. The Nilsson SU3 diagrams to be used are plotted in Figures  \ref{su3}, \ref{quasi-2}, and \ref{pseudo-4}. The quadrupole moment of the doubly magic configuration of  $^{68}$Ni is obviously zero.  A 4p-4h neutron configuration
  and 2p-2h proton configuration will be penalised by about four times the N=40 gap plus two times the N=28 one. Naively, this amounts to about 20 MeV, albeit two-body monopole terms (CD-SE) reduce it to about 12~MeV. However, these configurations have large quadrupole moments in the SPQR limit, and therefore large quadrupole-quadrupole energy gains, which,  at leading order, are proportional to Q$_0^{\pi}$ x Q$_0^{\nu}$, with non negligible contributions of Q$_0^{\pi}$ x Q$_0^{\pi}$ and Q$_0^{\nu}$ x Q$_0^{\nu}$ (Notice that in this case the mass quadrupole moments should be used). Therefore, the usual competition between normal and intruder states is to be expected already at  $^{68}$Ni. For Z$<$28 two effects favour energetically the intruder (deformed) states; the presence of valence protons in the normal configurations and the reduction of the N=40 gap approaching Z=20. Everything seems to point to another IoI. We shall promptly see that the experimental data and the SM-CI calculations borne out this surmise. A very important aspect to underline here is that whereas SM-CI calculations in valence spaces which do not incorporate the quadrupole partners of the  0g$_{9/2}$ orbit, 1d$_{5/2}$, and 2s$_{1/2}$, can cope with the spectroscopy of the normal states of the Nickel isotopes, they fail completely in the description of the deformed intruders and {\it a fortiori} they do not predict the occurrence of the fourth IoI \cite{kaneko2008}.

\bigskip
\noindent
Assuming a 4p-4h  configuration for the neutrons and a 2p-2h one for the protons, the SPQR predictions for the mass quadrupole moments at N=40 are:

\begin{itemize}
      \item  Z=28 ; Q$_0^{\nu}$ = 220~fm$^2$; Q$_0^{\pi}$ = 99~fm$^2$
      \item  Z=26 ; Q$_0^{\nu}$ = 220~fm$^2$; Q$_0^{\pi}$ = 110~fm$^2$
      \item  Z=24  ; Q$_0^{\nu}$ = 220~fm$^2$; Q$_0^{\pi}$ = 117~fm$^2$
      \item  Z=22  ; Q$_0^{\nu}$ = 220~fm$^2$; Q$_0^{\pi}$ = 85~fm$^2$
        \item  Z=20 ; Q$_0^{\nu}$ = 220~fm$^2$; Q$_0^{\pi}$ = 0~fm$^2$
\end{itemize}

\noindent
It is seen that the maximum deformation is expected at mid proton shell in $^{64}$Cr. However a substantial proton contribution
occurs for all the isotopes except, obviously, for $^{60}$Ca. Whether these intruder configurations turn out becoming dominant
in each ground state depend, as already discussed of the competition between their monopole energy losses and their quadrupole
quadrupole energy gains, which we proceed to discuss in the following sections.

\subsection{Flagship experimental and theoretical results} 
\subsubsection{$^{68}$Ni}
The N=40 sub-shell closure has been actively investigated in the recent years, especially in the region of {\bf $^{68}$Ni}. 
The excitation energy of the 2$^+$ was first measured at 2033 MeV higher than neighbouring Ni isotopes, from a deep inelastic measurement where the identification of the emitting nucleus was based on a shell-model-educated guess of the existence of a 5$^-$ isomer and a $5^- \rightarrow 2^+ \rightarrow 0+$ cascade  \cite{broda1995}, and then confirmed from $\beta$ decay \cite{mueller2000} and Coulomb excitation \cite{sorlin2002}. The corresponding electromagnetic transition probability B(E2;$0^+ \rightarrow 2^+$) of the first excited state of $^{68}$Ni was measured to be relatively low at 255$\pm$60 e$^2$ fm$^4$ from intermediate-energy Coulomb excitation  \cite{sorlin2002}, confirmed later on by a safe-energy Coulomb-excitation measurement with reduced statistics \cite{bree2008}. When compared to those of neighboring Ni isotopes, they suggest that the N~=~40 gap between the $\nu$-pf shell and the $\nu$-g$_{9/2}$ orbital stabilizes the nucleus in a spherical shape. It has been suggested that N~=~40 may be magic far away from stability~\cite{sorlin2002}, whereas other interpretation disputed this conclusion ~\cite{langanke2003}. Experimentally, one does not observe a stabilizing effect from the pf-g$_{9/2}$ gap in other neutron-rich N~=~40 nuclei, as can be seen in the 2$^+$ energies of even-mass Fe and Cr isotopes.

\medskip
\noindent
Three 0$^+$ states have been observed so far in the low-energy spectrum of $^{68}$Ni, demonstrating the complex nature of this nucleus. The second 0$^+$ state is the first excited state of $^{68}$Ni with an excitation energy remeasured at 1604 keV \cite{recchia2013,flavigny2015,suchyta2014} with a half-life t1/2 = 270(5) ns \cite{sorlin2002}. The third 0$^+$ state at 2511~keV was first observed in a
$\beta$-decay experiment \cite{mueller2000} and confirmed later from in-beam gamma spectroscopy after multi-nucleon transfer between a $^{70}$Zn beam and $^{238}$U, $^{208}$Pb and $^{197}$Au targets \cite{chiara2012}. It was concluded from the non-observation of the third 0$^+$ in the reaction $^{66}Ni(t,p)$ that it has a significant proton-excitation component \cite{flavigny2019}, see also references \cite{rother2011,dijon2012}.

\begin{figure}[h]
\begin{center}
\includegraphics[trim=2cm 3cm 3cm 2cm,clip,width=0.8\textwidth]{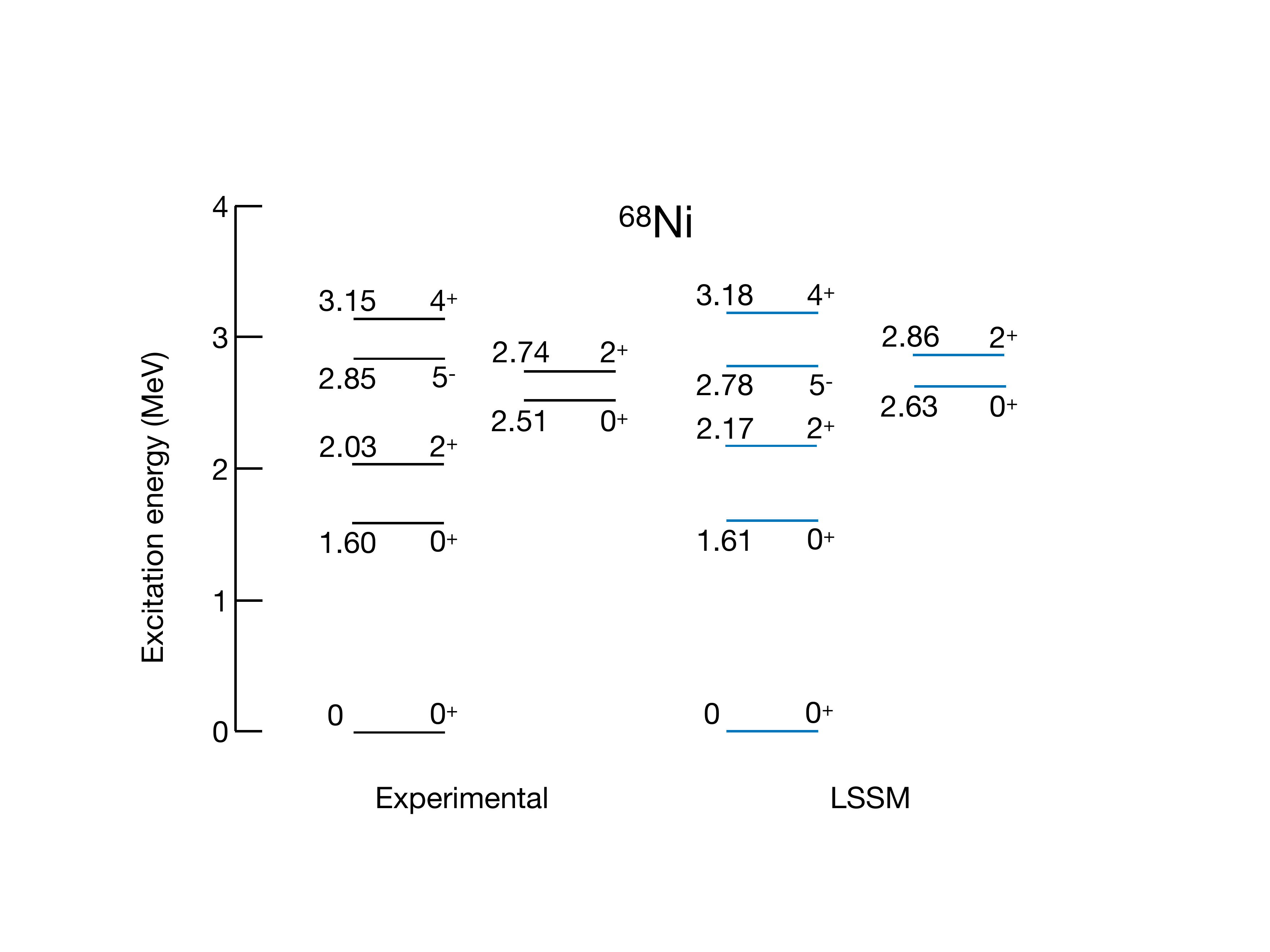}
\caption{Level scheme of $^{68}$Ni: shell-model results with the LNPS-U interaction compared with the experimental data. References in the text.\label{ni68}} 
\end{center}
\end{figure}

\medskip
\noindent
 The theoretical description in the SM-CI context is very satisfactory. The valence space includes the 1p$_{3/2}$, 0f$_{5/2}$, and 1p$_{1/2}$ orbits plus
 the $\Delta j$=2, $\Delta l$=2, Quasi-SU3 sequence 0g$_{9/2}$, 1d$_{5/2}$, for the neutrons and the full $pf$-shell for the protons. The effective interaction is
 a minimal evolution of the original LNPS \cite{lenzi2010}. Similar results have been also obtained in reference \cite{tsunoda2012}. Three structures show up
 prominently at low energy: The 0$^+$ ground state is dominated by the double closed shell configuration N=40, Z=28, at 60\%, thus one can say that it is
 doubly magic. The rest of the wave function is mainly composed of pairing like 2p-2h neutron excitations. The first excited state at 1604~keV is another
 0$^+$ which supports a band-like structure including the first  2$^+$ and  4$^+$ states.  It has been argued that it is an oblate band of 2p-2h neutron plus 1p-1h 
 nature. Indeed the calculations produce positive spectroscopic quadrupole moments. However, as discussed in reference \cite{poves2020}, when analysed in terms
 of the Kumar invariants, it appears that the existence of an associated intrinsic state is dubious, because it is very soft both in the $\beta$ and $\gamma$ degrees of freedom.
 The lowest negative parity state appears at about 3~MeV. Most interesting is the presence of a third 0$^+$ state which is the head of a prolate well deformed
 band with $\beta$=0.3 which is consistent in structure and E2 properties with the intruder state predicted by the SPQR model discussed above, with 
 Q$_0^{\nu}$ = 152~fm$^2$; Q$_0^{\pi}$ = 76~fm$^2$. These values are not far from the limit of full quadrupole dominance and provide a clear example of
 coexistence of a doubly magic ground state and a well deformed excited band. This is usually presented as  a case of shape coexistence (triple shape
 coexistence if we accept that the oblate states have an intrinsic shape). The caveat advanced in reference \cite{poves2020}
is that it is impossible to assign a shape to the  doubly magic ground state. We have plotted these results in Fig. \ref{ni68}. The agreement
with the experimental data is amazing and extends to the electromagnetic decay probabilities. Shape coexistence has been also reported in $^{66}$Ni
in references \cite{leoni2017,olaizola2017}.
 
\subsubsection{$^{64}$Cr. At the heart of the IoI}

The experimental systematics indicates a growing deformation when approaching N~=~40, which is interpreted as the inversion of the deformed intruder configurations
discussed previously and the normal ones corresponding to the N=40 closure. The excitation energy of the first 2$^+$ state in Iron and Chromium isotopes has been measured up to {\bf $^{72}$Fe} and {\bf $^{66}$Cr}~\cite{santamaria2015}, respectively. 
A sudden onset of deformation along the Fe chain has been inferred from an increase of transition probabilities B(E2; $0^+_1\rightarrow2^+_1$) from $^{62}$Fe to
$^{68}$Fe~\cite{rother2011,ljungvall2010,crawford2013}. The Chromium isotopes show the same
tendency with the lowest 2$_1^+$ energy measured in this mass region
at 386(10) keV ~\cite{santamaria2015} in $^{66}$Cr, slightly lower that then 420(7) keV excitation energy for first 2$^+$ state of $^{64}$Cr~\cite{gade2010}, and a B(E2)$\uparrow$ value similar in magnitude to $^{66,68}$Fe~\cite{crawford2013}. Those first observables demonstrate an increase of collectivity for both Cr and Fe isotopes as a function of neutron number beyond N=38, which is also supported by mass measurements~\cite{naimi2012}. See also references \cite{lunardi2007,gaudefroy2009,aoi2009,baugher2013,chiara2015,mougeot2018}.

\begin{figure}[h]
\begin{center}
\includegraphics[width=0.49\textwidth]{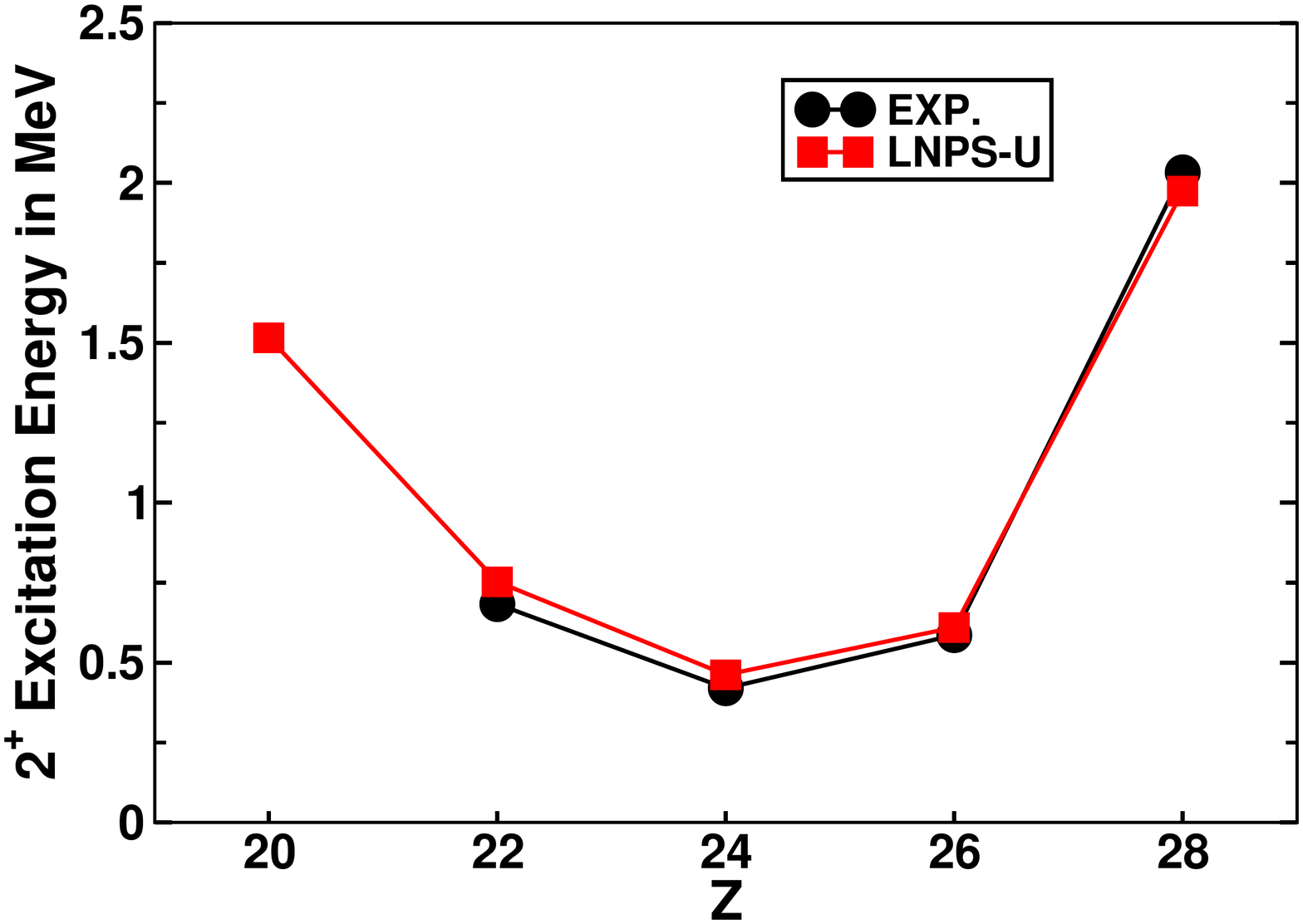}
\includegraphics[width=0.49\textwidth]{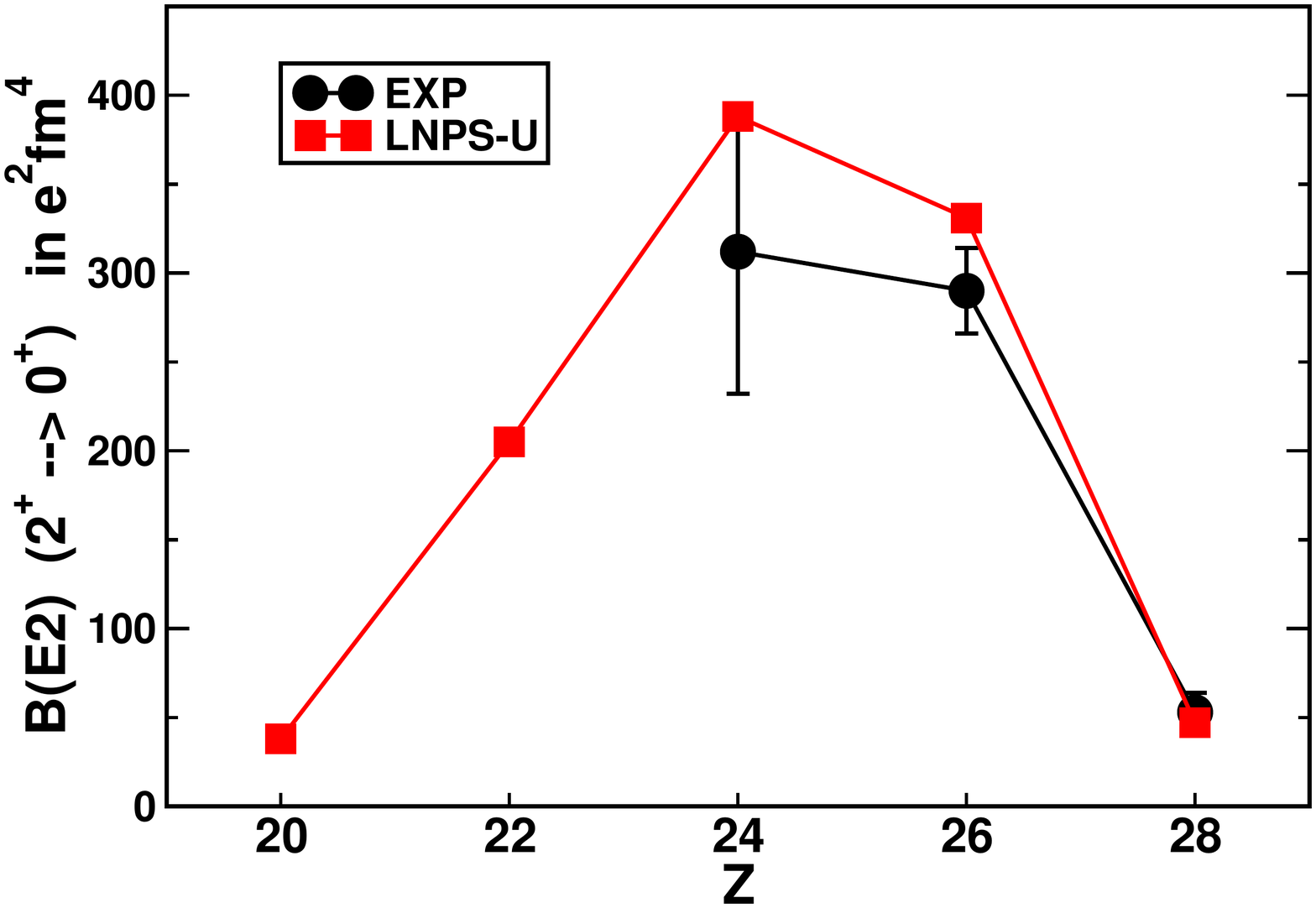}
\caption{The N=40 IoI as seen by the experimental and theoretical results for the excitation energies of the 2$^+$ states (left panel) and the
corresponding B(E2)'s (right panel). Notice the abrupt changes departing from $^{60}$Ca and  $^{68}$Ni. \label{n40-e2-be2}} 
\end{center}
\end{figure}

\noindent
On the theory side, the SM-CI calculations have done an excellent job again. In fact, reference \cite{lenzi2010} surmised the existence of the 4$^{th}$ IoI and made a lot of
predictions that were later on confirmed by experiment. The mechanism is simple; in the competition between the monopole field and the quadrupole correlations
the latter win and the deformed intruder states become the ground states below $^{68}$Ni. Indeed the excited deformed band in $^{68}$Ni is the precursor of the
phase transition leading to the IoI. The results for the N=40 isotopes are shown in Fig. \ref{n40-e2-be2}.
Notice the abrupt changes departing from $^{60}$Ca and  $^{68}$Ni, downwards in the 2$^+$ excitation energies and upwards in the B(E2)'s, fully consistent 
with the experiments and with the physical image of a transition to a rotational regime. In fact, a large share of the theoretical values 
in Fig. \ref{n40-e2-be2} were predictions later confirmed by experiment. Maximum quadrupole collectivity is achieved in  $^{64}$Cr, with values
Q$_0^{\nu}$ = 193~fm$^2$; Q$_0^{\pi}$ = 116~fm$^2$, which nearly saturate the SPQR limits quoted above.

\begin{figure}[H]
\begin{center}
\includegraphics[width=0.49\textwidth]{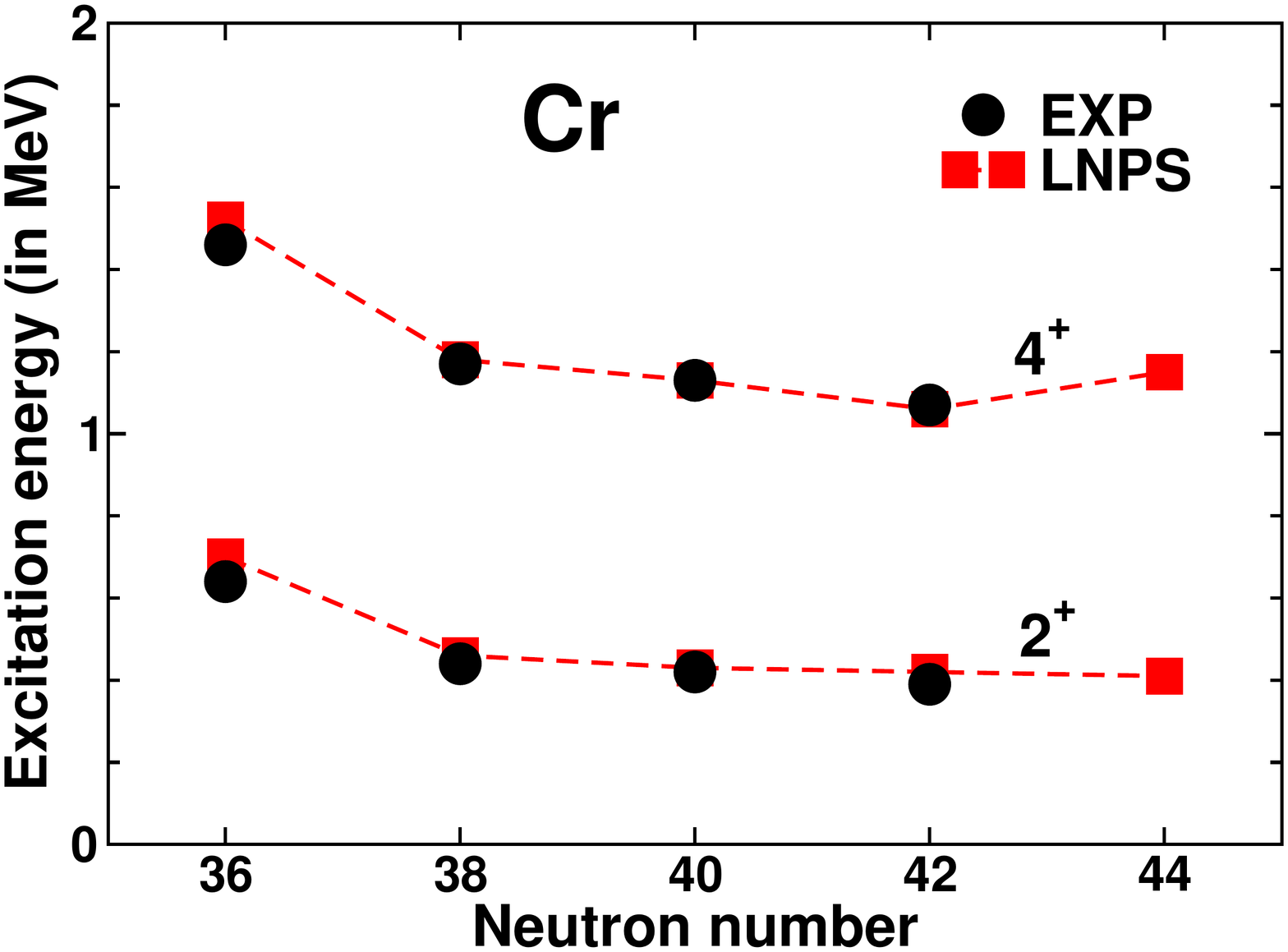}
\includegraphics[width=0.49\textwidth]{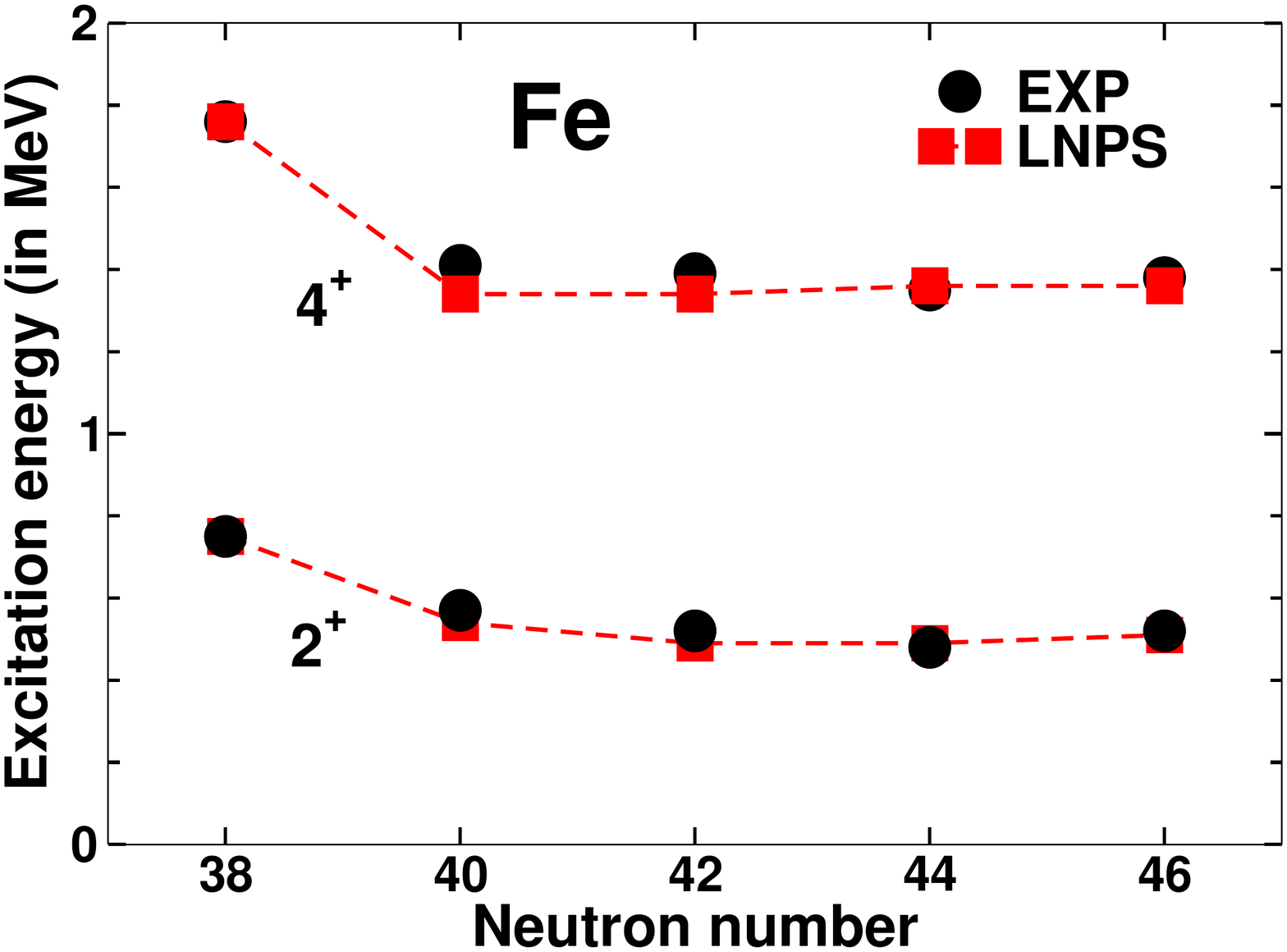}
\caption{Excitation energies of the 2$^+$ and 4$^+$ states in the Chromium isotopes (left) and in the Iron isotopes (right) compared with the SM-CI predictions with the effective interaction LNPS-U. \label{cr-fe}} 
\end{center}
\end{figure}

\medskip
\noindent
There has been a lot of recent experimental activity in the  4$^{th}$ IoI. In particular, the spectroscopy of the Iron and Chromium chain has been
extended up to N=46 and N=42 respectively \cite{santamaria2015}. We show in Fig. \ref{cr-fe} the systematics of the 2$^+$ and 4$^+$ experimental
excitation energies compared with the theoretical results. The agreement is again excellent, and it is seen that the transition to the IoI 
takes place at N=40 in 
the Iron chain and at N=38 in the Chromium's. New data beyond N=40 have been discussed in references \cite{cortes2018,gade2019}.

\medskip
\noindent
Moving farther away from stability the next milestone is $^{62}$Ti, whose spectroscopy was unknown till very recently. The measure of Cortes {\it et al.}
at RIKEN \cite{cortes2020},
which is at the edge of the present experimental capabilities,  has been able to obtain the excitation energies of the 2$^+$ at 683~keV, and of the 4$^+$ at 1506~keV, 
on top of the LNPS-U predictions. Clearly, whereas these states are intruders as well, the collectivity is not enough developed so as to produce the like of
static deformation. This is borne out by the SM-CI results that predict Q$_0^{\nu}$ = 152~fm$^2$; Q$_0^{\pi}$ = 65~fm$^2$, clearly smaller than the SPQR limit. The evolution of the  2$^+$ excitation energies does not show an abrupt change at N=40, but rather a smooth decrease from 1047~keV in N=36, 850~keV
in N=38 to  683~keV in N=40 \cite{gade2014}. The experimental results on  $^{61}$Ti \cite{wimmer2019} are consistent as well with the inclusion of 
this N=39 isotope in the IoI. Let's mention too the work of reference \cite{tarasov2009} which suggested that $^{62}$Ti might belong to the IoI and the
early comment of reference \cite{brown2001} in the same sense. Neither the beyond mean field calculations of references \cite{robledo2019,gaudefroy2009,rodriguez2016}, nor the VS-IMSRG approach of reference \cite{mougeot2018}, seem to be able to account for the
spectroscopy of $^{62}$Ti yet. Extra evidence of the intruder character of the ground state of $^{62}$Ti has been obtained from a very recent measurement
of its mass in reference \cite{michimasa2020}.

\begin{figure}
\vspace{2cm}
\includegraphics[width=0.49\textwidth]{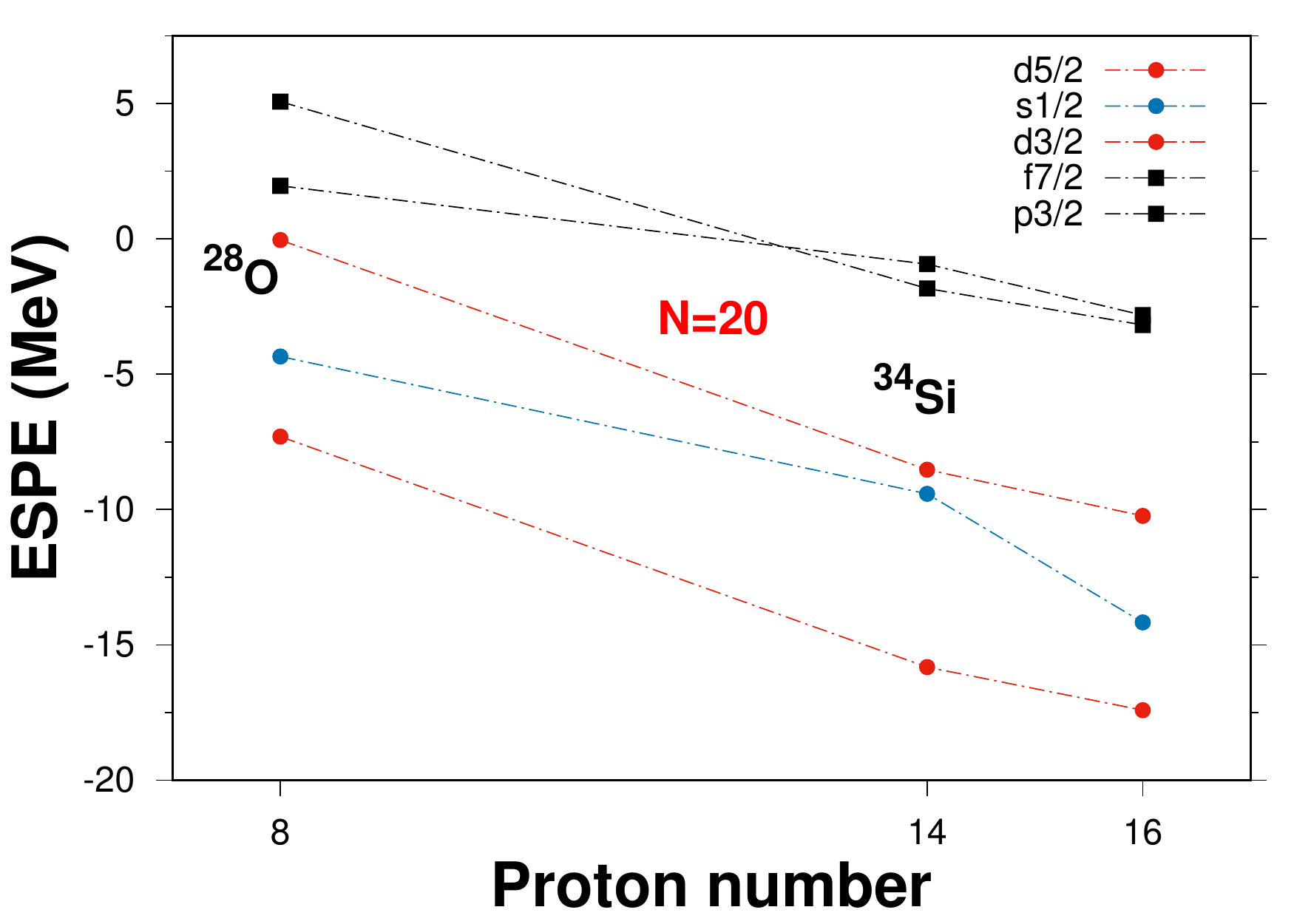}
\includegraphics[width=0.49\textwidth]{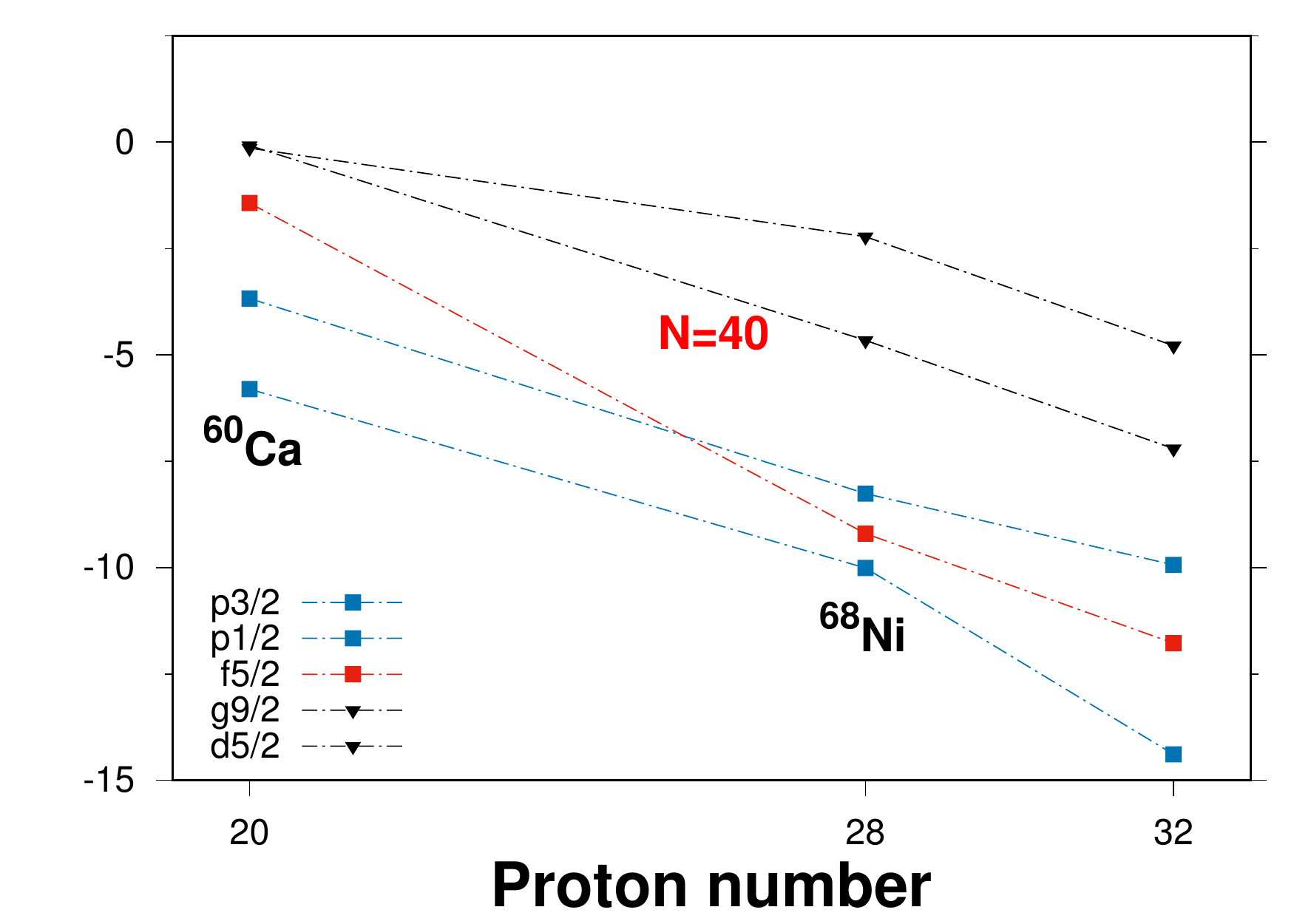}
\caption{(Left) Effective neutron single-particle energies at N=20 (left) and at N=40 (right) with the SDPFU-MIX and LNPS-U effective interactions.\label{comp-ESPE}}.
\end{figure}

\medskip
\noindent
An interesting point is the comparison between the mean-field situation occurring at N=20 and N=40 as show in Fig. \ref{comp-ESPE}.
In both cases, one observes a reduction of the N=20 and N=40 harmonic oscillator gaps as well as the close proximity of the 0f$_{7/2}$ - 1p$_{3/2}$ and 
0g$_{9/2}$ - 1d$_{5/2}$ pairs of orbits which are  the Quasi-SU3 neutron partners, and the inversion of their ordering  towards more neutron-rich cases.
The conjunction of these three factors  seems to be at play for the understanding of  the IoI's occurring far from stability.
It is worth to mention here that the orbital inversion seen through the Effective single-particle energies has been   traced back to N=20 in recent experimental works. On the other hand, the inversion of orbits at N=40, though not yet observed experimentally, is the one naturally proposed by the ab-initio Coupled Cluster calculations of reference \cite{hagen2012} for $^{60}$Ca. 

\subsubsection{$^{60}$Ca. A remarkable study case}

One can approach $^{60}$Ca following the Z=20 isotopic chain or the N=40 isotonic line. The usual valence space for the Calcium isotopes is the full
$pf$-shell, which indeed can provide a sound and consistent description of their low energy spectra. The $pf$-shell part of the LNPS-U interaction reproduces
nicely the $2^+$ excitation energies of the N=32 and N=34 isotopes as discussed in Section 6. The question is till which value of N this valence space makes sense,
{\it i. e.} when becomes necessary to take into account the excitations to the $sdg$-shell. In the $pf$-shell description, the  $2^+$ excitation energies of
$^{56}$Ca  and $^{58}$Ca should be similar and directly related to the diagonal pairing matrix element of the 0f$_{5/2}$ orbit, which in LNPS-U amounts to about 1.1~MeV.
Obviously, $^{60}$Ca is doubly magic and only its ground state belongs to the $pf$ space. New information on the N=36 and N=38 Calcium isotopes would be vital to understand
how the two valence spaces match.
The situation is under better control if we approach $^{60}$Ca from the heavier N=40 isotones, thanks to the (very recent) new experimental information about
$^{62}$Ti, which fully support the predictive power of the SM-CI calculations using the interaction LNPS-U. They produce, at the spherical mean field level,
a very small gap between the 0f$_{5/2}$ and the 1d$_{5/2}$ neutron orbits of 1.3~MeV. The 0g$_{9/2}$ orbit is just 60~keV above the 1d$_{5/2}$, something reminiscent
of the inversion of the 0f$_{7/2}$ and 1p$_{3/2}$ orbits at N=20 close to $^{28}$O. This inversion has been predicted as well by the Coupled Cluster calculations
of reference \cite{hagen2012}, although in this approach, which locates the drip line already at N=40, both orbits are unbound in $^{61}$Ca. The 1p$_{1/2}$ and 1p$_{3/2}$
orbits are respectively 2~MeV and 4~MeV more bound than the 0f$_{5/2}$.
Let's examine now our  predictions for the low energy spectrum of $^{60}$Ca: The ground state 0$^+$ has on average more than three neutrons above the N=40 closure,
with a doubly magic component of about 5\%. Hence it belongs to the IoI. In view of the quasi degeneracy of the  0f$_{5/2}$, 1d$_{5/2}$ and 0g$_{9/2}$ neutron
orbits a superfluid behaviour could be anticipated, but we will show that the situation is more complex. The  2$^+$ comes at 1.52~MeV followed by a triplet
of states;  0$^+$ at 2.36~MeV (the one state with the larger amount of closed N=40 configuration at 23\%),  2$^+$ at 2.68~MeV and 4$^+$ at 2.93~MeV, at odds with the superfluid expectations.  

\medskip
\noindent
To substantiate this issue we have made a model calculation in the LNPS valence  space with the monopole plus pairing (M+P) component of the 
LNPS interaction. The results show the fully paired ground state (0$^+$ seniority zero), and all the seniority two states degenerated at 
an excitation energy of 3.4~MeV, a level scheme very different from the realistic one, even if re-scaled in energy.  We compare in Table \ref{60Ca-occ} the occupation numbers of the
three  dominant orbits in the full LNPS and in the monopole plus pairing calculations.
And we find with some surprise that the LNPS-U occupancies are not that far from those of the
monopole plus pairing limit. Another surprise is that if we analyse the wave functions of the 0$^+$ and 2$^+$ states in both calculations in a
seniority scheme we (obviously) find  100\% $\nu$=0 and 100\%  $\nu$=2 for the 0$^+$ and 2$^+$ states in the M+P case, and 
90\% $\nu$=0 and 83\%  $\nu$=2 for the 0$^+$ and 2$^+$ in the realistic calculation.

\begin{table}[tbh]
\begin{center}
  \caption{Occupation numbers of the orbits with the LNPS-U interaction and in the monopole plus pairing case (M+P)(see text). \label{60Ca-occ}}
   \begin{tabular*}{\linewidth}{@{\extracolsep{\fill}}|c|c|ccc|}\hline
 Interaction & J$^{\pi}$  &  0f$_{5/2}$ &  0g$_{9/2}$ & 1d$_{5/2}$ \\ 
\hline
 LNPS-U & 0$^+$ & 3.05 & 2.38 & 1.00 \\
 M+P &  & 3.42 & 2.11 & 0.83   \\
 LNPS-U & 2$^+$ & 2.60 & 2.47 & 1.36  \\
 M+P &  & 3.40 & 1.76 & 1.10   \\
 LNPS-U & 4$^+$ & 2.57 & 2.46 & 1.32  \\
 M+P &  & 3.40 & 1.76 & 1.10   \\
\hline
    \end{tabular*}
 \end{center}
\end{table}

\medskip
\noindent
Indeed, the spectrum obtained with LNPS-U resembles more to that of a (distorted) quadrupole vibrator. But things are not that simple.
From the total E2 sum rule of the ground state of the LNPS-U calculation it is possible to extract the like of an intrinsic mass quadrupole
which amounts to Q$_0$ = 176~fm$^2$, which is not so far from the SPQR limit discussed above. On the other hand the ratio
\mbox{B(E2)(4$^+$ $\rightarrow$ 2$^+$)}/\mbox{B(E2)(2$^+$ $\rightarrow$ 0$^+$)}
is equal to 1.52, close to the Alaga rule 1.43 \cite{alaga1955}. In addition, the intrinsic quadrupole moments extracted from the 
spectroscopic quadrupole moments of the 2$^+$ and 4$^+$ states are very similar as well, even if a 20\% smaller that those extracted from the B(E2)'s.
Hence, whereas the level scheme suggest a vibrational regime, the E2 properties point rather to a rotational scheme.
If we naively compute an intrinsic
mass quadrupole moment from the \mbox{B(E2)(2$^+$ $\rightarrow$ 0$^+$)} in the monopole plus pairing calculation,
we get 61~fm$^2$; and the Alaga ratio of B(E2)'s,  is compatible with zero.
This suggest that the quadrupole-quadrupole T=1 interaction among the neutrons dominates over their pairing interaction.
We want to underline that this behaviour  is in full contradistinction with the common wisdom -and textbook lore- about the dominance of the pairing interaction when the nucleus has
only neutrons in quasi degenerated orbits on top of a closed shell. The origin of this anomaly is surely linked to the presence of a
Quasi-SU3 set of orbits at the Fermi surface.
Hence, why the spectrum of $^{60}$Ca (and by the way that of $^{62}$Ti) is not rotor-like? The provisional answer is that the quadrupole-quadrupole interaction is not dominant enough
because its T=1 channel is less attractive than the T=0 (neutron-proton) one. What determines the physics is the competition between the pairing and the quadrupole 
correlation energies, that in the latter case will be roughly proportional to: $$ \lambda_{2(T=1)} \; ((Q_0^{\pi})^2 + (Q_0^{\nu})^2) +  \lambda_{2(T=0)} \; (Q_0^{\pi}) \times  (Q_0^{\nu}) $$
where the $\lambda$'s are the isovector and isoscalar quadrupole coupling constants of the {\it in medium} effective
nucleon-nucleon interaction and
$\lambda_{2(T=1)} \approx  \frac{1}{3}\lambda_{2(T=0)}$ \cite{dufour1995}. Putting numbers in this expression it is easily seen that the amount of quadrupole correlation energy in 
$^{64}$Cr is about three times that of $^{60}$Ca. In the former case it seems to be enough to make the effect of pairing in the rotational spectrum perturbative, 
whereas in the latter, pairing can distort severely the spectrum, blurring any kind of J(J+1) behaviour. But, the question now is, can we give an established name to the
collectivity that $^{60}$Ca shows? It would be tempting to call it vibrational, in view of its spectrum, however it would be a new class of vibration until now overlooked.
One should remember that the microscopic description of vibrational spectra in nuclei  is based in the Brown-Bosterli model \cite{brown1959}, that assumes a closed core (in our  case the doubly magic configuration N=40,  Z=20 of $^{60}$Ca) and diagonalizes the quadrupole-quadrupole interaction in the space of the 1p-1h excitations. If the interaction is coherent
with the particle-hole excitations, a single state will gain a lot of energy. We call it the
collective state, or the nuclear phonon, aka the vibration. The pairing interaction plays no role in this game. Nothing less germane to what we have been discussing for $^{60}$Ca.
In fact, the particle-hole matrix elements of the quadrupole operator are just null in our case study, thus only negative parity phonons could be produced.
And, obviously, there is not such a thing as a closed shell state upon
which to build the phonon. Therefore, the  vibrational image does not hold, and the occurrence of a vibrational-like spectrum must be accidental.
The picture that emerges is that of a pairing frustrated rotor. The more so if we examine which would be the structure of $^{60}$Ca calculated with a pure
monopole plus quadrupole-quadrupole interaction. To make the comparison meaningful, we use the ESPE of the LNPS-U interaction in the calculation with a quadrupole-quadrupole interaction scaled to give excitation energies  consistent with the full interaction results. The solution of lower energy corresponds
to the 6p-6h configuration. The level scheme shows an yrast band that follows the J(J+1) sequence almost exactly,
 As for the E2 properties, the intrinsic quadrupole moment saturates the SPQR limit, Q$_0$=300~fm$^2$. The Alaga rule ratio is 1.37, close to its asymptotic limit 1.43. This quasi-perfect rotor retains most of its E2 properties when pairing enters the game, meanwhile, the rotational spectrum is washed out, and a vibrational pattern accidentally emerges. This justifies our description of $^{60}$Ca as a pairing frustrated rotor. One could expect a similar physics in $^{28}$O, the neutron rich end of the
 N=20 isotone chain, given the presence of  equivalent orbits around the Fermi Level. However, the newest experimental data on the neutron rich Fluorine isotopes
 and its interpretation in the SM-CI context in reference \cite{revel2020}, suggest that the N=20 gap in $^{28}$O is larger that the N=40 gap in $^{60}$Ca. Therefore
 the ground state of $^{28}$O contains a substantial amount of the doubly magic configuration, 
 and the final physical picture is not germane to that of $^{60}$Ca.

\section{N=50; the quest for  $^{78}$Ni}

\subsection{Experimental highlights}
The study of the isotope $^{78}$Ni has been presented as a major goal for all the new-generation RIB facilities, such as the RIBF, FRIB or FAIR. {\bf $^{78}$Ni} may be the only neutron-rich candidate for a doubly magic nucleus which was lacking, up to recently, spectroscopic information on excited states. Its first spectroscopy was measured at the RIBF and revealed an excited  state at 2.60(3) MeV excitation energy \cite{taniuchi2019} decaying via gamma transition to the ground state and assigned to its first 2$^+$ state. This state shows a strong rise of the 2$^+$ excitation-energy systematics along the Ni isotopic chain and provides a signature for a shell closure at N=50 in $^{78}$Ni. This shell closure is consistent with several previous measurements. The Coulomb excitation of neutron-rich Zn isotopes~\cite{vandewalle2007} was measured at REX-ISOLDE by use of the MINIBALL Germanium array detector \cite{miniball}. The recoil beam and target-like particles were detected in coincidence with the prompt gamma rays. The measurement led to a low B(E2;$\uparrow$) of 0.073(15) $e$b$^2$ for $^{80}$Zn and a systematics for N=50 isotones from Z=42 to Z=30 well reproduced by shell-model calculations considering a N=50 shell closure, provided the use of a rather large proton effective charge. The spectroscopy of Ni isotopes up to 
$^{76}$Ni was first performed at the NSCL from the $\beta$ decay of Co isotopes~\cite{mazzocchi2005}. In particular the four-fold gamma cascade following the decay of a 8$^+$ isomer in $^{76}$Ni was successfully reproduced and interpreted as in a neutron $fpg_{9/2}$ valence space above a $^{56}$Ni core, indicating a shell closure at N=50. The $\beta$ decay lifetime measurements of $^{78}$Ni~\cite{hosmer2005,xu2014} are also consistent with a $N=50$ shell closure. Also mass measurements down to $Z=30$ are consistent with shell closures at $N=50$ and $Z=28$~\cite{hakala2008}. In particular, the high-precision mass measurement of the neutron-rich copper isotopes $^{75-79}$Cu at ISOLTRAP by use of a combination of Penning-trap and time-of-flight spectrometry \cite{welker2017}, indicate a doubly-closed shell structure for $^{78}$Ni while nucleon excitation above the $Z=28$ and $N=50$ within the shell model are necessary for a fine reproduction of the mass surface.

\begin{figure}[h]
\begin{center}
\includegraphics[trim=0cm 4cm 0cm 4cm,clip,width=1.0\textwidth]{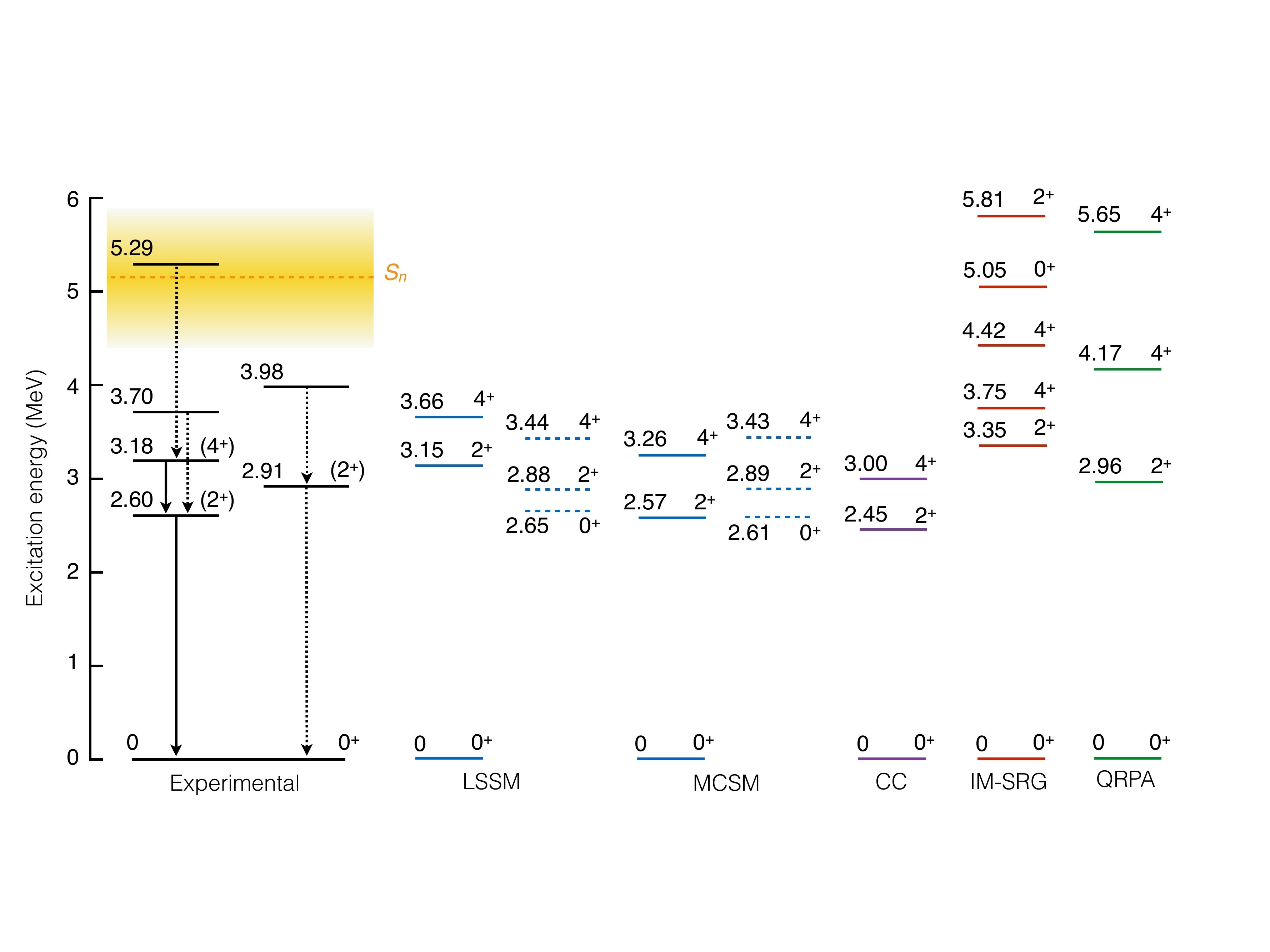}
\caption{Experimental spectrum of  $^{78}$Ni, compared with different theoretical calculations. Adapted from ref.~\cite{taniuchi2019}.\label{Ni78spec}}
\end{center}
\end{figure}

\medskip
\noindent
Conversely, experimental studies of $^{66}$Cr and $^{70,72}$Fe with the setup composed of the high-efficient NaI DALI2 array \cite{takeuchi2014} and the MINOS system combining a thick liquid hydrogen target and a vertex tracker \cite{obertelli2014,seastar2018}, revealed constantly low 2$^+$ and 4$^+$ that question the $N = 50$ shell closure 
for element numbers $Z=24,26$~\cite{santamaria2015}. This scenario is supported by large-scale shell-model calculations that predict 
deformed ground states below $Z = 28$~\cite{nowacki2016}, and therefore a breakdown of the $N=50$ shell closure, predicting similar low-lying intruder states in $^{78}$Ni. These predictions are confirmed by the observation of a gamma transition at 2.91(4) MeV in $^{78}$Ni which has been tentatively assigned to a second 2$^+$ state since it decays directly to the ground state. Interestingly enough, this state is not observed in the $^{78}$Ni spectrum when produced from $^{79}$Cu$(p,2p)$ but is significantly populated compared to the first 2$^+$ state when $^{78}$Ni is produced from $^{80}$Zn$(p,3p)$. Although there is so far no microscopic formalism that allows to predict $(p,3p)$ exclusive cross sections, this observation was used as a indication that the state at 2.9 MeV has a significant n-particles n-holes configuration and weak one-particle-one-hole configuration, consistent with an intruder state (see Fig. \ref{Ni78spec} and the discussion in the next section).   

\medskip
\noindent
Concerning the proton orbits, spectroscopic studies of even-odd Copper isotopes have shown a lowering of $5/2^{-}$ states towards $^{79}$Cu \cite{franchoo1998}, interpreted as a reduction of the $Z=28$ proton shell gap due to the strong proton-neutron tensor force~\cite{sieja2012}. Further support for the weakening of the $Z = 28$ proton shell gap has been provided by 
the Coulomb excitation of $^{76,78,80}$Zn~\cite{vandewalle2007}. On the other hand, the recent spectroscopy of $^{79}$Cu appears consistent with a doubly-magic structure of $^{78}$Ni~\cite{olivier2017}. In this study performed at the RIBF, {\bf $^{79}$Cu} was populated from $^{80}$Zn$(p,2p)$ and its spectroscopy was performed from gamma spectroscopy with the DALI2 array and a typical energy resolution of 10\%. Most of the transitions could not be resolved but significant strength originating from the removal of a $f_{7/2}$ proton was identified at an excitation energy below 2.2 MeV, indicating a significant $Z=28$ gap. A new high-resolution measurement dedicated to the spectroscopy of $^{79}$Cu would be necessary for more quantitative conclusions about its structure.

\subsection{A new Harmonic Oscillator valence space}
\begin{figure}
\vspace{2cm}
\includegraphics[width=0.47\textwidth]{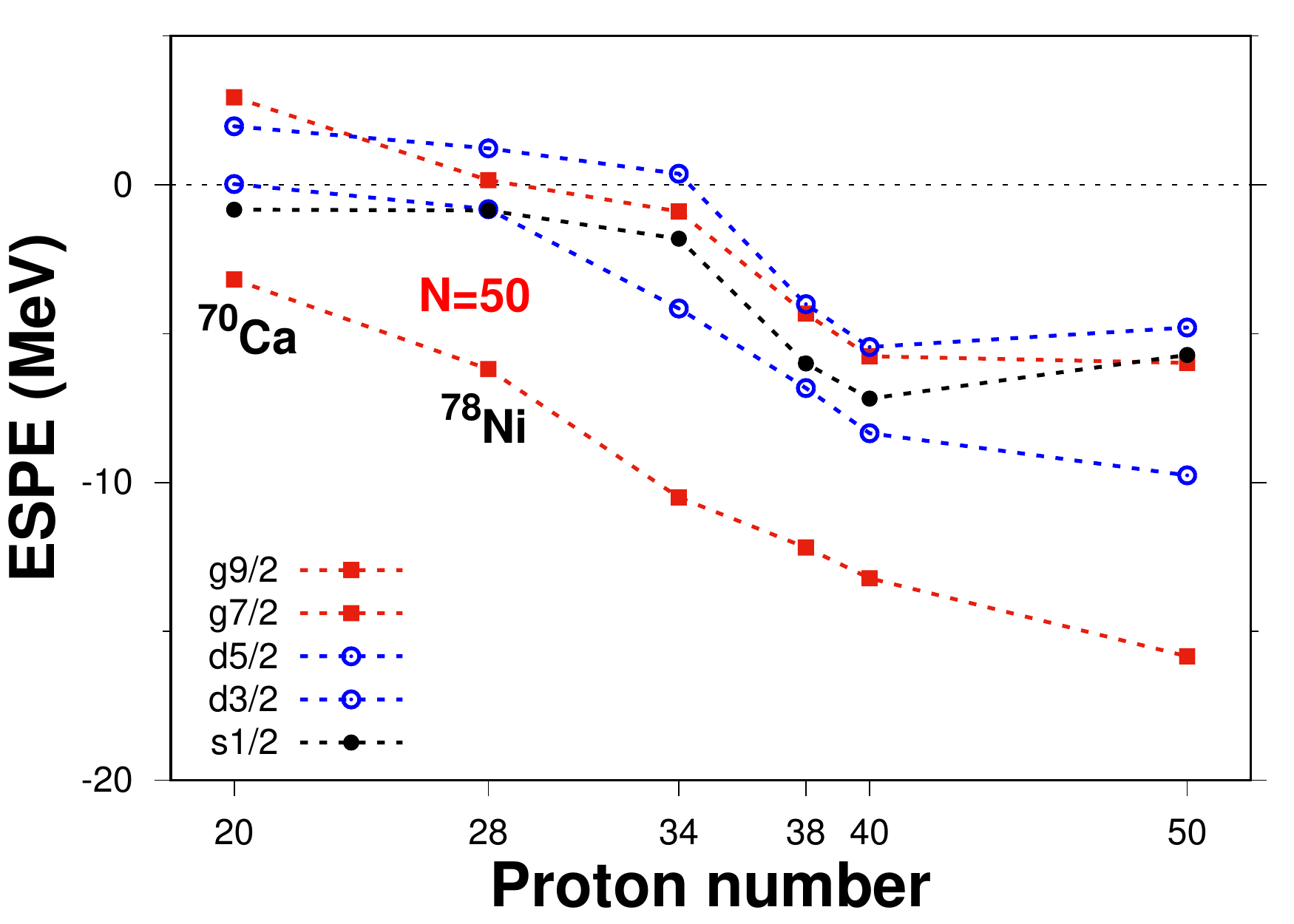}
\includegraphics[width=0.47\textwidth]{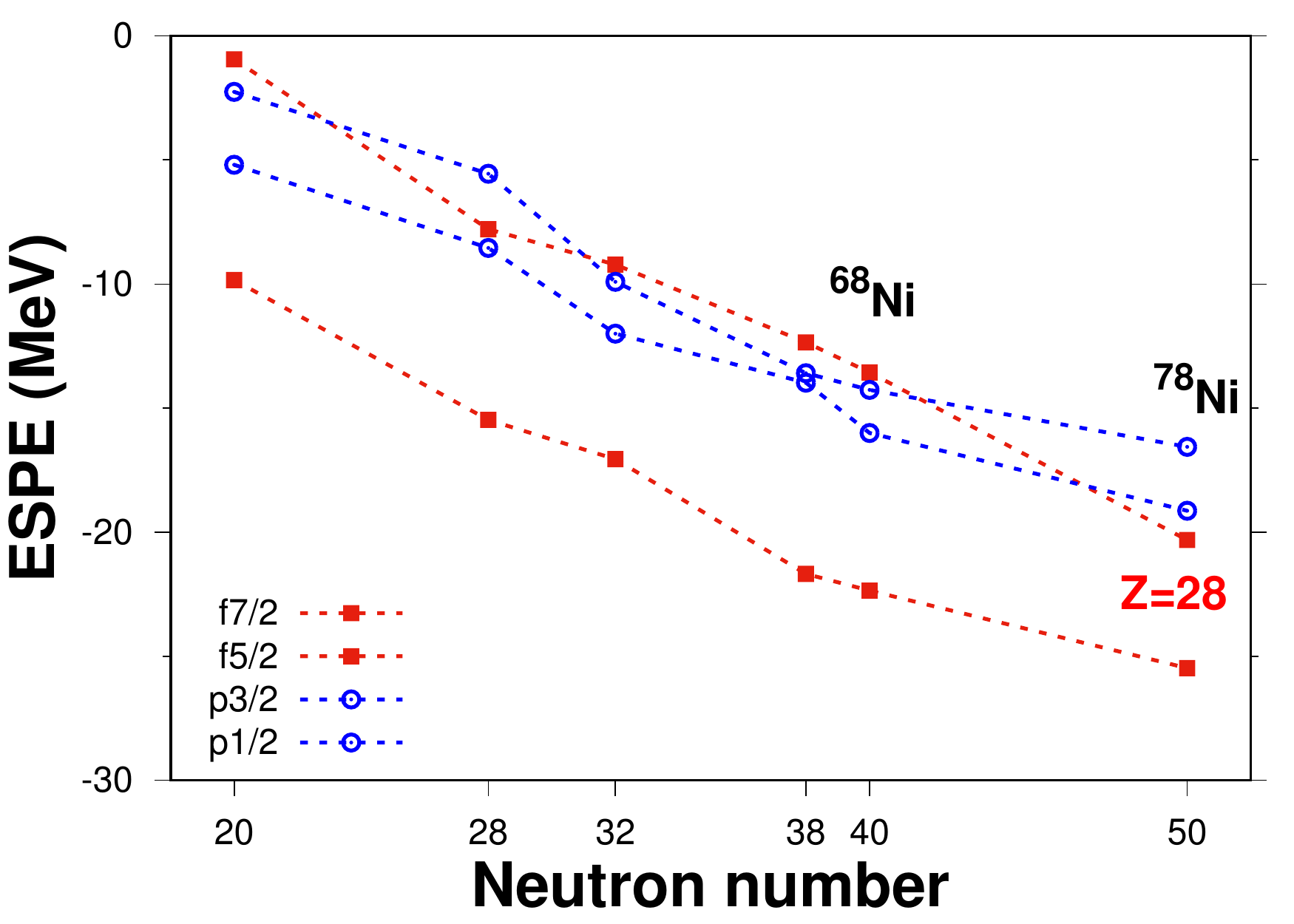}\\
\caption{(Left) Effective neutron single-particle energies at N=50. (Right) The same for the protons. PFSDG-U interaction.\label{gap.nu}} 
\end{figure}

This wealth of experimental information, which in some cases supports the magicity of N=50 and in many others the presence of collective behaviors reminds us of the
physics that we have studied at the IoI's N=20 at the sd-pf interface and at N=40. As before,
the understanding of the physics at the vicinity of $^{78}$Ni requires valences spaces which fully contain
the degrees of freedom necessary for the description of the competition between the spherical 
mean field and the nuclear correlations. In nuclei with several neutrons in excess of N=40, the
neutron excitation's from the $pf$ to the $sdg$ orbits are blocked, and the quadrupole collectivity of the neutrons that until this moment
was based in the Pseudo+Quasi SU3 scheme that we  have
described in detail in the previous section, changes when approaching N=50. In this case the  collectivity of the particle-hole excitations across N=50
is of single orbit (the 0g$_{9/2}$ nature for the holes and Pseudo-SU3 in the $sdg$ shell nature for the particles (SP in our SPQR scheme).
Therefore, the valence space of the PFSDG-U interaction \cite{nowacki2016} contains two full harmonic oscillator shells, the $pf$ shell for protons and the $sdg$ shell for neutrons.
The matrix elements of the PFSDG-U interaction are the of ones of realistic interactions whose proton-neutron monopoles have been constrained to reproduce proton and neutron shell gaps  at Z=28 and N=50 respectively (in particular using the $^{82}$Zn mass measurement \cite{wolf2013}) . The corresponding evolution of the Effective single-particle energies is presented in Figure~\ref{gap.nu}. The PFSDG-U interaction is the natural extension of  LNPS-U for the proton $pf$ neutron $sdg$ valence space.

\begin{figure}
\vspace{3cm}
\hspace{0.5cm} \includegraphics[width=0.75\textwidth]{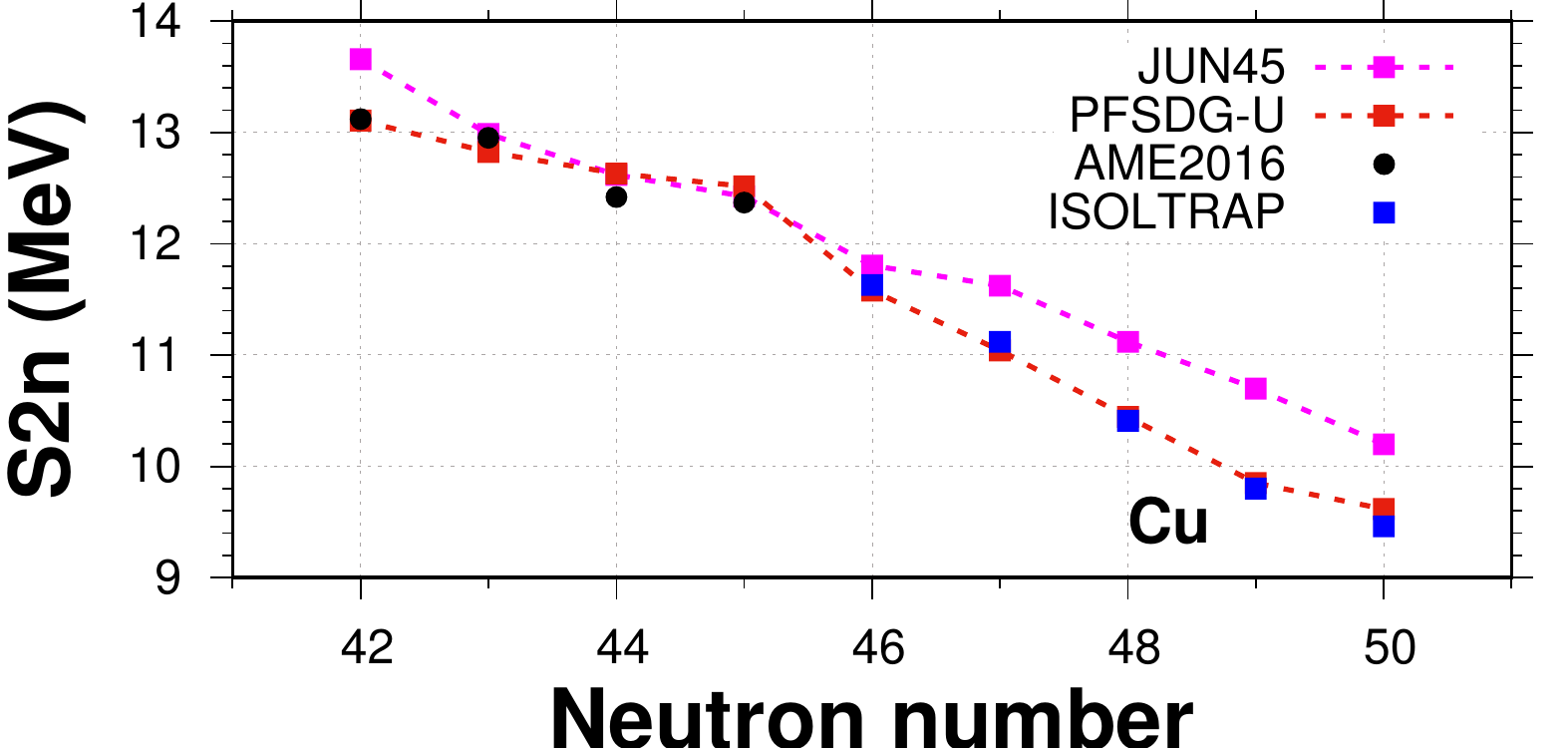}
\caption{Two neutron separation energies in the neutron-rich Copper isotopes. Experimental data from ref. \cite{welker2017} and  the Atomic Mass Evaluation (AME2016), compared with with the 
SM-CI results in the valence space $r3g$ with the JUN45 interaction
\cite{honma2009}, and in the $pf-sdg$ space with the PFSDG-U interaction.\label{cumasses} }
\end{figure}

\medskip
\noindent
The need of core excitations across the Z=28 and N=50 gaps is illustrated in Figure~\ref{cumasses} for the two-neutron separation energies in the copper chain up to N=50. The convincing agreement  with the experimental data from reference \cite{welker2017}, obtained by the calculations using the PFSDG-U effective interaction is at odds with the discrepancies that show up starting at N-46 with the calculations in the valence space $r3g$ 
(1p$_{3/2}$, 1p$_{1/2}$, 0f$_{5/2}$, 0g$_{9/2}$) and  the JUN45 interaction \cite{honma2009}.


\medskip
\noindent



\subsection{$^{78}$Ni and the fifth IoI}

 In view of the successful comparisons with experiment that we have just examined, we can submit that the valence space which encompasses the major oscillator shells,
 N=3 for the
 protons and N=4 for the neutrons, together with the PFSDG-U interaction, provides a reliable framework to describe the physics around $^{78}$Ni. Therefore, we
 can safely discuss now our predictions for this emblematic nucleus. Before that, is it convenient to recall which are 
 the substructures which support  the quadrupole collectivity in this valence space. We are familiar with the doings of SU3 in  a major oscillator shell,
 and its restrictions Pseudo and Quasi-SU3. Below Z=28 the protons can approach a Quasi-SU3 regime, whereas at Z=28 and beyond, the particle hole excitations 
 (or just the particles in the latter case) would rather
 combine Quasi-SU3 or single orbit (0f$_{7/2}$) quadrupole coherence for the holes, with Pseudo-SU3 for the particles. The rules of the game have been abundantly
 explained in previous sections, and can be made quantitative with the help of the figures for the intrinsic quadrupole moments already available.
 For the neutrons, the situation is similar with the 0g$_{9/2}$ orbit playing the role of the 0f$_{7/2}$. As in all the other cases discussed before,
 the location of the intruder states of 
 n-particles n-holes across N=50 and Z=28 character, depends critically on the evolution of the spherical mean field; in the first place of the gaps at Z=28 and N=50
 (see Fig. \ref{gap.nu}) and secondly on the splitting of the orbits above these two closures. It is seen in Fig. \ref{gap.nu} that the Z=28
proton gap increases slowly from 3~MeV at Z=20 to 4~MeV at Z=28. In addition, the splitting of the corresponding Pseudo-SU3 triplet is compatible with
the build-up of quadrupole coherence as shown in reference \cite{zuker2015}. 
One should notice the crossing of the f5/2 and p3/2 orbitals (and the close proximity of the three orbitals above Z=28) as well as a reduction of the Z=28 shell gap.
On the neutron side the variation of the N=50 gap with Z is more steep, changing  from 5~MeV at Z=28 to 2~MeV at Z=20. As in the proton side, the orbits of the Pseudo-SU3
quadruplet are close enough, hence the np-nh neutron excitations can develop quadrupole coherence as discussed above.  In addition, the N=50 gap is affected by
the configuration driven shell evolution (CD-SE) mechanism, {\it i. e.} by the fact that the effective spherical mean field is configuration dependent. In the present
case it amounts to a reduction of the N=50 gap in the 4p-4h configurations of about 1.5~MeV with respect to the values in  Fig. \ref{gap.nu}. The splitting of the
quadruplet is similarly reduced. Both shifts contribute to make the intruder states more bound than naively expected.

\subsubsection{The SM-CI predictions: Coexistence}

\begin{figure}[h]
\begin{center}
\includegraphics[width=0.49\textwidth]{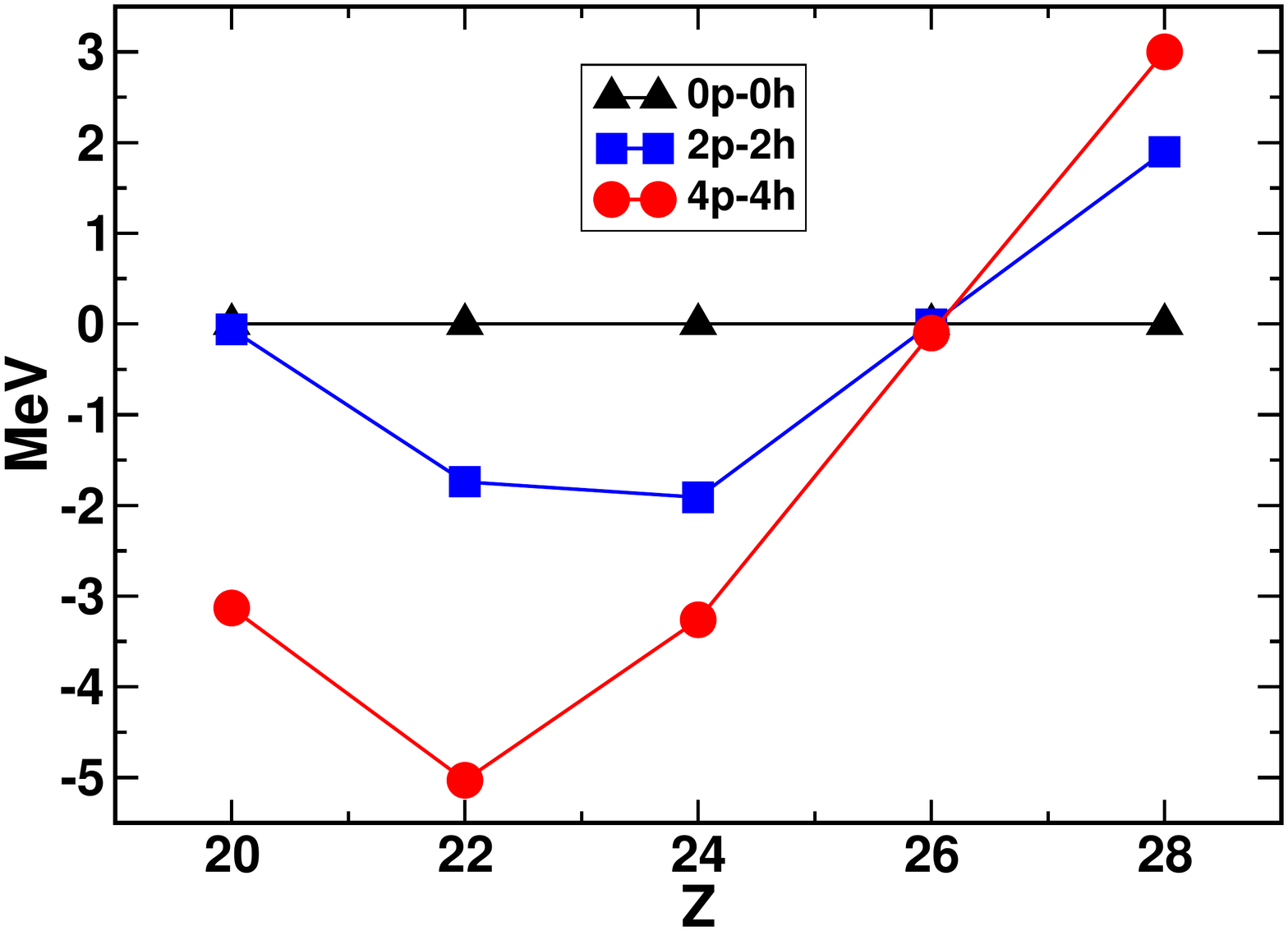}
\includegraphics[width=0.49\textwidth]{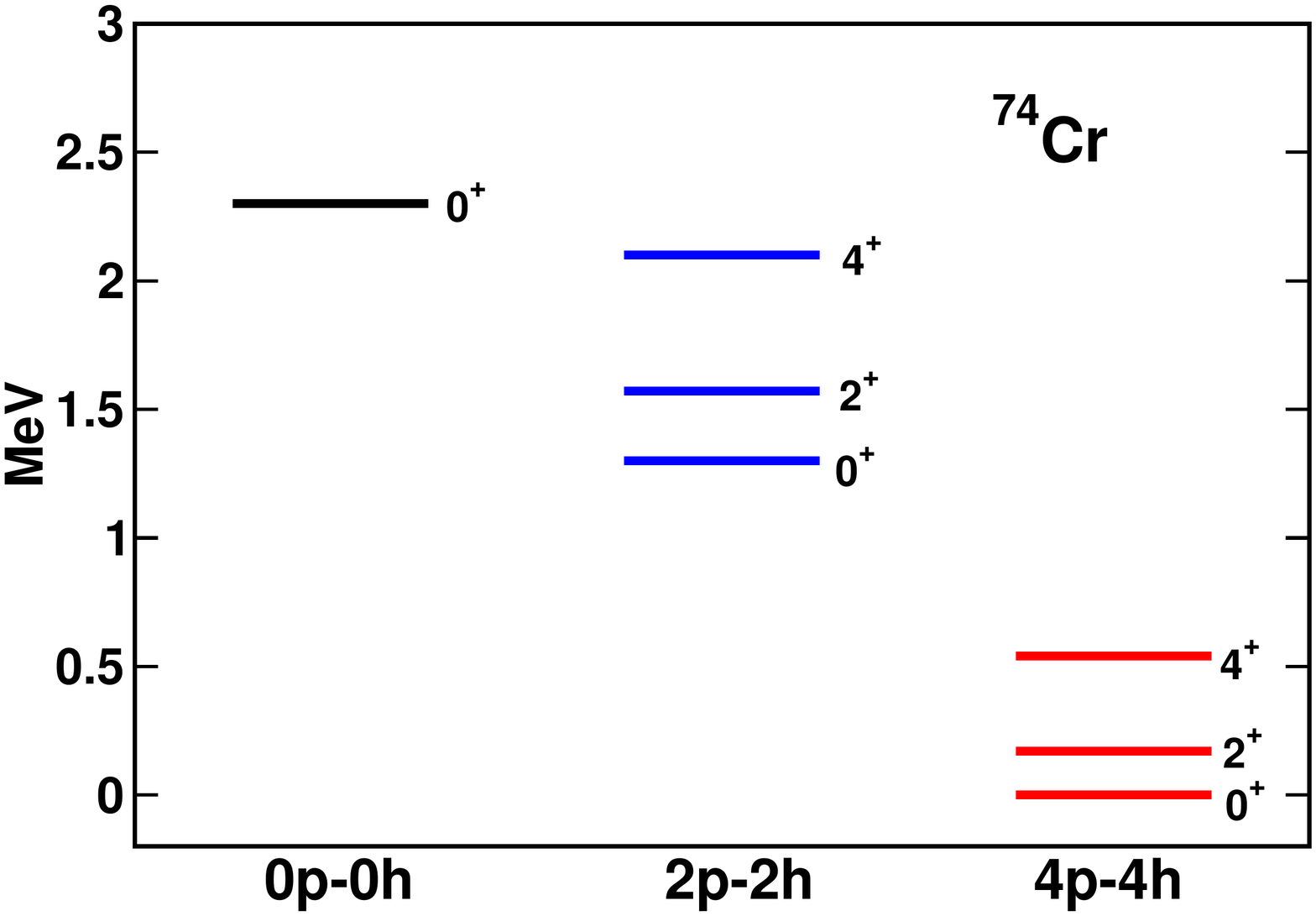}
\caption{(Left) Relative energies of the np-nh neutron configurations at N=50. (Right) Level schemes at fixed np-nh configurations in $^{74}$Cr. \label{e.npnh}} 
\end{center}
\end{figure}

\noindent
 We shall now put at work the SPQR heuristic to locate the most favourable configurations from the point of view of a schematic
 monopole plus quadrupole-quadrupole Hamiltonian,  in the limit of maximal  quadrupole dominance. For the N=50 isotones  
 the protons which are in the Quasi-SU3 space (0f$_{7/2}$, 1p$_{3/2}$) contribute  the following values 
 to the intrinsic mass quadrupole moment (bare masses are used): 
 $Q_0^{m}(\pi)$=11.5, 12.5, 14.0,  9.4 and 0~b$^2$, for  Z=28, 26, 24, 22, and 20 respectively.
 The neutron 0p-0h configurations have zero quadrupole moment. For the 2p-2h configurations  the  two neutrons are in
 the Pseudo-SU3  space (0g$_{7/2}$, 1d$_{5/2}$, 1d$_{3/2}$, 2s$_{1/2}$). Their contribution to   $Q_0^{m}(\nu)$ is 14.7~b$^2$. The
 two neutron holes in 0g$_{9/2}$ add 8~b$^2$. For the 4p-4h neutron configuration the values are   22.2~b$^2$ and 
 10.7~b$^2$, respectively (b is the HO length parameter). The total intrinsic quadrupole moments that we should use to compute the
 mass deformation parameter do need effective "masses" which we take equal to the standard isoscalar charge 
 \mbox{(q$_{\pi}$(1.31)+q$_{\nu}$(0.46)=1.77)}. Doing so we get typical values $\beta_m$=0.24  for the 2p-2h configurations and
 $\beta_m$=0.30  for the 4p-4h ones. 
 Let's now compute the energies of the np-nh configurations using the following expression  adapted from ref.~\cite{zuker2015}:
 
 \begin{equation}
  E_{n}=   G^{mp}_n(50)  -\hbar \omega \kappa\left
  (\frac{\langle Q_0^{m}(\pi)\rangle}{15 \: b^2}+\frac{\langle
     Q_0^{m}(\nu) \rangle}{23 \: b^2}\right)^2
\label{eq:su3}      
\end{equation}

 \noindent
 with $ \hbar \omega \kappa$=3.0~MeV. $G^{mp}_n(50)$ contains the monopole gaps plus a pairing correction:

  \begin{equation}  
   G^{mp}_n(50)  = {\rm n} \: (\frac{3.0}{8} n^{\pi}_f    +   {2.25})+ \Delta({\rm n})  + \delta_p({\rm n})   
 \label{eq:gap}
  \end{equation}

 \noindent
 The term $\Delta({\rm n})$  takes into account the quadratic monopole contribution to the np-nh configurations which
 amounts to --0.9 and --4.4 MeV for the 2p-2h and 4p-4h cases, and $ \delta_p({\rm n})$ gives an estimation of the extra pairing energy of the
 np-nh configurations (--1~MeV and --2~MeV for n=2 and n=4).
 

\medskip
\noindent
We have checked the results given by Eq.~(\ref{eq:su3}) via direct diagonalization of the quadrupole-quadrupole interaction with the same strength,
and the agreement with the analytic expression
proposed in ref.~\cite{zuker2015} is excellent.  Once the gaps calculated with Eq.~(\ref{eq:gap}) are plugged in equation Eq.~(\ref{eq:su3})
we obtain the relative position of the np-nh band heads in the isotopes of interest, that we  gather in the left panel of Fig.~\ref{e.npnh}.
In $^{78}$Ni  the 0p-0h configuration, the N=50 closed shell, is the lowest with both 2p-2h and 4p-4h deformed states 
appearing at low energy.  It is seen also in the figure that very rapidly the deformed 4p-4h
band becomes   yrast in $^{74}$Cr,  and  $^{72}$Ti and remains so even in   $^{70}$Ca.  Notice that whereas the
maximum gain of quadrupole energy of the 4p-4h deformed configuration takes place in $^{74}$Cr, the maximum total energy gain occurs at 
$^{72}$Ti, due to its smaller neutron gap.
Therefore, the SPQR heuristic
suggests a doubly magic ground state in $^{78}$Ni  with  coexisting low-lying deformed  bands and deformed yrast bands
of (mainly) 4p-4h nature in the remaining isotopes. 
 In Fig. \ref{e.npnh} (right panel) we have plotted the 
spectra of the different  np-nh configurations in the case of $^{74}$Cr with the PFSDG-U interaction (but the structure of the 2p-2h and 4p-4h bands is similar
for $^{78}$Ni, $^{76}$Fe and $^{72}$Ti).
It is seen that the np-nh spectra tend to become rotational, the more so in the 4p-4h case. The E2 properties are consistent with the presence 
of a well deformed intrinsic state too.

\begin{figure}[h]
\begin{center}
\includegraphics[width=0.8\textwidth]{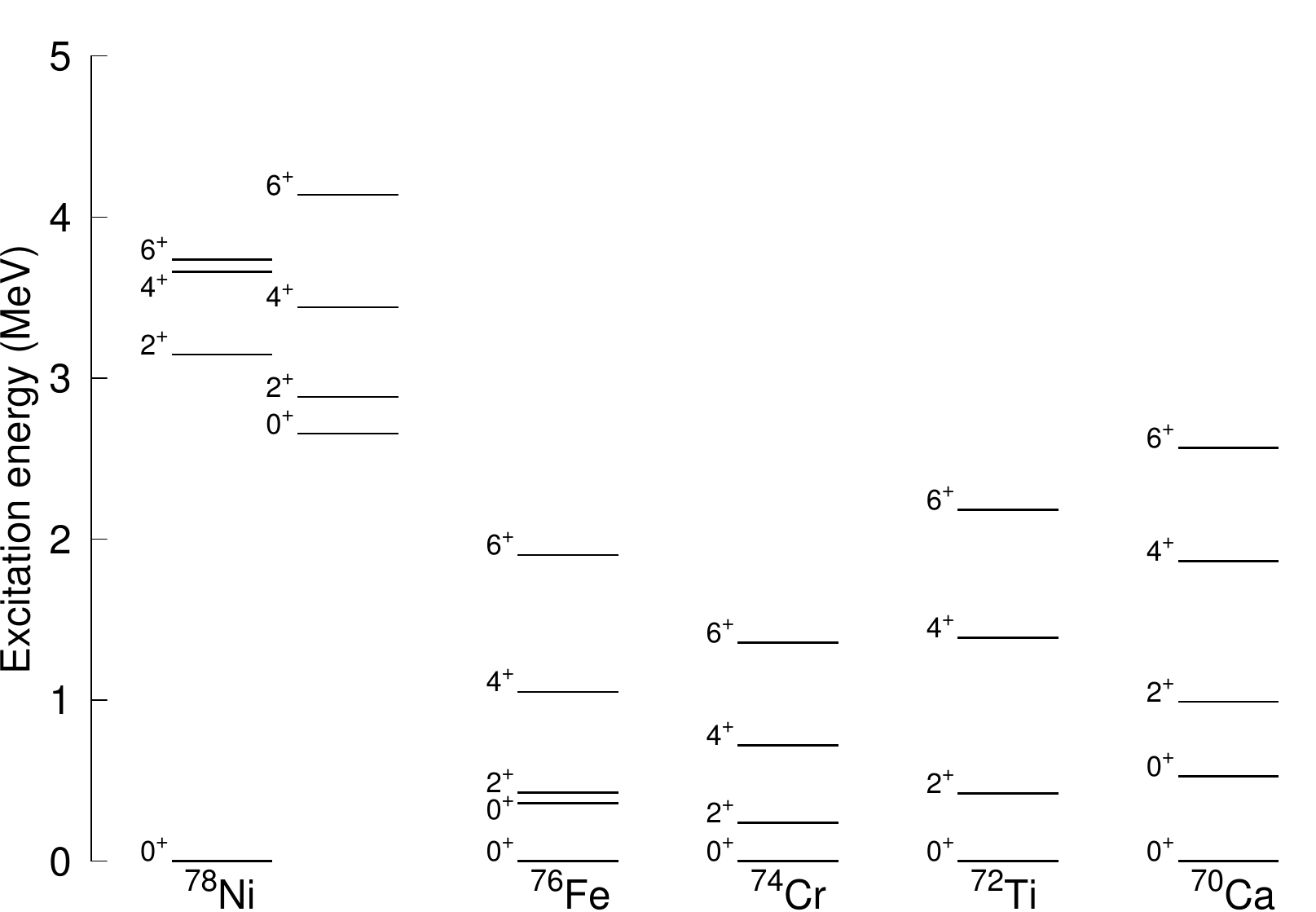}
\caption{From coexistence in $^{78}$Ni (left)  to the fifth Island of Inversion (right). Theoretical predictions with the PFSDG-U interaction.\label{N50spec}} 
\end{center}
\end{figure}

\medskip
\noindent
In fact the full SM-CI calculations with the PFSDG-U interaction align with these schematic results. Let's first focus our attention  in $^{78}$Ni, 
at the left side of Fig. \ref{N50spec}. Two structures emerge; A doubly magic ground state (albeit strongly correlated, because the closed N=50 Z=28 configuration
amounts just to 70\%). On top of it we find several states of correlated neutron particle-hole nature whose total angular momentum correspond to the coupling
of a hole in 0g$_{9/2}$ orbit and a particle in the remaining $sdg$ orbits. Among all the possible couplings, the lowest one is a 2$^+$ at about 3~MeV
(without any particular 
physical motivation) followed by a triplet,  4$^+$, 5$^+$, 6$^+$, a few hundreds keV higher. What is more exotic is the appearance of another, completely
different, structure of clear rotational nature based in a 0$^+$ state of multi-particle hole nature (on average four neutrons above N=50 and two protons above Z=28)
with a perfect J(J+1) spectrum and rotational E2 properties. This 0$^+$ state is the first excited state according to the calculation. The  2$^+$ state of this coexisting band 
lies at 2.9 MeV (see ref. \cite{nowacki2016} for more details). These predictions have been borne out by the experiment of Taniuchi {\it et al.} at Riken 
of reference \cite{taniuchi2019} that we have discussed in section 7.1. In Fig. \ref{Ni78spec} the experimental results are compared with several theoretical calculations:
The Coupled Cluster (CC) calculation of reference \cite{hagen2016} accounts well for the doubly magic nature of the ground state of $^{78}$Ni and predict the excitation energy of the
particle-hole like 2$^+$  reasonably well. It is for the moment beyond the reach of the CC approach to produce an  intruder coexisting band.
The QRPA calculation from the Bruy\'eres le Ch\^atel group, using the Gogny force, does not produce the intruder deformed band, as it is the case for the
VS-IMSRG results of  Men\'endez, {\it et al.}. Finally, the MCSM of the Tokyo group, adopting the $pf-sdg$ valence space and an effective interaction which is an extension of their
A3DM, produces results very similar to those of reference \cite{nowacki2016}, but there is neither a published account of the details of the calculation,
nor a more complete systematics in the region yet. 
To wrap this section up, we conclude that the coarse grained traits of the structure of $^{78}$Ni are well understood in the framework of the SM-CI description,
although more experimental data and refinements of the theory are still awaiting us in the next future.

\subsubsection{The Fifth IoI}
  We have often surmised  that coexistence in a very neutron-rich doubly-magic nucleus is the natural portal to an Island of Inversion. Indeed, this is the case 
  when we move further away from the stability in the N=50 isotones of lower Z. The full SM-CI calculations shown in Fig. \ref{N50spec} follow closely what
  was anticipated by the results at fixed number of particle-hole  excitations of Fig. \ref{e.npnh}. Let's first examine the case of $^{74}$Cr:  The ground state
  band is that of a perfect prolate rotor, $\beta$=0.32, with a (nearly)  J(J+1) spacing and fully consistent E2 properties (Alaga ratio 1.45). 
  It is dominated by 4p-4h neutron excitations
  at 30\% plus 6p-6h at 20\%. The closed N=50 configurations have negligible amplitudes. A straightforward example of transition from the doubly magic ground state of  
  $^{78}$Ni to the heart of the deformed IoI. Notice in Fig. \ref{N50spec} the blatant similarity between the coexisting excited band of $^{78}$Ni and the yrast 
  band of $^{74}$Cr. What is the status of $^{76}$Fe? In Fig. \ref{e.npnh} it is seen that the 0p-0h, 2p-2h and 4p-4h 0$^+$ band-heads are degenerate. Therefore we
  expect a lot of mixing that should diminish with J. And that is what shows up in the full calculation of  Fig. \ref{N50spec}. Two 0$^+$, the ground state and the first 
  excited state, followed by the higher J yrast states, with a roughly J(J+1) spacing on top of the 2$^+$. Let's have a closer look into the structure of these states.
  The ground state has 28\% 0p-0h, therefore it is intruder dominated, hence it its fair to conclude that $^{76}$Fe belongs to the IoI. The first excited 0$^+$ is 
  what remains of the closed shell at 45\%. The mixing reduces  the B(E2) of  the 2$^+$ to the ground state and increases its excitation energy perturbing the J(J+1)
  behaviour of the band that, however, is recovered at higher spins as well as the prolate rotor E2 properties. 
   The intruder components are dominantly of 2p-2h nature. The spectrum of  $^{72}$Ti is not really rotor-like
   but its (prolate) E2 properties are. As hinted in Fig. \ref{e.npnh}  the 4p-4h
   excitations are dominant, hence it belongs as well to the IoI. In $^{70}$Ca the calculation produces a ground state which is intruder dominated with only 20\%
   of closed N=50 configuration. The rest of the wave function is evenly spread in the np-nh configurations, as it is also the case for the other yrast states. Therefore
   it belongs to the IoI too. The first excited 0$^+$ is  the doubly magic configuration at 55\%. Comparing with Fig. \ref{e.npnh} we see that the degeneracy of the
   0p-0h state with the 2p-2h band-head enhances its mixing and somehow blocks the expected dominance of the 4p-4h configuration in the ground state.

\begin{figure}[h]
\begin{center}
\includegraphics[width=0.475\textwidth]{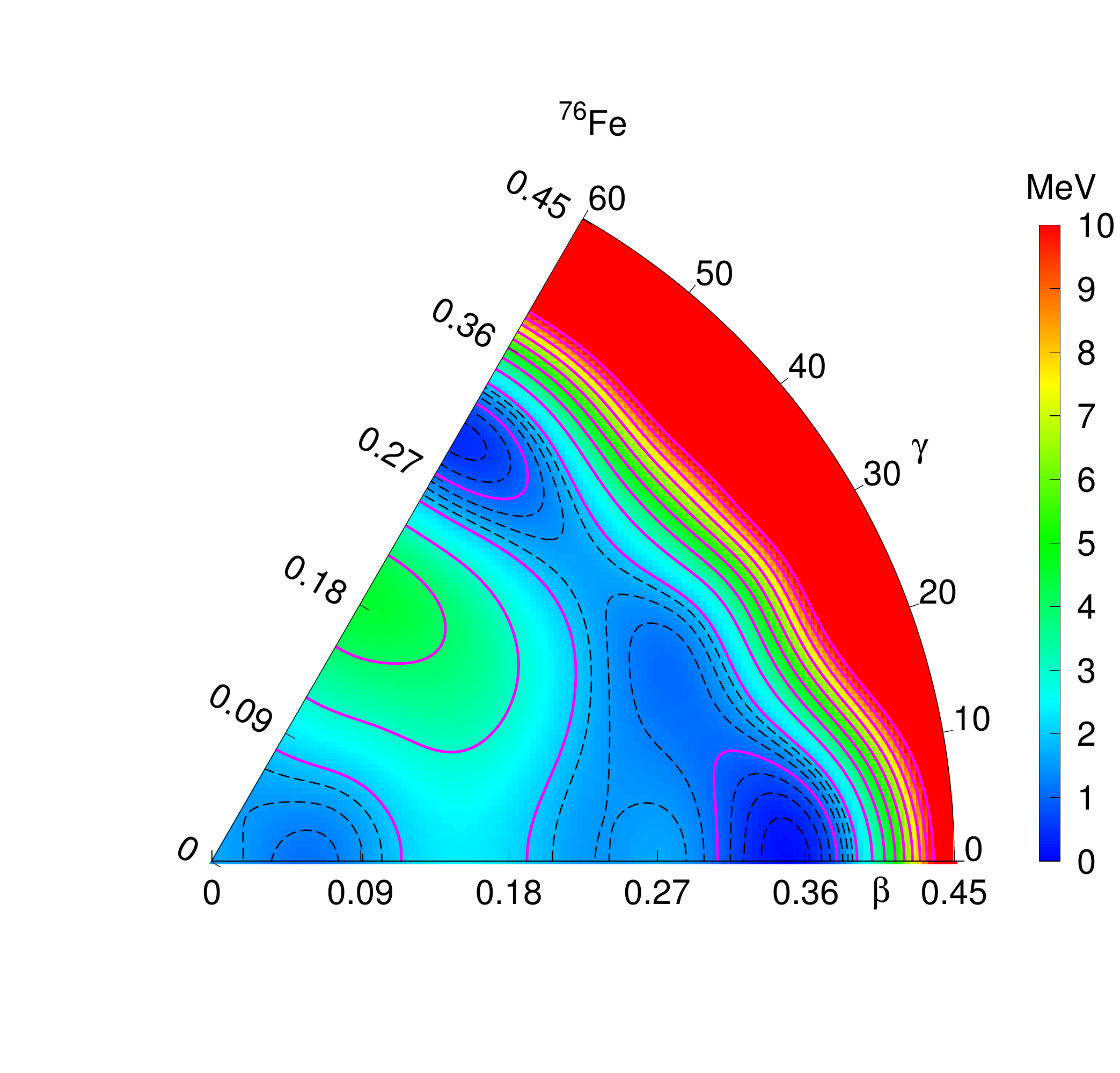}
\includegraphics[width=0.475\textwidth]{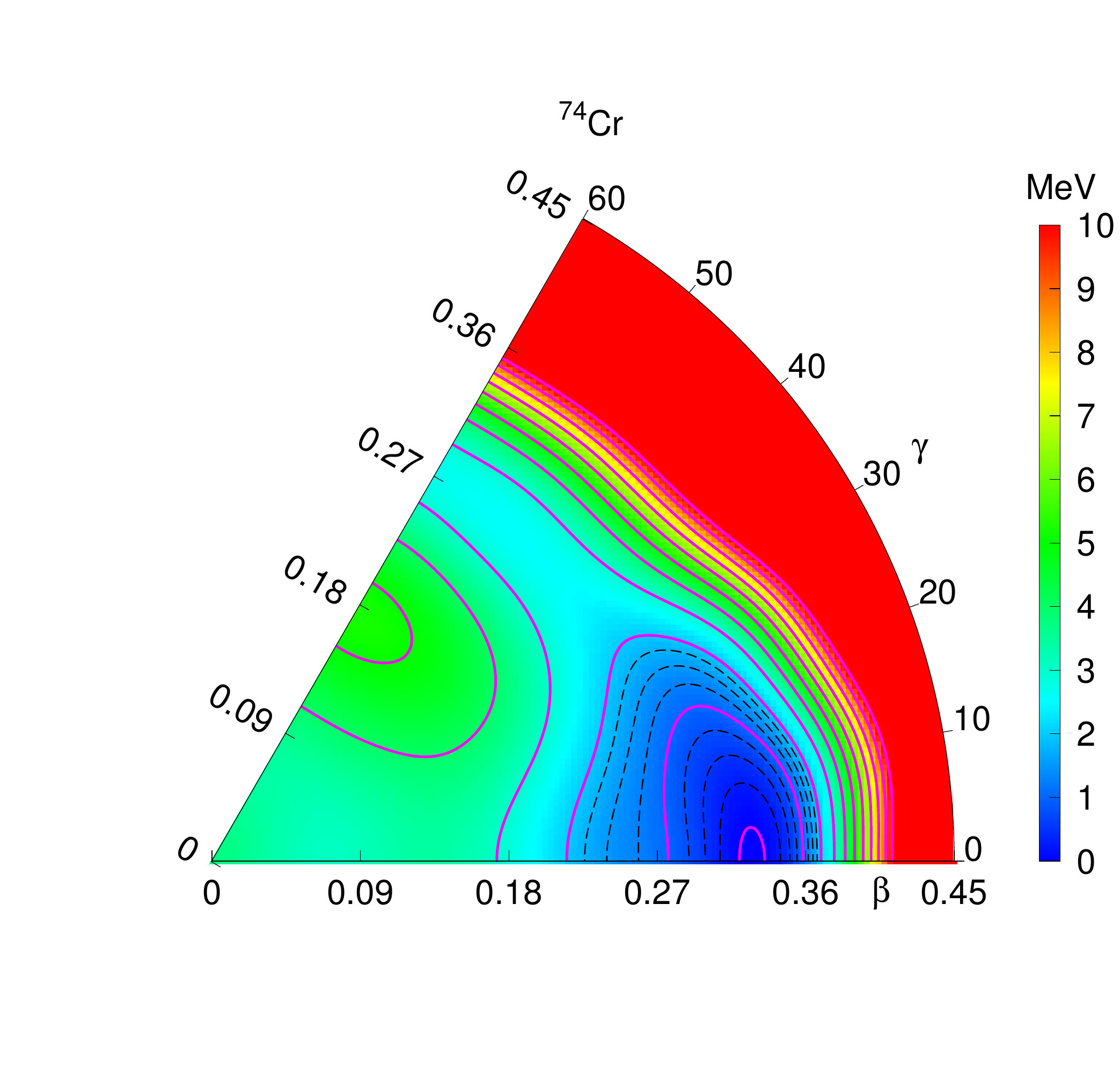}
\caption{Potential Energy Surfaces (PES) in $^{76}$Fe (left)  and  $^{74}$Cr (right). Theoretical predictions with the PFSDG-U interaction.\label{N50PES}} 
\end{center}
\end{figure}

\bigskip
\noindent
These results can be also analysed in the intrinsic frame, performing a deformed Hartree-Fock calculation in the same valence space and with the same interaction.
We have plotted the potential energy surfaces that are obtained for $^{76}$Fe and $^{74}$Cr in Fig. \ref{N50PES}.
Going from $^{78}$Ni towards  $^{70}$Ca, there is a rapid transition from sphericity to a collective regime. In $^{76}$Fe there are three minima two deformed,
prolate and oblate, and
one spherical. The shape coexistence in the intrinsic frame leads after full mixing to the situation described above in which the shapes of the two 0$^+$'s
states are somehow blurred and only a prolate structure remains at higher spins in  $^{76}$Fe.  The spectroscopy of $^{76}$Fe should be accessible at the RIBF within a week of beam time when a $^{238}$U primary beam intensity of 500 pnA will be reached. Indeed, simple comparison with the $^{78}$Ni spectroscopy by R. Taniuchi and collaborators gives that the secondary-beam production cross section will be lower by two orders of magnitude while, if $^{76}$Fe first excited states are low lying, the detection efficiency should be increased by a factor 5. If the $\gamma$-array resolution is considered identical for both measurements, an increase factor of $\sim$20 in primary beam intensity would be necessary. The high resolving power of GRETA and the high primary beam intensities soon available at FRIB could make this physics case accessible in a nearer future via in-beam gamma spectroscopy. 
In $^{74}$Cr, a single minimum at a large prolate deformation is seen in the PES, making it the first $N=50$ isotope inside the island of inversion below $^{78}$Ni with a well developed deformed ground state band. Unfortunately, this nucleus is out of experimental reach at current facilities and will probably not be accessed in the coming two decades. When accessible, its spectroscopy will give a unique benchmark to assess and quantitatively describe the onset of deformation at $N=50$. 

\subsubsection{The N=40 and N=50 IoI's merge}
\begin{figure}[h]
\vspace{2cm}
 \includegraphics[width=0.7\textwidth]{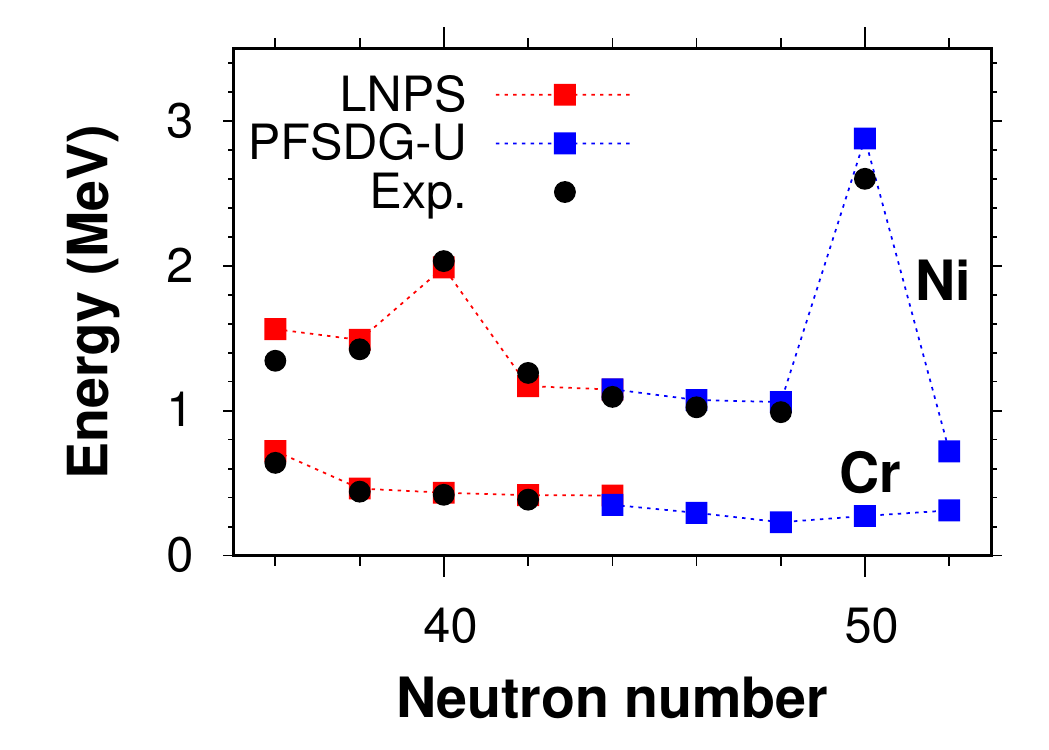}
\caption{Evolution of the excitation energy of the first 2$^+$ state in the isotopes of Nickel and Chromium. Theory vs. experiment where available. \label{ni.cr}} 
\end{figure}

  We have seen that the different IoI's share  many common features, that make it possible to consider them as as universal behaviour. We
  saw in Fig. \ref{mg_20-28} (right panel) that for the Magnesium isotopes  the N=20 and N=28 IoI's merge. If we make a similar plot
  for the neutron rich Nickel and Chromium isotopes as in Fig. \ref{ni.cr}, we observe that in the Nickel chain the 2$^+$ excitation energies have two
  clear peaks
  at N=40 and N=50, revealing their magic nature. On the contrary, in the Chromium isotopes, the 2$^+$ excitation energies are  not affected by
  none of the two neutron closures. We have not drawn the
  Iron results not to make the figure too busy, but its behaviour is identical to that of the Chromium's.
  Therefore, we can say that the N=40 and N=50 IoI's merge in the Iron and Chromium isotopes.
 The same conclusion can be reached by examining  Table \ref{tab:IOI50} where we have gathered some quadrupole properties of 
 the chromium and iron chains for N=46, 48 and 50. It is seen that the maximum absolute deformation corresponds to  $^{74}$Cr with
 $\beta^m$=0.39. Its lighter isotopes are nonetheless deformed as well. In the iron chain, where the deformation is smaller as discussed previously, the more deformed nucleus is again its N=50 isotope $^{76}$Fe.

\begin{table}[h]
    \centering
\begin{tabular}{|r|ccccc|}
    \hline \hline \\ [-4pt]
                &  E*(2$^+_1$)   & Q$_s$     &  BE2$\downarrow$ & $\overline{Q^m_0}$ & $\beta^m$     \\ [2pt] 
                &  (MeV)            & (e~fm$^2$) &  (e$^2$~fm$^4$)    & (fm$^2$)&       \\ [5pt]  
    $^{70}$Cr    &   0.30       &   -41     &   420             &   340   &   0.26      \\
    $^{72}$Cr    &   0.23       &   -48     &   549             &   407   &   0.30      \\
 $\mathbf{^{74}}$Cr   &  0.24&  -51& 630& 552   & 0.39      \\[5pt]
    $^{72}$Fe    &   0.44       &   -36     &   316             &   289   &   0.21      \\
    $^{74}$Fe    &   0.47       &   -39     &   330             &   308   &   0.22      \\
    $^{76}$Fe    &   0.35       &   -39     &   346             &   320   &   0.25      \\
\hline \hline 
 \end{tabular}
    \caption{Evolution of the collectivity in the Chromium and Iron chains approaching N=50}
    \label{tab:IOI50}
\end{table}

\subsection{N=40-50; above Nickel, miscellaneous results} 
\begin{figure}[h]
\vspace{2cm}
\includegraphics[width=0.45\textwidth]{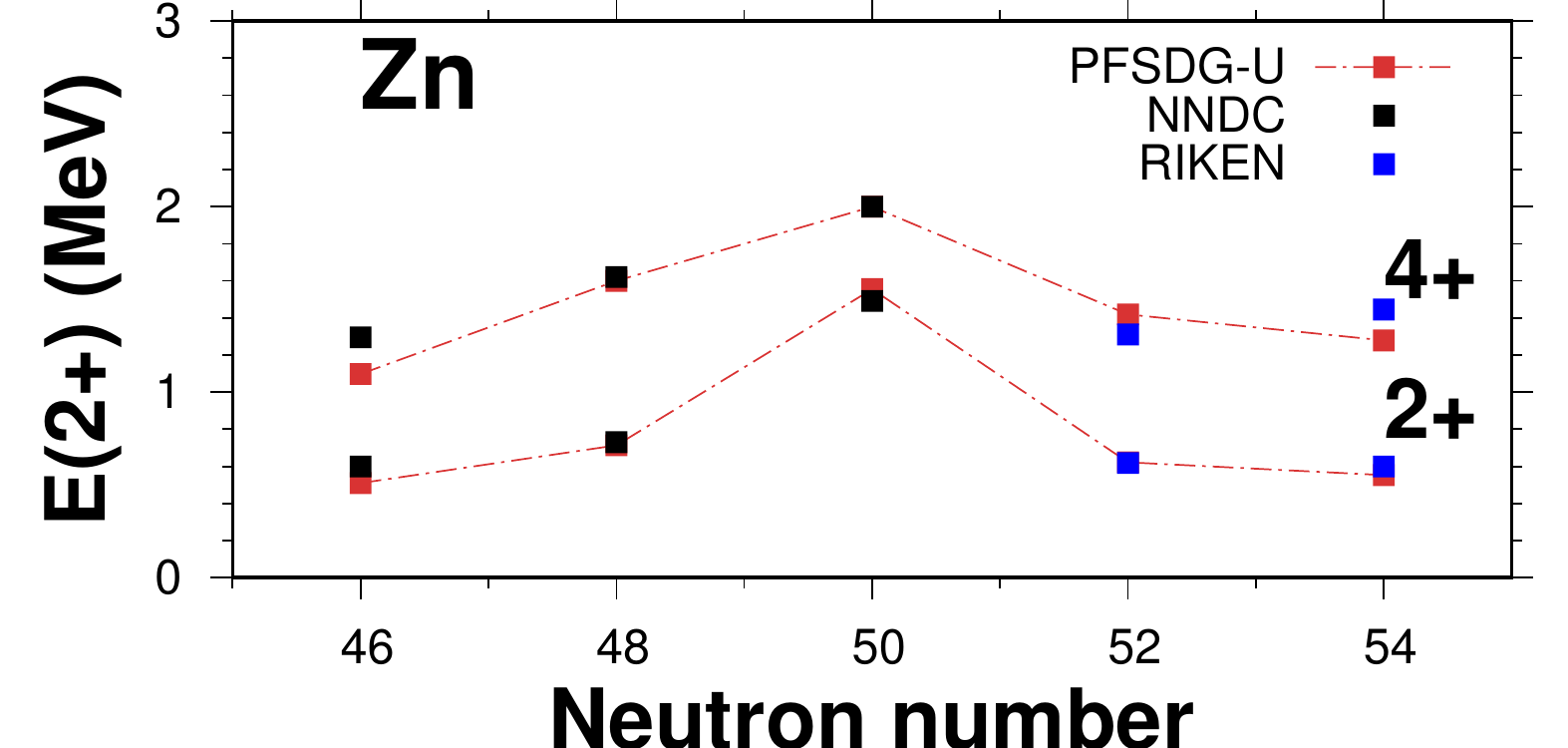} \hskip 20pt
\includegraphics[width=0.45\textwidth]{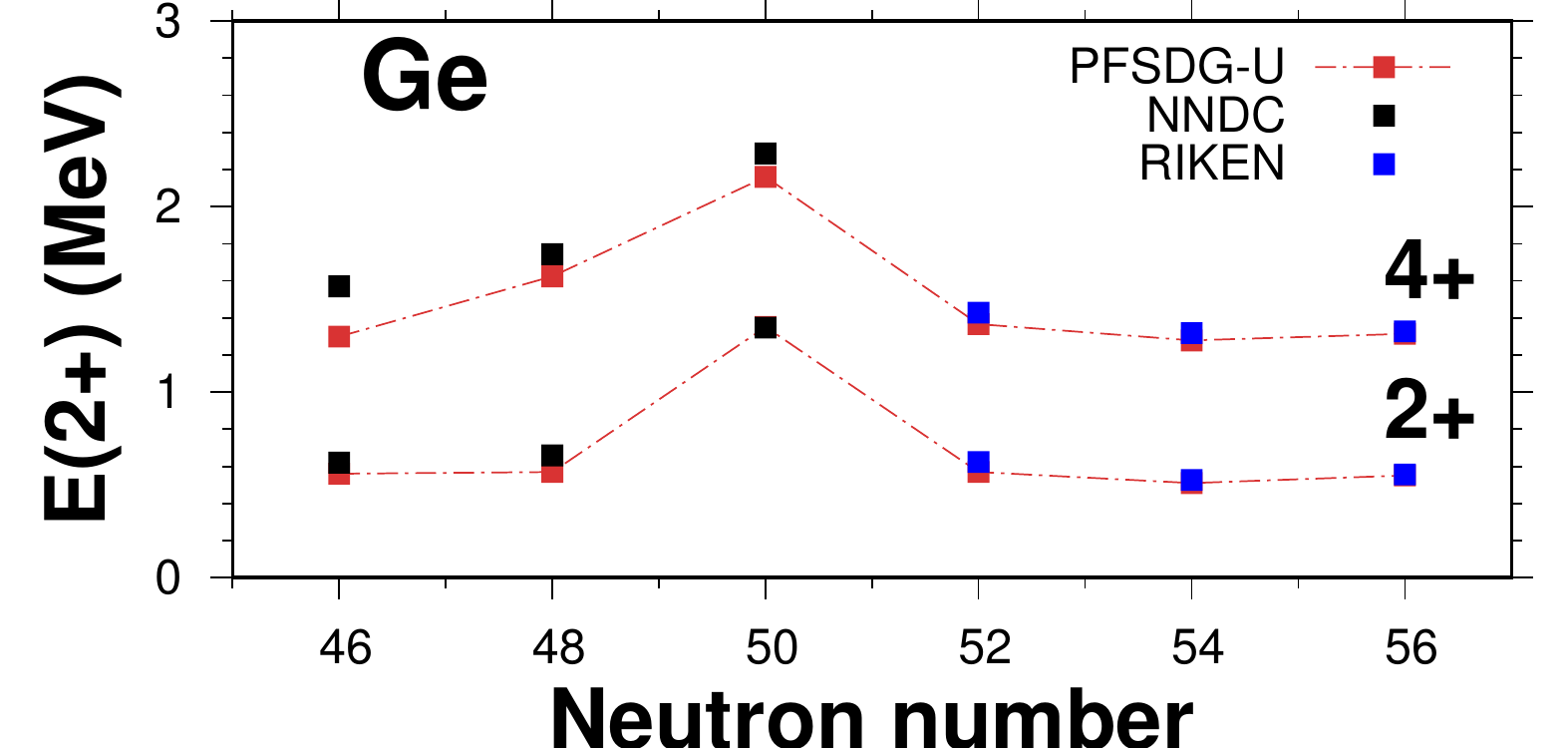}
\caption{Excitation energies of the 2$^+$ and 4$^+$ states of the Zinc and Germanium  isotopes. Experimental results from references \cite{shand2017,wraith2017, lettman2017} compared with
the SM-CI calculations with the PFSDG-U interaction. \label{Germaniums}}
\end{figure}

The spectroscopy along several isotopic chains reflects also the need of enlarging the neutron valence space beyond the 0g$_{9/2}$ orbit. We  present
in Fig. \ref{Germaniums} a comparison of the evolution of the 2$^+$ and   4$^+$ excitation energies predicted by the PFSDG-U calculations in the neutron-rich
Zinc isotopes with the experimental results from \cite{shand2017,wraith2017} and a similar one for the Germanium isotopes with the measures of reference \cite{lettman2017}.
The agreement is remarkable, and data and theory offer solid hints of the persistence of the N=50 neutron shell closure beyond Z=28.

\bigskip

\begin{figure}[h]
\begin{center}
\includegraphics[width=0.5\textwidth]{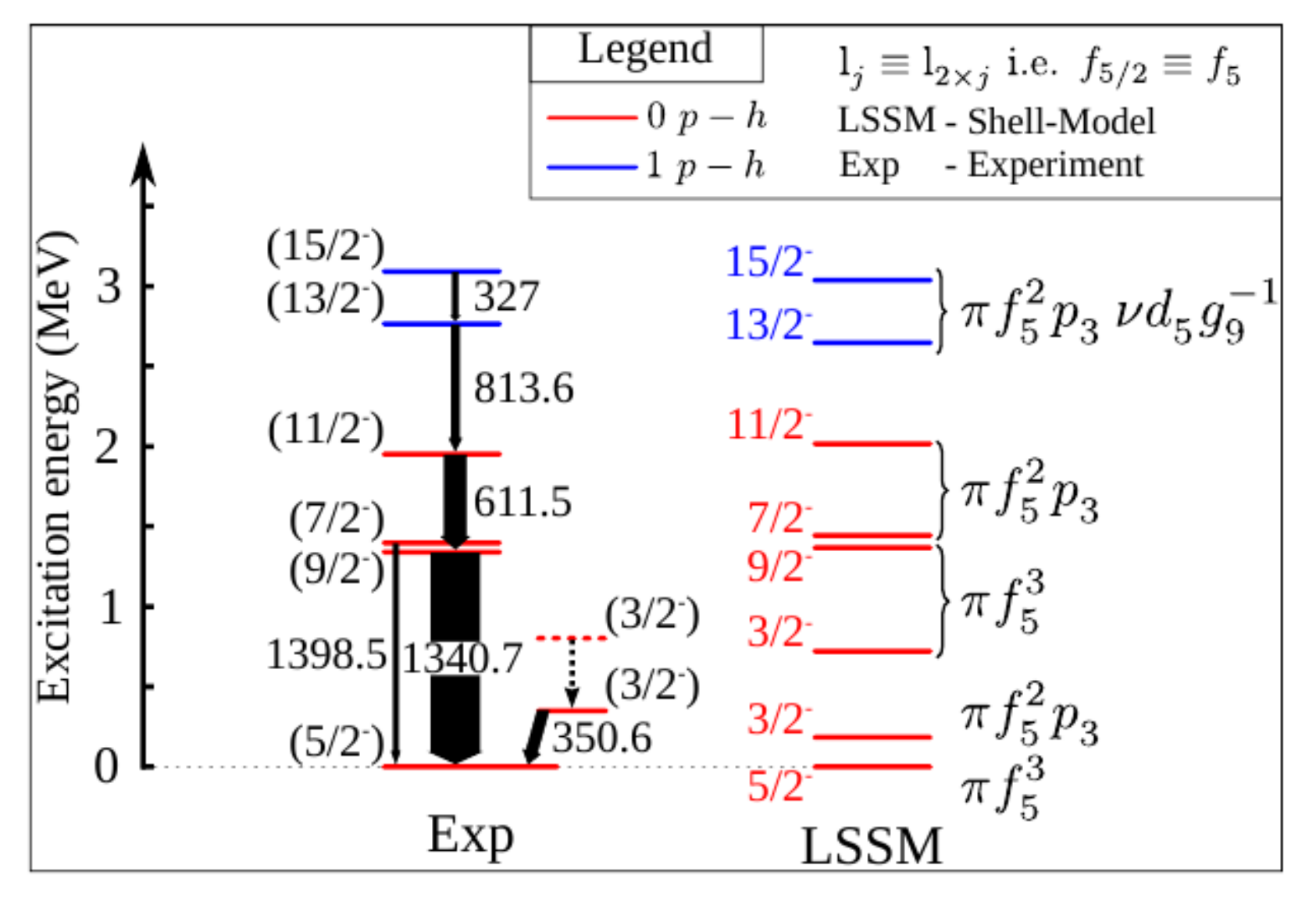} 
\caption{Experimental spectrum of  $^{81}$Ga \cite{dudouet2019} compared with the SM-CI results.\label{ga81} }
\end{center}
\end{figure}

\noindent
Another example of the excellent performance of the PFSDG-U interaction is provided by the comparison with the detailed level scheme of $^{81}$Ga, obtained recently at GANIL~\cite{dudouet2019}, that we show in Fig.~\ref{ga81}.
A study carried out at the ALTO facility focused on the $\beta$-delayed electron-conversion spectroscopy of photofission products selected on mass A=80 ~\cite{gottardo2016}.  It was shown earlier that $^{80}$Ga was the main isobar in the beam. The experiment showed a clear electron-conversion peak at 648 keV, corresponding to the decay of the 2$_1^+$ known excited state of $^{80}$Ge at an excitation energy of 659 keV after correction from the 11.1 keV binding energy of K electrons in Ge isotopes. A second transition, with less statistics, was observed at  628 keV. From this observation, it was proposed that the first excited state of $^{80}$Ge is an intruder 0$^+$ at 639 keV, located just below the excited 2$_1^+$ state. The SM-CI calculation does not produce such a state. A recent dedicated $\beta$-decay experiment at TRIUMF does not confirm the existence of this hypothetical low-lying 0$^+$ excited state~\cite{garcia2020}, contradicting the scenario of the appearance of low-lying intruders already at Z=32.  The calculation predicts the excited 0$^+$'s a bit below 2~MeV both in $^{80}$Ge and $^{82}$Ge. 

\noindent
A recent lifetime measurement of short-lived states in the region of neutron-rich Ge isotopes was performed at GANIL using the recoil distance Doppler-shift (RDDS) method combined with the AGATA detector. Nuclei were produced via fusion and transfer-fission of $^{238}$U beam particles impinging on a Be target. Reaction residues were unambiguously identified. The isotope $^{84}$Ge could be reach in the measured, although with low statistics. Only the 4$^+_1$ state could be observed as a feeder of the 2$^+_1$ state. The low-statistics data were analysed to extract s lifetime for the 2$^+_1$ state after correction from the observed feeding. A lifetime of 13.8$^{+7.9}_{-9.8}$ ps was determined, leading to a reduced transition probability of $B(E2;2^+\rightarrow 0^+)=621.2^{+1522.0}_{-226.2}$ $e^2$fm$^4$ \cite{delafosse2018}. Based on this result and assuming that the obtained central value is the correct one, C. Delafosse and collaborators concluded a shape transition along the N=52 isotonic chain from a triaxial shape in $^{86}$Se to a prolate shape in $^{84}$Ge, at variance with shell-model predictions. The low statistics of the measurement and the rather weak control of feeding call, first, for a precise measurement to confirm the $B(E2;2_1^+\rightarrow 0_1^+)$ value before any conclusion can be drawn on the collectivity and shapes of $^{84}$Ge's low-lying states. A Coulomb-excitation measurement at intermediate energies would minimize the effect of feeding. 
In general, the evolution of the excitation energy and wavefunction of intruder states as one approach Z=28 along the N=50 and 52 isotonic chains is important to pin down the details of the fifth island of inversion and constrain predictions for nuclei beyond $^{78}$Ni. In the near future, this mass-region should be investigated best at FRIB with the high-resolving power GRETA array.

\begin{table}[H]
    \centering
\begin{tabular}{|r|cccccc|}
    \hline \hline \\
 & E*$_{exp.}(2^+)$ &  E*$_{theo.}(2^+)$   & Q$_s$     &  BE2$\downarrow$  & $\beta^m$  & $\gamma^m$   \\ [5pt] 
    $^{80}$Ni    &  -    & 0.71           &   -20     &  187   & 0.19     &  30$^{\circ}$   \\
    $^{82}$Ni    &  -    & 0.54       &   +31     &   421  &   0.21   &  37$^{\circ}$   \\
    $^{84}$Ni  &   -    &  0.28       &   +54     &  723  &  0.26   & 46$^{\circ}$ \\[5pt]
    $^{82}$Zn    &  0.62 & 0.62       &   -36     &   316  &   0.20       &  28$^{\circ}$ \\
    $^{84}$Zn    &  0.60 & 0.55       &   -39     &   330  &   0.18       & 33$^{\circ}$  \\
    $^{86}$Zn    &  -    & 0.59       &   -39     &   346  &   0.17       & 39$^{\circ}$   \\[5pt]
    $^{84}$Ge    &  0.62 & 0.56       &   -28     &   352  &   0.19   &  28$^{\circ}$ \\
    $^{86}$Ge    &  0.53 & 0.51       &   +24     &   560  &   0.22   &  33$^{\circ}$  \\
    $^{88}$Ge    &  0.56 & 0.55       &   +39     &   400  &   0.22   &  41$^{\circ}$    \\   
\hline \hline 
 \end{tabular}
    \caption{Collectivity in  Nickel, Zinc and  Germanium  at N=52,  54 and 56 (units as in Table \ref{tab:IOI50}).}
    \label{tab:ni-zn-ge}
\end{table}

\noindent   
In Table \ref{tab:ni-zn-ge}, we present the predictions of the PFSDG-U interaction for the Nickel,
Zinc and  Germanium chains at N=52,  54 and 56. Where experimentally know,
the agreement with the 2$^+$ excitation energies is outstanding. 
The E2 properties indicate the onset of a regime of  (moderate) deformation for N$>$50. In the Zinc and Germanium isotopic chains,  the deformation originates from the underlying Pseudo-SU3/Quasi-SU3 symmetries as already discussed in 
\cite{sieja2013}. In the Nickel isotopes, the calculations show  the migration of an excited 
deformed structure in ${^{78,80}}$Ni with 0$^+$ band-heads lying at 2.5 MeV and 1.2 MeV respectively, to become yrast in   ${^{82,84}}$Ni.
 In particular,  ${^{84}}$Ni has the largest deformation (rather oblate) with $\beta$=0.28 and a very low 2$^+$ at 0.28~MeV
which indicates that intruder components are dominant. A similar behaviour is found in   $^{86}$Ge, although the calculated  value of $\gamma$ makes it close to triaxial. Experimental data are taken from references \cite{wolf2013,yang2016,shiga2016,yang2018}.

\section{The fate of other shell closures approaching the neutron drip line. N=70, N=82 and beyond} 

 Heavier candidates to doubly-magic nuclei (adjacent or not to new IoI's) involve the HO closures 70 and 110 and 
 the spin orbit ones 82 and 126. The combination with Z=28 is excluded because $^{98}$Ni is surely unbound. Hence,
 our lighter goal would be {\bf $^{\bf 110}$Zr}. The question
 about the possibility of a re-emergence of the HO shell closures has been in the air for some time. In particular, a quenching of the N=82 shell gap in favor of an increased N=70 gap in neutron-rich nuclei below $^{132}$Sn was proposed as a candidate explanation for the abundance of r-process nuclei around mass $\sim$ 110 \cite{kratz2005}. We have already shown that our calculations do not predict that to be the case for the other combination of two HO magic numbers Z=20 and N=40 at $^{60}$Ca, whereas most BMF calculations did.
A recent in-beam $\gamma$ spectroscopy experiment \cite{paul2017} has shown that $^{110}$Zr is strongly deformed, based on the assignment of two low-lying transitions to the decay of the first $2^+$ and $4^+$ states in {\bf $^{\bf 110}$Zr}. The inclusion of these data in the systematics of $2^+_1$ excitation energy and of the so-called $R_{42}=E(4^+_1)/E(2^+_1)$ ratio along the N=70 isotonic chain is unambiguous and reveals an increase of collectivity as protons are removed. 
It is worth to devote some time to
 explain the broad traits of the Zirconium isotopes with the heuristic tools developed in this review. A schematic view of
 the relevant orbits is drawn in Fig. \ref{vsp-schem}. SM-CI calculations of the Zr isotopes up N=58 were reported in ref.~\cite{sieja2009}, and extended with the MCSM techniques up to $^{110}$Zr in ref.~\cite{togashi2016}.
 
 \medskip
\noindent
At the Z=40 (N=40) HO closure, np-nh configurations are favored by the quadrupole interaction, because the particles occupy  the Quasi-SU3 block
of the $sdg$-shell and the holes the Pseudo-SU3 block of the $pf$-shell. They will become yrast, or will not, depending on the size of the N=40 gap.
It has been experimentally known since long that  $^{80}$Zr is strongly deformed \cite{lister1987} with $\beta$ $\sim$ 0.4. The SPQR model
for the 8p-8h configuration predicts Q$_0$= 350~efm$^2$ in very good agreement with the $\beta$ value deduced from the excitation energy of the
2$^+$ state. The addition of neutrons in the 0g$_{9/2}$ orbit, blocks the neutron excitations and increases the proton Z=40 gap, producing
a sort of local double magicity in $^{90}$Zr. Beyond N=50, the neutrons occupy the Pseudo-SU3 block of the $sdg$ shell and the collectivity
increases slowly till $^{96}$Zr. As seen in the right panel of Fig. \ref{pseudo-4}, six neutrons exhaust the collectivity of this space; adding
up to six more leaves the quadrupole moment unchanged, and adding more it starts diminishing. Therefore, it becomes more favorable to occupy 
the 0h$_{11/2}$ orbit. As discussed in reference \cite{togashi2016} this reduces drastically
the Z=40 gap (an example of CD-SE) and the 4p-4h proton configuration becomes yrast in $^{100}$Zr, producing a sharp shape transition. The
SPQR calculation predicts a large deformation with Q$_0$= 365~efm$^2$ which will remain approximately constant until (at least) $^{110}$Zr,
as it is the case experimentally, in view of the very similar excitation energies of the 2$^+$ states between N=60 and N=70 \cite{grodzins1962}. In the present 
estimations, we have not considered the possibility of a lowering of the 1f$_{7/2}$ and 2p$_{3/2}$ neutron orbits to reconstruct the Quasi-SU3 block
of the $pfh$ shell. If this  happens, the neutrons can increase the total deformation to $\beta$ $\sim$ 0.6, which doesn't seem to be the case in
 $^{110}$Zr. The next milestone nucleus is the hypothetical candidate to doubly magic  {\bf $^{\bf 122}$Zr}. With the models that we have explored so far,
 one can expect a competition between closed shell and deformed or even superdeformed solutions. Production this nucleus from fission is out of reach at the current facilities or upgrades to come in the next years. This nucleus can be in principle reachable by the removal of
 ten protons from a not-yet existing high-intensity fast beam of the doubly magic {\bf $^{\bf 132}$Sn}. A related open question is whether another IoI  might occur closer to the stability at $^{126}$Ru or $^{124}$Mo.

\begin{figure}[h]
\includegraphics[width=0.50\textwidth]{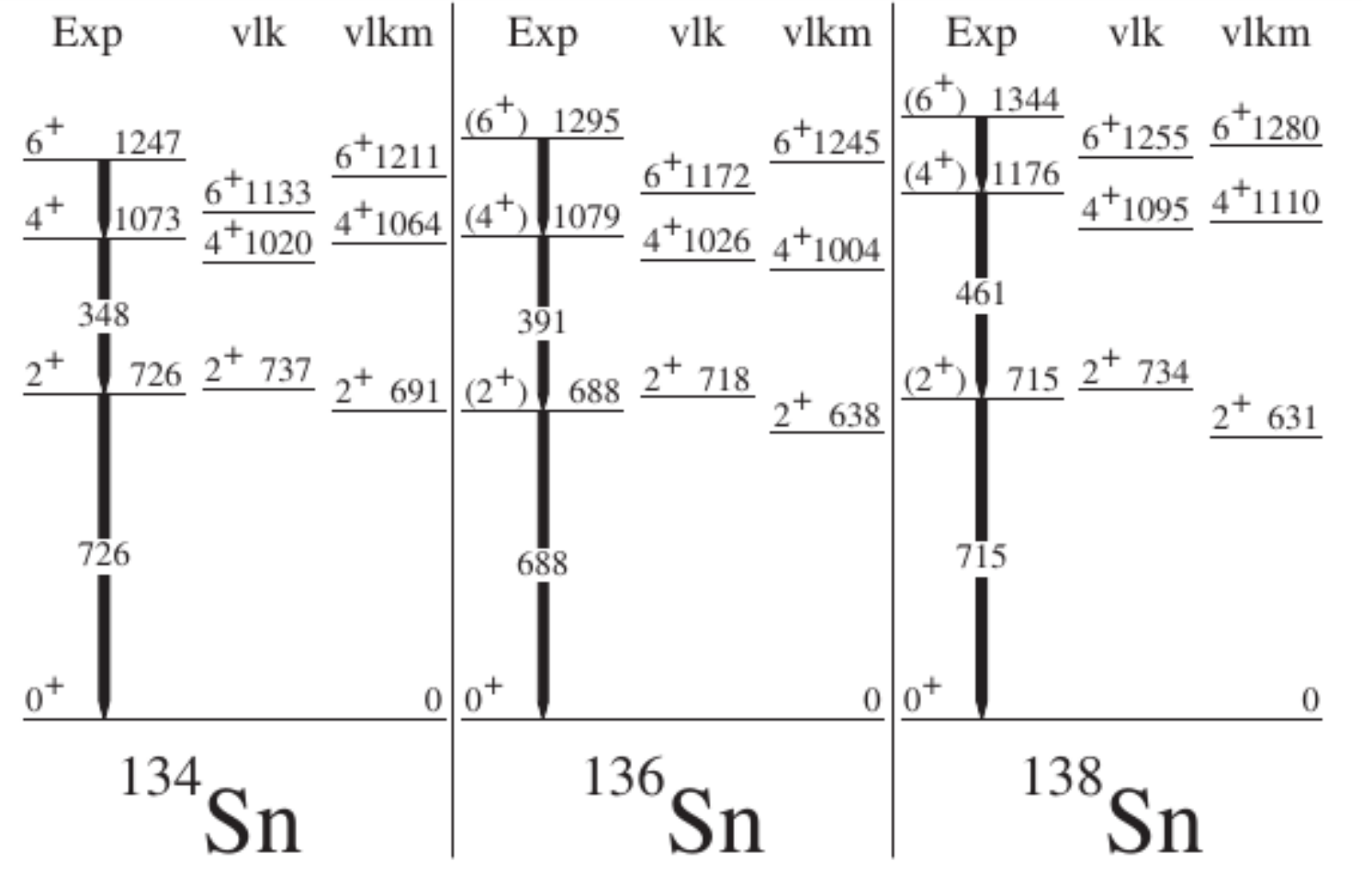} \hskip 20pt
\includegraphics[width=0.45\textwidth]{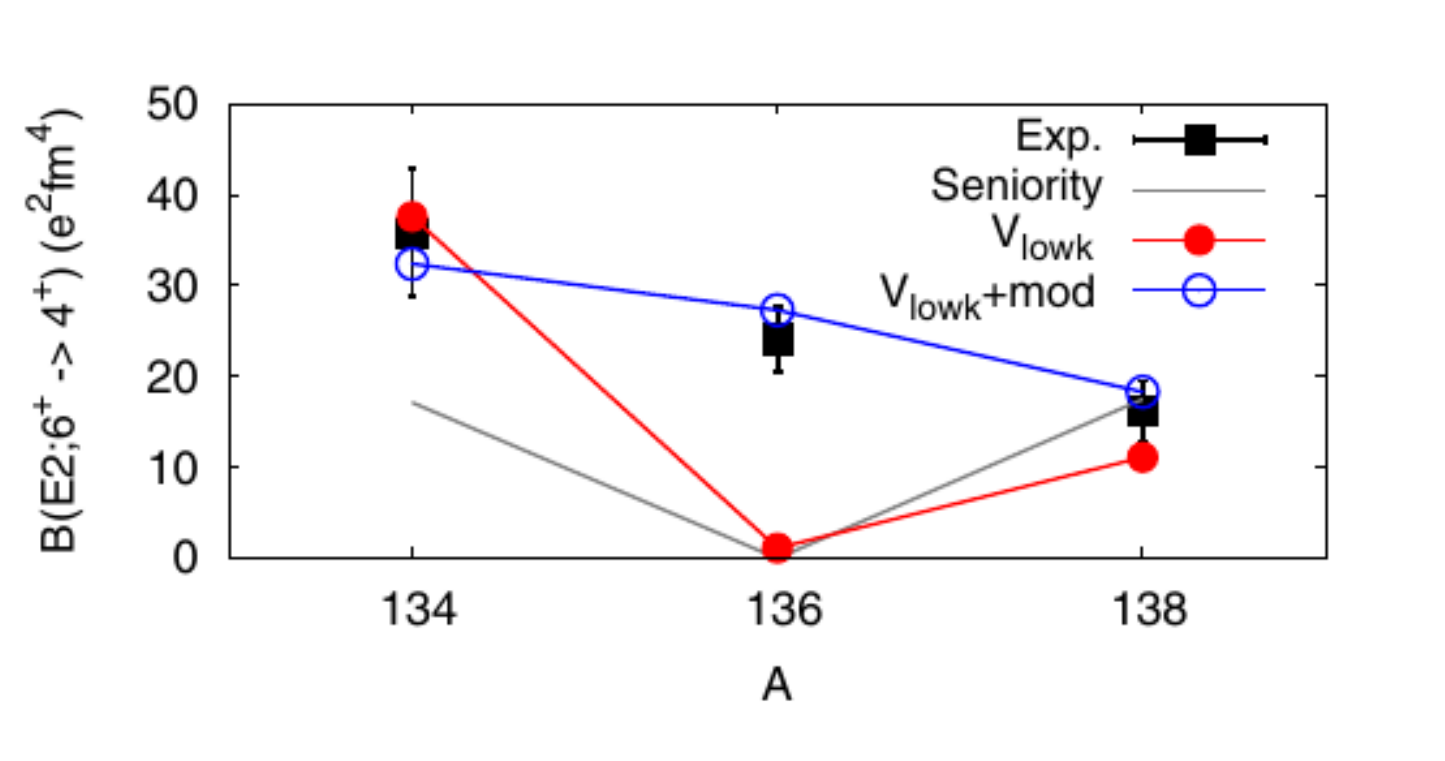}
\caption{(Right) Low-lying states of $^{134-138}$Sn, experiment compared with the LSSM calculations (see text). (Left) Comparison of the
seniority predictions for the $B(E2, 6^+ \to 4^+)$  in the same nuclei with the experimental results and the LSSM calculations\label{Sn-BE2}} 
\end{figure}

\medskip
\noindent
The close similarity of the mechanisms that govern the competition between the spherical mean field "normal filling states" and the intruder states,  emphasized in Fig.~\ref{vsp-schem}, can be extended to the whole neutron-rich regions at N=8-14, 20-28, 40-50 and 70-82, where the geometrical properties of the valence space exhibit the same degrees of freedom for the interplay between mean-field regularities (manifested in the shell closures) and the collective regime.
On the same line, as in previous $(sd)^\pi(pf)^\nu$ and $(pf)^\pi(sdg)^\nu$ descriptions, we will address the physics 
of the mass range $ 100 < A < 140$ in the $(sdg)^\pi(pfh)^\nu$ valence space in particular for the description of the
N=82 isotones and the core excitations across Z=50 and N=82 around $^{132}$Sn. 
Like in the lighter mass regions, the description of the core excitations in the region of double magic $^{132}$Sn requires two harmonic oscillator valence spaces: 
$0h_{11/2},1f_{7\
/2},0h_{9/2},1f_{5/2},2p_{3/2}, 2p_{1/2}$ orbitals for neutrons and $0g_{9/2},0g_{7/2},1d_{5/2},1d_{3/2},2s_{1/2}$ orbitals for protons above a closed $^{110}$Zr core. This model space allows in $^{132}$Sn to 
incorporate directly $0\hbar\omega$ quadrupole particle-hole excitations by opening the $\nu 0h_{11/2}$ and $\pi 0g_{9/2}$ orbits.
The effective interaction was derived from CD-Bonn potential, renormalized following the so-called $V_{low-k}$ approach, and adapted to the model space by means of many body perturbation theory. A few  monopole constraints were enforced into the realistic $V_{low-k}$ interaction to obtain the experimental single-particle energies of $^{133}$Sn, $^{133}$Sb, and
the $N=82$ and $N=83$ isotones, as well as  to  reproduce  the neutron and proton gaps evolution around $^{132}$Sn and $^{120}$Sn, incorporating the recent mass measurements in neutron-rich Cd isotopes \cite{Manea2020}. 
These modifications are  aimed  to incorporate the monopole part of the "real" $3N$ forces which are
not explicitly taken into account in the calculations.
In addition, the  calculated $B(E2, 2^+ \to 0^+)$  of   {\bf $^{\bf 132}$Sn} displayed in Fig \ref{Sn-BE2}, provides a sensitive test
of the effective interactions. In $^{134-138}$Sn, the breaking of the seniority scheme, as shown in Fig. \ref{Sn-BE2} (right)
can be interpreted as the fingerprint of the presence of core excitations. The excellent reproduction of the 
experimental B(E2) trend of the Tin isotopes \cite{simpson2014} by the LSSM calculations provide a very strong test of the effective interaction 
and make it possible to monitor the   cross shell excitations. The LSSM predictions 
for $^{132}$Sn are in very good agreement (see Fig. \ref{Sn-BE2}) with the recent coulomb excitation data from reference \cite{rosiak2018}.   
Although core excitations appear to be present and manifest in heavier Tin isotopes, the
properties of {\bf $^{\bf 132}$Sn} make evident its strong magic character. Contrary to the
$^{78}$Ni case, the $2^+$ lies very high in energy, it is of particle-hole character, and its E2 decay rate is relatively 
weak compared to the $^{78}$Ni one. Besides, no low-lying  deformed structure has been observed (or predicted). 

\bigskip

\hskip -15pt

\begin{figure}
    \centering
\begin{minipage}[t]{0.47\linewidth}
\vspace{20pt}
\centering
\includegraphics[width=1.0\textwidth]{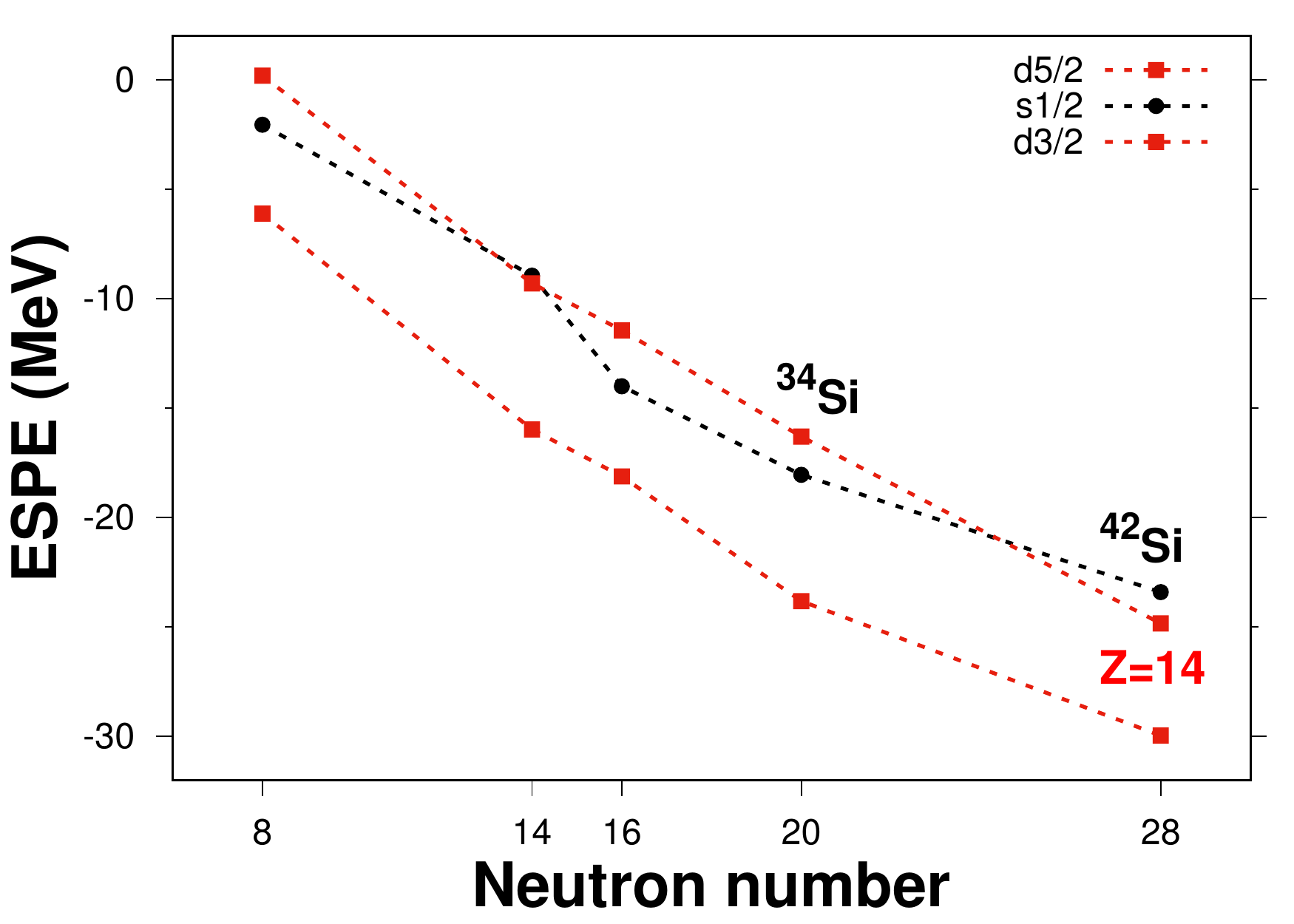}
\end{minipage}
\hskip 15pt
\begin{minipage}[t]{0.47\textwidth}
\vspace{0pt}
\centering
\begin{table}[H]
    \centering
 \begin{tabular}{rrrr}
    \hline \hline \\ [-2pt]
                &  G$_{matrix}$      &  SDPF-U       &   diff.      \\ [2pt] 
Total  &  -3.15         &   -2.38  &  +0.77       \\ [2pt]
  Central    & -+0.24     &  -0.11  &  -0.35   \\ 
Vector        &    -0.27      & +0.11  &   +0.22     \\ 
 LS       & -0.11    & +0.11  & +0.60  \\ 
 ALS      & -0.16     & +0.44  & +0.12   \\ 
 Tensor    &  -2.65     & -2.77  &  +0.12    \\ 
\hline \hline 
 \end{tabular}
\end{table} 
\end{minipage}

\vskip -15pt

\begin{minipage}[t]{0.47\linewidth}
\vspace{20pt}
\centering
\includegraphics[width=1.0\textwidth]{ESPE-Z28.pdf}
\end{minipage}
\hskip 15pt
\begin{minipage}[t]{0.47\textwidth}
\vspace{0pt}
\centering
\begin{table}[H]
    \centering
  \begin{tabular}{rrrr}
    \hline \hline \\ [-2pt]
    & G$_{matrix}$      &  PFSDG-U       &   diff.      \\ [2pt] 
Total  &          -3.21         &   -3.70  &  -0.49       \\ [2pt]
             -0.28     &  -0.23  &  +0.05   \\ 
  Vector        &            -0.081      &  -1.09  &   -1.01     \\ 
             -0.081    & -0.65  & -0.57  \\ 
             +0.00     & -0.44  & -0.44   \\ 
 Tensor    &              -2.84     &  -2.38  &  +0.46    \\ 
\hline \hline 
 \end{tabular}
\end{table} 
\end{minipage}

\vskip -15pt

\begin{minipage}[t]{0.47\linewidth}
\vspace{20pt}
\centering
\includegraphics[width=1.0\textwidth]{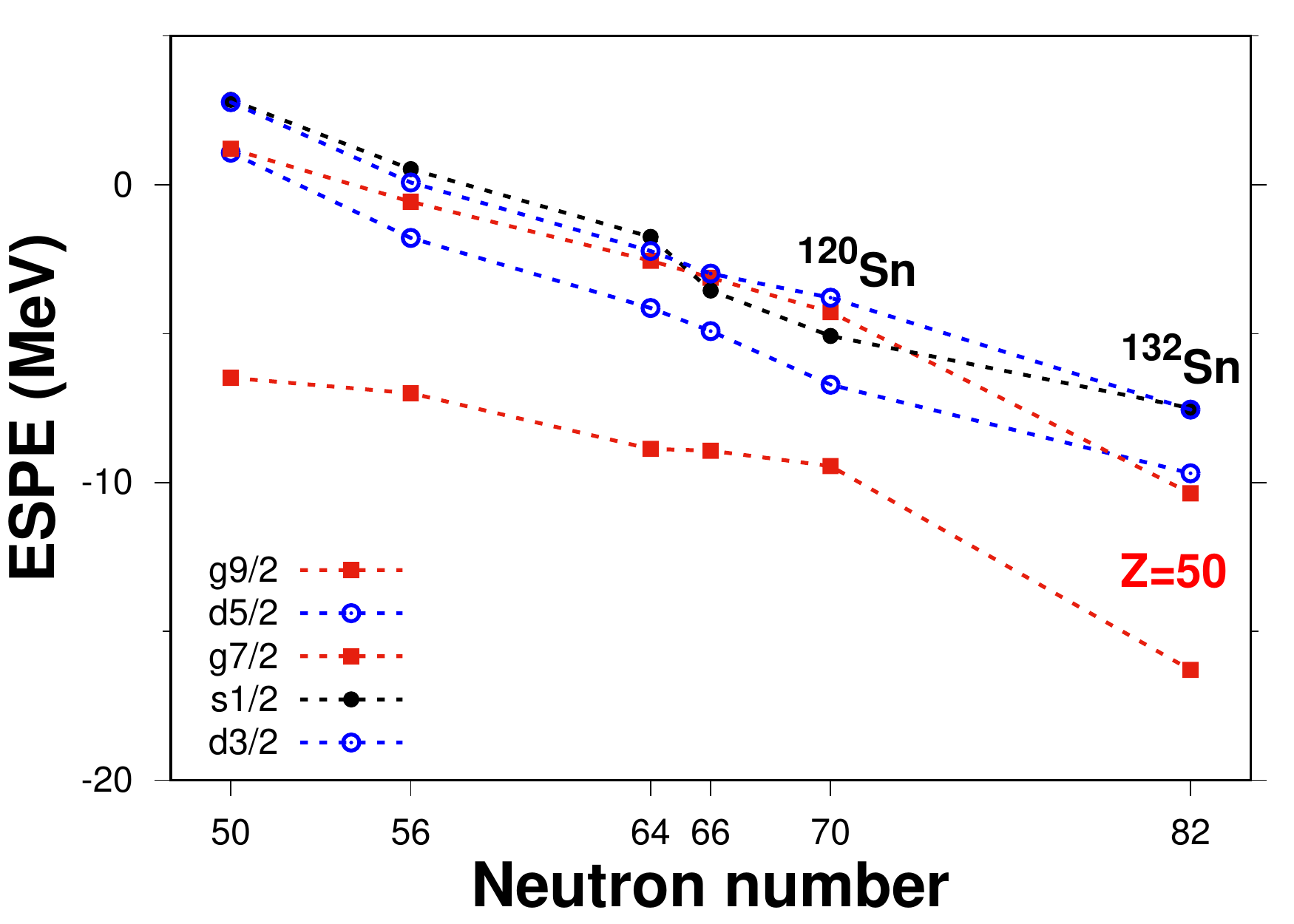}
\end{minipage}
\hskip 15pt
\begin{minipage}[t]{0.47\textwidth}
\vspace{0pt}
\centering
\begin{table}[H]
    \centering
 \begin{tabular}{rrrr}
    \hline \hline \\ 
   &  V$_{lowk}$      &  NNSP       &   diff.      \\ [2pt] 
 Total  &  -2.15         &     +0.89 &  +3.04       \\ [2pt]
 Central    &  -0.034     &  +0.30  &  +0.33   \\
Vector        & +0.12      & +1.55  &   +1.43     \\ 
LS & -0.038    & -0.06  & -0.022  \\ 
ALS &  +0.49     & +1.61  & +1.12   \\ 
Tensor    &  -2.30     &  -0.96  &  +1.34    \\ \hline \hline 
 \end{tabular}
\end{table} 
\end{minipage}
    \caption{(Top) Effective proton single-particle energies at Z=14 with the SDPF-U effective interaction (left) and the $\Delta$(0$d_\frac{5}{2}$-0$d_\frac{3}{2}$) with the filling  of the 0$f_\frac{7}{2}$ orbital between $^{34}$Si and $^{42}$Si (right). (Middle) Effective proton single-particle energies at Z=28 with the PFSDG-U effective interaction (left) and the $\Delta$(0$f_\frac{7}{2}$-0$f_\frac{5}{2}$) with the filling  of the 0$g_\frac{9}{2}$ orbital between $^{68}$Ni and $^{78}$Ni (right). (Bottom) Effective proton single-particle energies at Z=50 with the NNSP effective interaction (left) and the $\Delta$(0$g_\frac{9}{2}$-0$g_\frac{7}{2}$) with the filling  of the 0$h_\frac{11}{2}$ orbital between $^{120}$Sn and $^{132}$Sn (right).}
    \label{fig:gap}
\end{figure}

\noindent
At this point, it is interesting to notice the different regimes which develop in 
"Harmonic Oscillator" nuclei at the neutron-rich side: $^{42}$Si appears to be deformed, $^{78}$Ni exhibits
shape coexistence while $^{132}$Sn shows strong magicity.
In order to understand the physical differences which show up in these three cases,
we have plotted in Fig. \ref{fig:gap} the monopole behaviour of the 
effective interactions introduced so far. In the proton ESPE plots as a function of the neutron filling, one observes a common feature: an orbital  crossing  the Fermi level, resulting in the  Z=14, Z=28 and Z=50 gaps formed from the spin-orbit splitting (namely $0d_\frac{5}{2}$-0$d_\frac{3}{2}$,  $0f_\frac{7}{2}$-0$f_\frac{5}{2}$ and  $0g_\frac{9}{2}$-0$g_\frac{7}{2}$ ) for the three regions. The crossing of these proton  orbits has been experimentally demonstrated by extensive  
studies of the spectroscopy of the Potassium, Copper and Antimony isotopes
 in Refs. \cite{flanagan2009,degroote2017,sahin2017,vajta2018}. 
 On the other hand the variation of the proton gaps with the neutron filling differs significantly in the three cases. The Z=14 and Z=28 gaps get reduced with the filling of the  $0f_\frac{7}{2}$ and $0g_\frac{9}{2}$ orbitals respectively, while the Z=50 gap gets increased with the filling of the $0h_\frac{11}{2}$ orbital.
 
\medskip
\noindent
Further insight on the mechanisms at play can be obtained analysing the  effective interactions in terms of their central, vector 
and tensor parts, through the spin-tensor decomposition, that we present in the tables inserted in Fig. \ref{fig:gap}.
The different physical outcomes  can be traced back to the result of the competition between the
spin-orbit and tensor components of the effective interaction as function of the mass region.
In light systems, the tensor part dominates the gap evolution, as it is the main ingredient to reduce the spin-orbit splitting up to $^{42}$Si. On the other hand, the vector  component appears to be a major player in mid  and heavier mass regions around A$\sim$80 and 
A$\sim$132. In the latter region, its amplitude is as large as the tensor component and even  counterbalances the tensor mechanism, 
producing a net increase of the proton gap in $^{132}$Sn, responsible for its observed strong magic character.

 \medskip
\noindent
 Indeed, after this analysis one can also address the issue of the N=82 shell quenching, as already observed for the N=28 and N=50 isotones. This issue has been debated in the past because the structure of  the waiting-point nuclei affects the r-process abundance distribution \cite{kratz2005,dillman2003,  zhi2013}. 
 The experimental information show the N=82 isotopes remain spherical and semi-magic for $^{130}$Cd and $^{128}$Pd with a high 2$^+$ excitation energy \cite{jungclaus2007}.
Using the NNSP effective interaction, deformed Hartree-Fock minima sit at the spherical point  for the lighter $^{126}$Ru, $^{124}$Te  and $^{122}$Zr isotopes, a situation appearing to be at variance with lighter mass regions  and the N=82 shell closure and the associated spherical regime seem to dominate at the mean-field level. Further investigations with beyond mean-field approximations and beta decay half lives evaluation will provide a detailed assessment of its astrophysical impact.
 
\newpage 
\section{Conclusions and outlook}
 We have examined in this review the physics of the most neutron-rich nuclei up to mass $\sim 150$.
 As our anchor points in this navigation of the nuclear chart, we have selected the neutron-rich doubly magic nuclei, $^{34}$Si, $^{68}$Ni, and
 $^{78}$Ni, which show coexisting deformed intruder states, and appear to be the portals of the Islands of Inversion which have been documented in their lower Z isotones.
 At present, the theoretical models which can reproduce (or predict) the spectroscopic properties of large sets of nuclei are dominated by the Shell Model with Configuration Interaction at Large Scale, which appears in the literature under the acronyms SM-CI, LSSM, ISM, MCSM, often simply shortened as SM, which is the main theoretical content of the review. The other family
 of models is rooted in the mean field description with density dependent interactions (or energy density functionals) which in order to become real spectroscopic tools must go beyond the mean field approximation, including the correlations via symmetry breaking. In a next step, they require the restoration of the broken symmetries, and the re-mixing of these solutions through the Generator Coordinate Method (GCM) before comparing with the experimental data. This procedure comes under the name SCCM (Symmetry Conserving Configuration Mixing) in a clear attempt to converge with the SM-CI approach \cite{robledo2019}.

 \medskip
 \noindent
 All these methods are based in the use of effective
 "in medium" nucleon-nucleon interactions. These range from the fully empirical SM-CI interactions, whose one and two-body matrix elements are fitted to a large set of
 experimental excitation energies, to the "ab initio" approaches aiming at a rigorous contact with the bare nucleon-nucleon interaction.  The 
 "ab initio" approaches have, as of now, several shortcomings: i) The connection with the fundamental theory of the strong interaction is made via Chiral perturbation
 theory which naturally induces many body interactions in the perturbation series. This requires, at this very basic level, the definition of a scale, and the
 presence of several free parameters which must be fitted to the few-body experimental data. Up to now there is not a unique choice even of these starting parameters,
 and often the good choice for the few body systems fails when applied to the standard nuclear many-body problem. ii) In addition, two other steps are necessary to
 make the description applicable to finite nuclei, to regularize the short range repulsion of the nucleon-nucleon interaction and to project properly
 the resulting one into the valence space (the same problems that Kuo and Brown faced fifty years ago). In these steps, induced many-body interactions
 and new cut-offs enter into the calculations. Therefore, when the VS-IMSRG method finally produces spectroscopic results to compare with the experiment,
 via a conventional SM-CI calculation, it requires to make a non negligible number of choices among the interactions and cut-offs.
 Advances in the "ab initio" techniques should improve the situation in the near future  \cite{stroberg2019}.
 A mixed approach is advocated in this review, which
 is based in the monopole multipole separation of the effective interactions. The latter can be computed "ab initio" without any problem, whereas the monopole
 part, that  should eventually incorporate all the many body components either "real" or  arising in the renormalization procedure, is,  for the time being,  fitted to a few, 
 simple, reference nuclei (typically doubly magic ones plus or minus one nucleon).
  
 \medskip
 \noindent
 The monopole plus multipole  strategy is described in full detail, with particular emphasis on the heuristics of the quadrupole-quadrupole channel
 of the interaction (SPQR)  which is the driving force of the nuclear dynamics in most situations. In particular it fuels the multi-particle-hole intruder states which are
 responsible for the appearance of the Islands of Inversion nearby doubly magic neutron rich nuclei. Let's mention here a very  recent review more focused in the
 shell evolution (see reference \cite{otsuka2020}). The IoI's at N=20, 28, 40 and 50 are studied in depth both
 in their experimental manifestations and the techniques that have made it possible to access them,  in  the theoretical interpretation of the data, and in some cases, even in the theoretical predictions, that culminate with the scenario of coexistence in the long sought doubly magic $^{78}$Ni.
 Our overarching purpose has been to unveil the general physical mechanisms which are common to the different regions explored, and in addition to
 single out the particularities that provoke differences in  physical behaviour among them. Indeed, this allows to prospect in which other regions  could the same kind of physics show up. This is the case for the next harmonic oscillator and spin-orbit magic neutron closures far from stability at N=70 and N=82 that we have briefly touched upon.
 
\begin{figure}[H]\hskip 0pt
\includegraphics[height=0.25\textwidth]{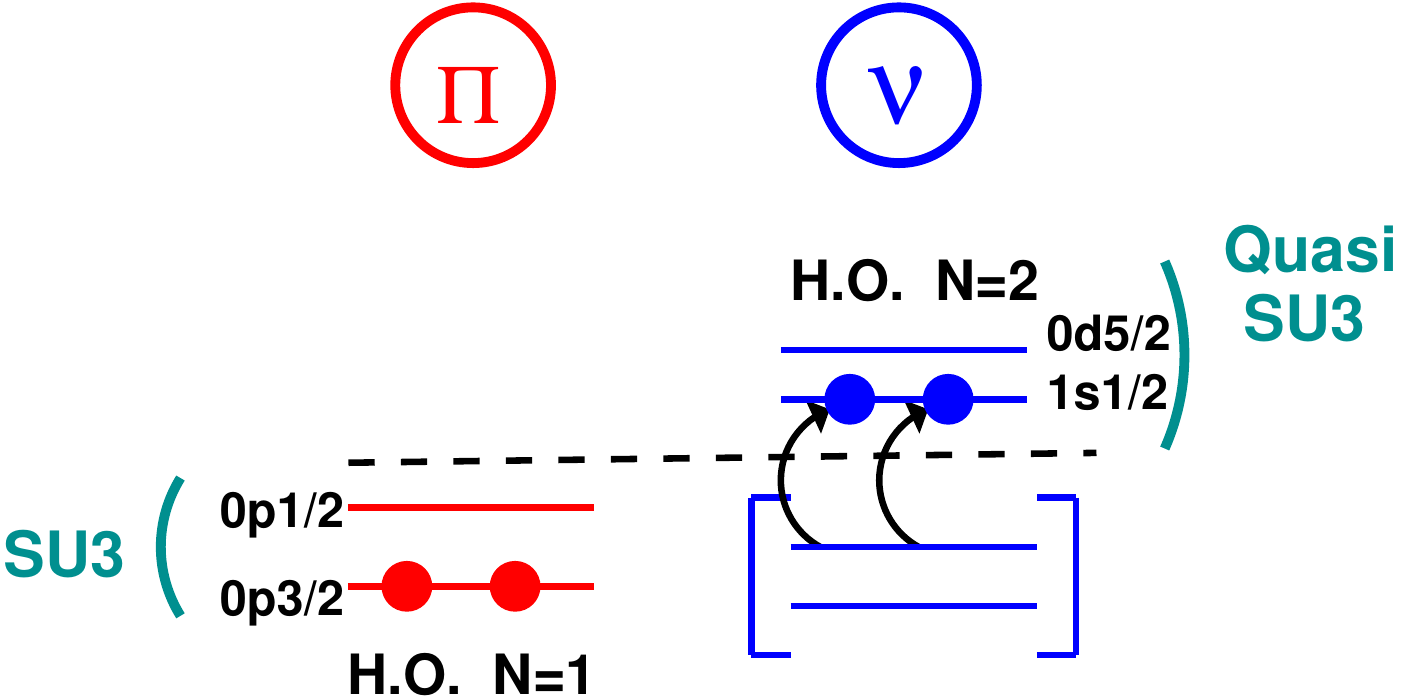} 
\hskip 20pt
\includegraphics[height=0.25\textwidth]{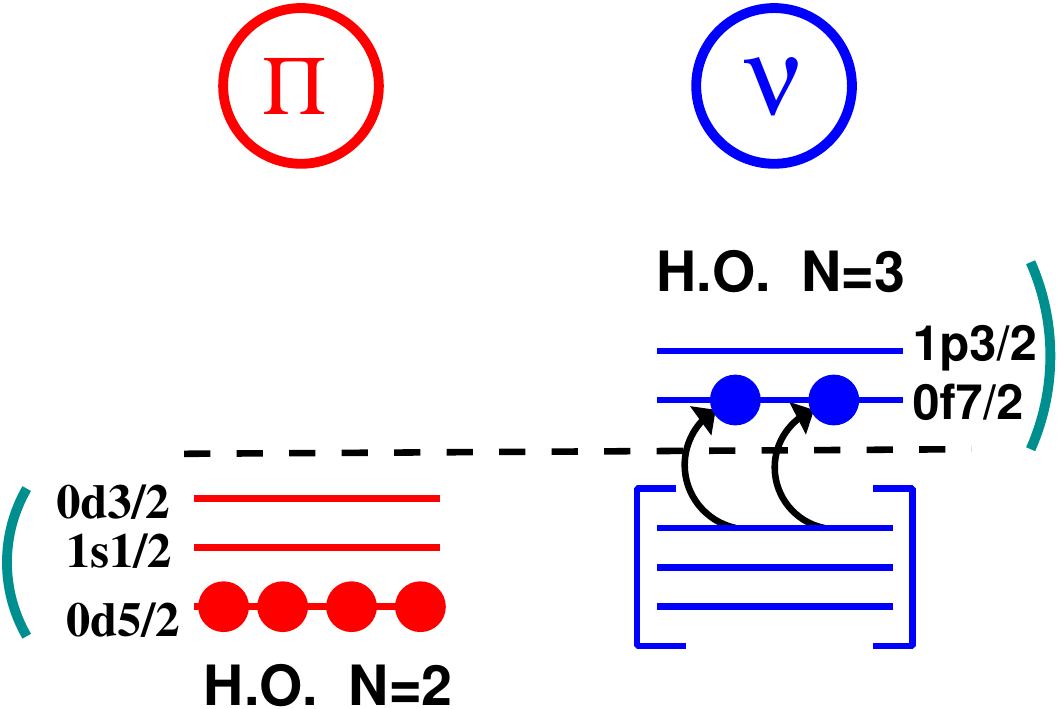}
\vskip 40pt
\includegraphics[height=0.22\textwidth]{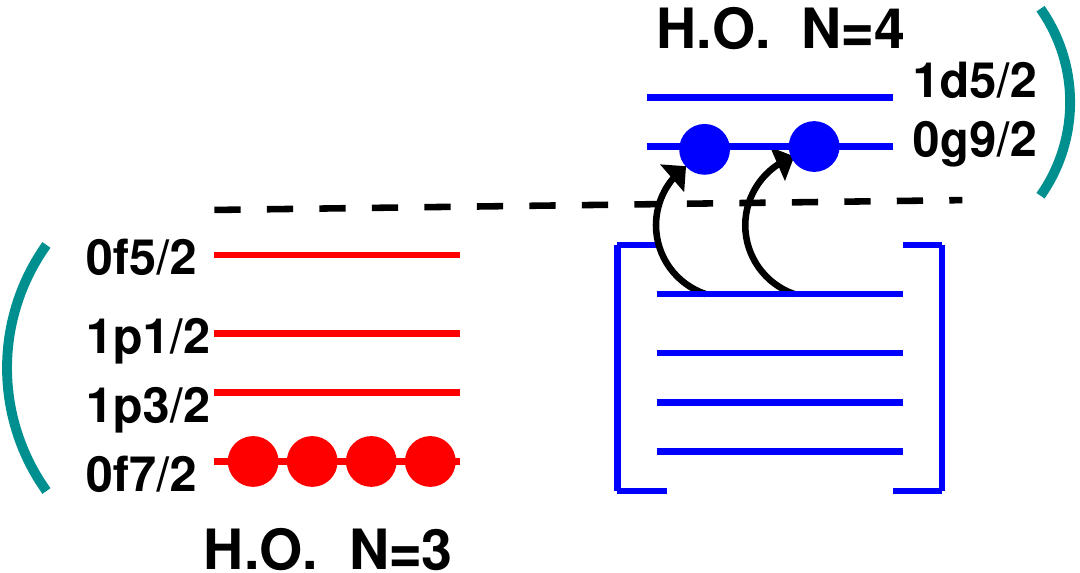} 
\hskip 40pt
\includegraphics[height=0.225\textwidth]{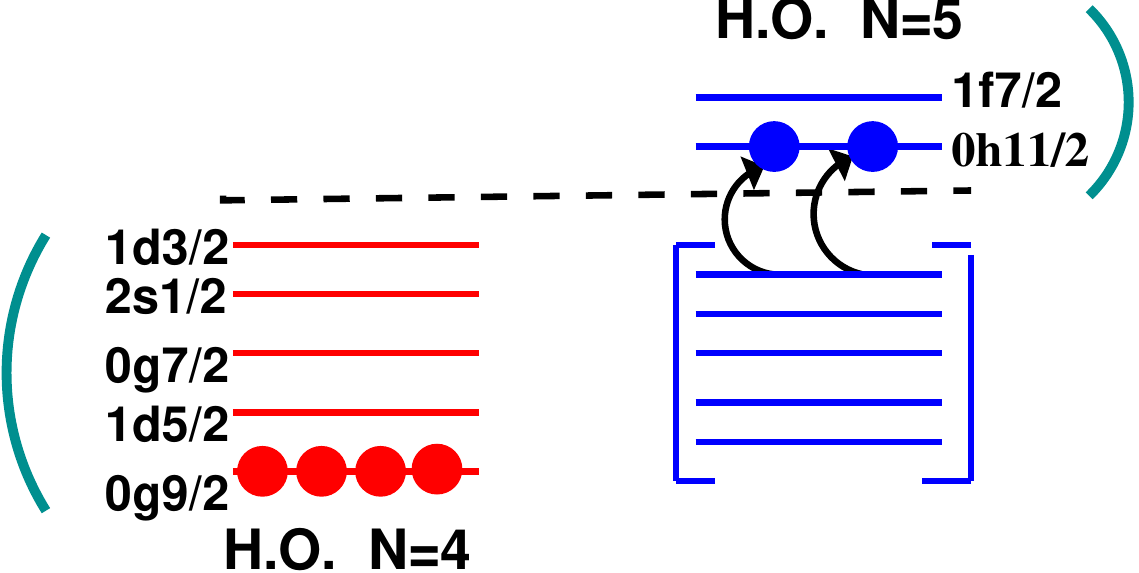}
\caption{Schematic view of the valence spaces at N=8, 20, 40 and 70. The intruder configurations that develop quadrupole collectivity are
          highlighted.\label{vsp-schem}} \end{figure}

\begin{figure}[h]\hskip 20pt
\includegraphics[height=0.25\textwidth]{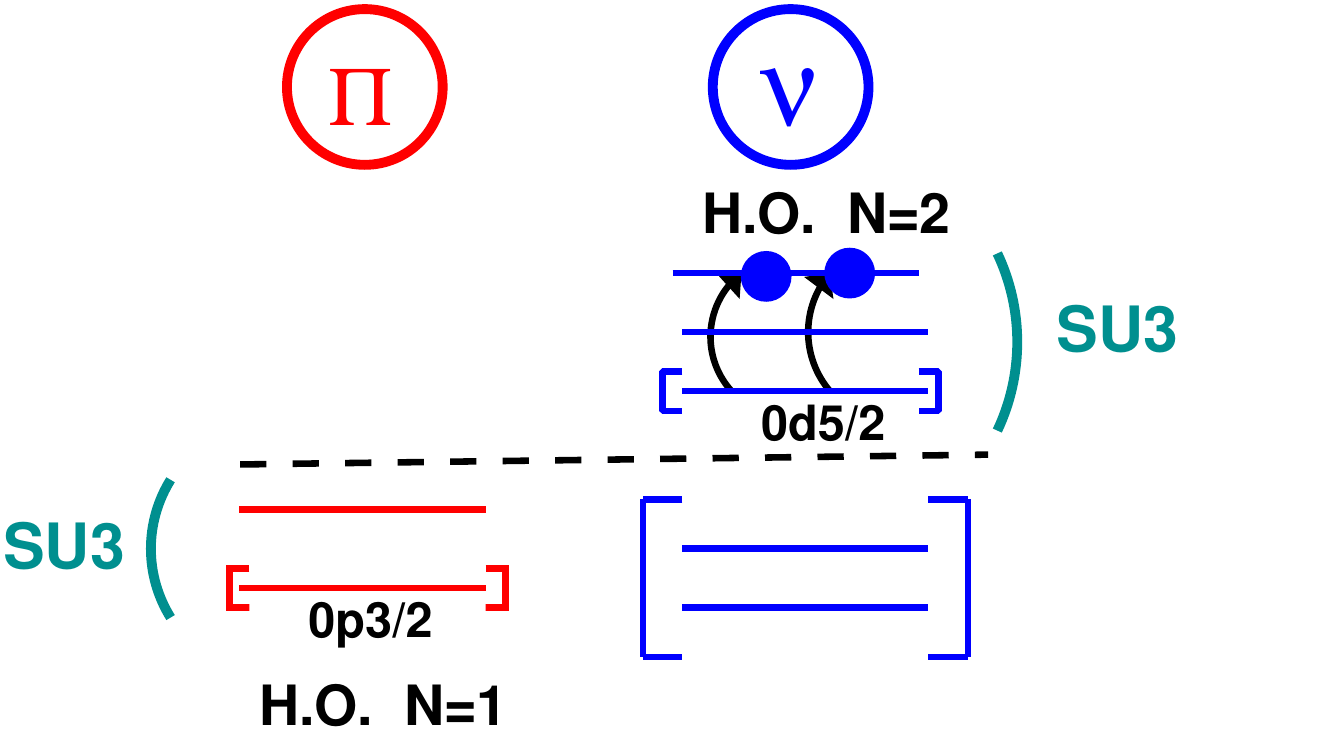} 
\hskip 20pt
\includegraphics[height=0.25\textwidth]{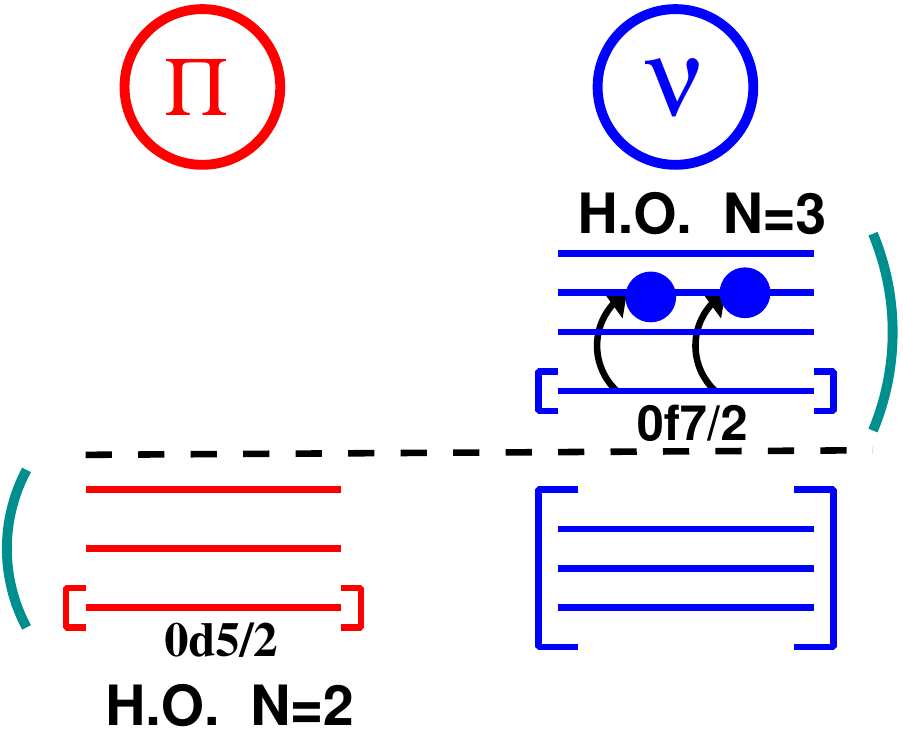}
\vskip 40pt
\hskip 20pt
\includegraphics[height=0.25\textwidth]{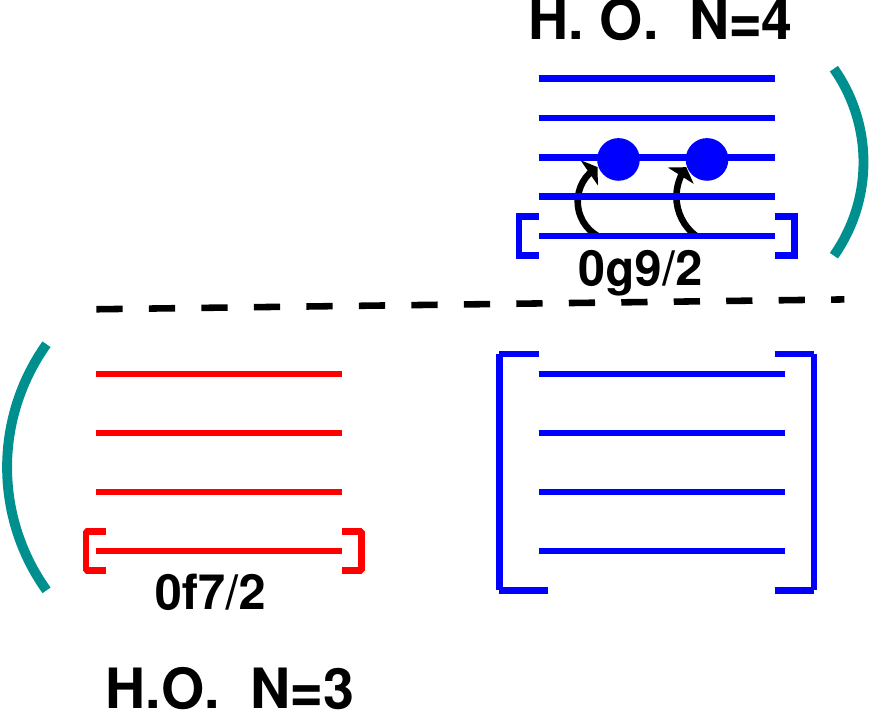} 
\hskip 40pt
\includegraphics[height=0.25\textwidth]{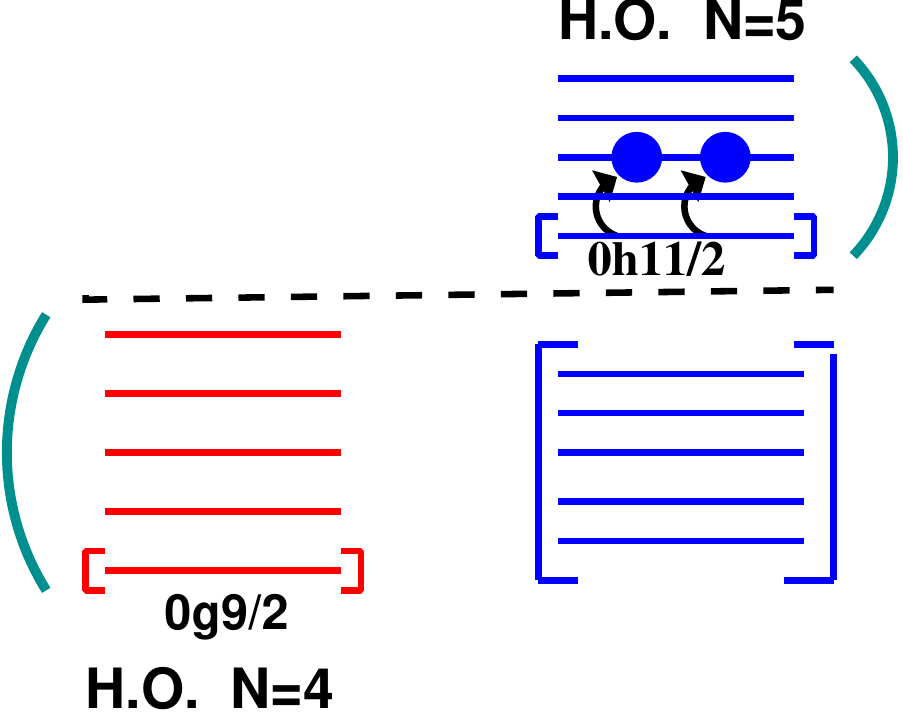}
\caption{Schematic view of the valence spaces at N=14, 28, 50 and 82. The intruder configurations that develop quadrupole collectivity are
          highlighted.
 \label{vsp-schemII}} 
\end{figure}

\begin{figure}[h]
\begin{center}
\includegraphics[width=1.0\textwidth]{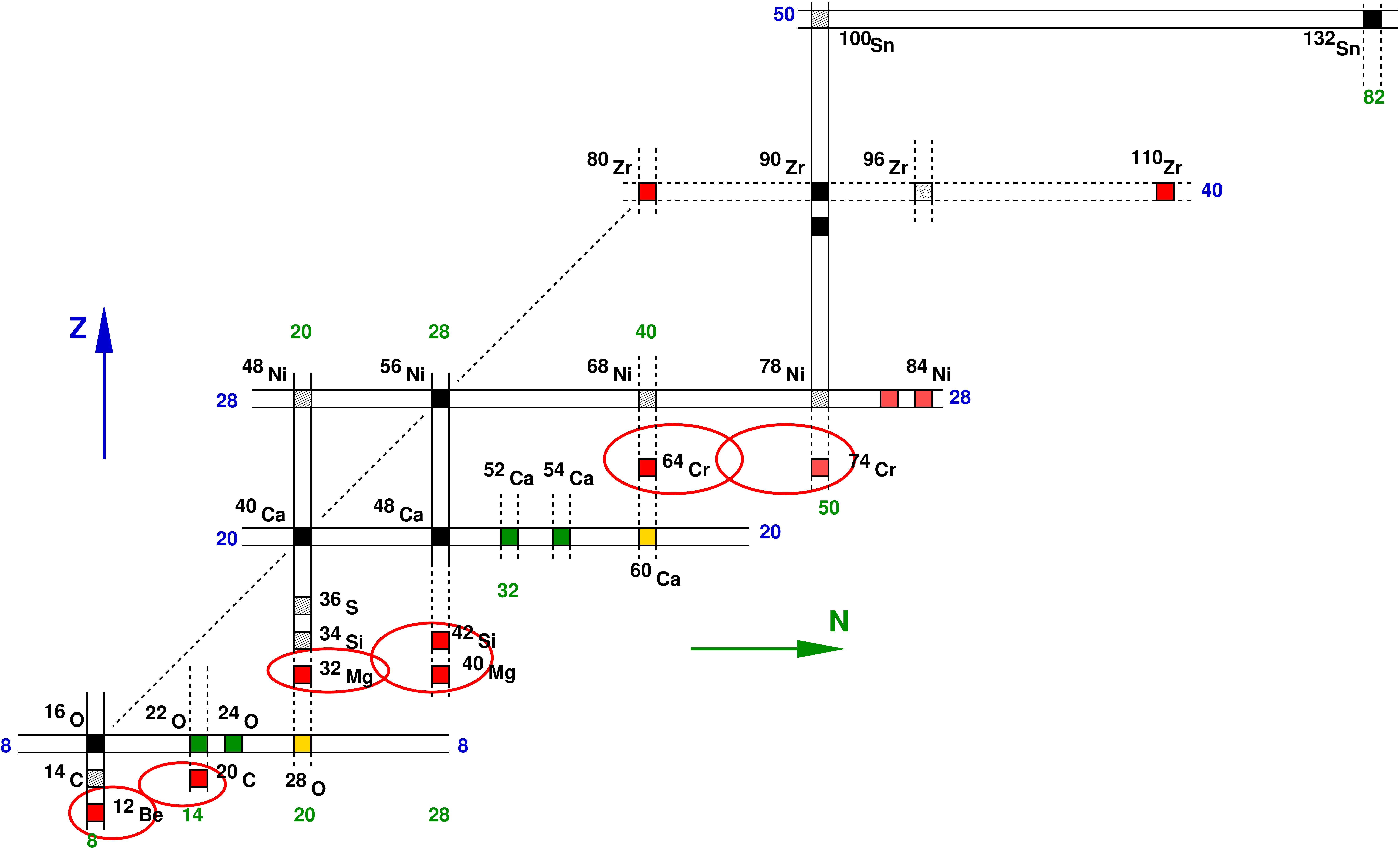} 
\caption{Landscape of the light and mid-mass part  of the Segr\'e chart. The Islands of Inversion (Deformation)  are enclosed by  red
ellipses. Notice that for certain elements, a couple of adjacent IoI's merge. We lack of enough experimental information about the very neutron 
rich Beryllium  isotopes. Indeed, $^{18}$Be is neutron unbound, as it is the first excited state of $^{14}$Be. However it is tempting to submit
that the N=8 and N=14 IoI's might merge too in some isotopic chains. The color code is the same as in Fig. \ref{landscape-ini}.
\label{landscape-fin}}
\end{center}
\end{figure}

\medskip
 \noindent
The common features of the deformation regions and islands of inversion  we have addressed 
throughout this review can be  graphically  summarized  in Figures~\ref{vsp-schem} and \ref{vsp-schemII}.
Physically sound valence spaces can made that include a single harmonic oscillator shell for protons,  and two contiguous harmonic oscillator
shells for neutrons, but algebraic symmetries further reduce the leading orbitals involved  at the Fermi surface to minimal
Pseudo-SU3 or Quasi-SU3 subspaces. This is the case for the 
physics developing at the neutron 
HO magic numbers N=8, 20, 40 and 70, that results from the competition between  the normal filling and the intruder configurations 
which produce deformed yrast solutions or shape coexistence, the motto of  this review. It is illustrated for  the three central cases 
in the IoI's below doubly magic $^{14}$C, $^{34}$Si and $^{68}$Ni,whereas
$^{114}$Ru is included for completeness, although its potential doubly magic reference, $^{110}$Zr, is 
itself strongly deformed.
Another kind of template for deformation is summarized 
in Fig. \ref{vsp-schemII} for the spin-orbit magic numbers N=14, 28, 50 and 82, highlighting the nuclei that have a spin-orbit magic
proton number as well; Z=6 in $^{20}$C, Z=14 in $^{42}$Si, Z=28 in $^{78}$Ni and Z=50 in $^{132}$Sn. As discussed in the text, different behaviors emerge
driven by the evolution of the spherical mean field; producing, in turn,  a well deformed (oblate) $^{42}$Si, a robust doubly magic $^{132}$Sn, and the 
coexistence of a doubly magic ground state with a low lying prolate deformed band in $^{78}$Ni.
In addition, the physics developing at these neutron-rich edges overlaps in several cases, as shown in Figure~\ref{landscape-fin}, providing striking similarities and a kind of universal paradigm for light and medium mass neutron-rich nuclei.


  \medskip
 \noindent
 Some of the most exciting nuclei that we have dealt with in this review are either poorly known or even completely unknown. We have spent some space speculating on the doings of $^{60}$Ca, whose spectroscopy should be at the reach of the forthcoming upgrades or new radioactive-ion facilities world-wide, which may
 as well explore the shores of the fifth IoI nearby $^{78}$Ni. These studies represent a real challenge on the isotope production side, and call for the development of detectors or experimental techniques with increased sensitivity. In particular, the exploration of the key regions discussed above lead to experiments with statistics of only few counts, requiring optimum signal over background. The need for the highest sensitivity is a the heart of the physics with rare isotopes, as already expressed by C. D\'etraz in 1979 when commenting about spectroscopy measurements with exotic nuclei:
 (sic) {\it Il arrive que lorsqu'on a plus d'un \'ev\'enement par signataire de l'article, on consid\`ere qu'il s'agisse d'un travail avec une bonne statistique} (It occurs that when one reaches more than one event per co-author of the article, the work is considered of good statistics) \cite{detraz1979a}. 
 
\newpage
\noindent
{\bf\Large Glossary.}
\begin{itemize}
    \item AGATA. Advanced GAmma Tracking Array.
    \item B(MF). Beyond (Mean field).
    \item CC. Coupled Cluster.
    \item CD-SE. Configuration driven shell evolution (aka Type II shell evolution).
    \item DWBA. Distorted wave Born approximation.
    \item EDF. Energy density functional.
    \item ESPE. Effective single-particle energies.
    \item FAIR. Facility for Antiproton and Ion Research
    \item FRIB. Facility for Rare Isotope Beams
    \item GANIL. Grand Acc\'el\'erateur National d'Ions Lourds
    \item GCM. Generator coordinate method.
    \item GRETA. Gamma Ray Energy Tracking Array.
     \item GRETINA. Gamma-Ray Energy Tracking In-beam Nuclear Array.
     \item GRIT. Granularity, Resolution, Identification and Transparency.
    \item HIAF. High Intensity heavy-ion Accelerator Facility.
    \item HO. Harmonic oscillator
    \item IoI.  Island of Inversion.
    \item IPM. Independent particle model.
    \item ISOL. Isotopic Separation On Line
    \item JUN45. Empirical effective interaction for the 1p$_{3/2}$, 1p$_{1/2}$, 0f$_{95/2}$, 0g$_{9/2}$, valence space.
    \item KB3(G). Monopole constrained realistic interactions for the $pf$-shell.
    \item LNPS-U. Monopole constrained realistic interaction for the $pf$-0g$_{9/2}$-1d$_{5/2}$ valence space.
    \item LSSM. Large-scale shell model.
    \item MBPT. Many body perturbation theory.
    \item MCSM.  Monte Carlo shell model.
    \item MINOS. MagIc Numbers Off Stability.
     \item NCSM. No core shell model.
     \item NNSP. Monopole  constrained realistic interaction for the $sdg$-$pfh$ shells.
    \item NSCL. National Superconducting Cyclotron Laboratory
    \item PES. Potential energy surface.
    \item PFSDG-U.  Monopole constrained realistic interaction for the $pf$-$sdg$ shells.
    \item QCD. Quantum chromodynamics.
     \item RIBF. Radioactive Ion Beam Factory
    \item RPA.  Random phase approximation.
    \item RRPA. Renormalized RPA.
    \item SM-CI.  Large scale shell model calculations, aka LSSM, ISM, or just SM.
    \item SCCM. Symmetry conserving configuration mixing.
    \item SDPF-U-MIX. Monopole constrained realistic interaction  for the $sd$-$pf$ shells.
    \item SPQR. Single orbit, Pseudo-SU3, Quasi-Su3 realization of the quadrupole-quadrupole interaction.
    \item USD(A-B). Empirical effective interactions for the $sd$-shell.
    \item VS-IMSRG. Valence space - in medium similarity renormalization group.
\end{itemize}

\vspace{2cm}
\noindent
{\bf\Large Acknowledgements.}

\medskip
\noindent
The SM-CI calculations pertaining to the Strasbourg-Madrid Shell Model collaboration
discussed in this manuscript have been performed with the ANTOINE and NATHAN shell-model codes, developed by
Etienne Caurier, a great physicist and friend which passed away in May 2020.  We dedicate this work to his memory.  

\medskip
\noindent
The results presented in this review stem from long lasting collaborative work and endless discussions with many colleagues, among them, and prominently, E. Caurier,  A. P. Zuker, S. M. Lenzi, and for the more experimental parts  H. Grawe($\dagger$2020) and  O. Sorlin.
The authors thank warmly M. Huyse for providing information on the history of the ISOL technique and RI production. AP's work is supported in part by the Ministerio de Ciencia, Innovaci\'on y Universidades (Spain), Severo Ochoa Programme SEV-2016-0597 and grant PGC-2018-94583. AO acknowledges the support by the Alexander von Humboldt foundation and by the Deutsche Forschungsgemeinschaft (DFG, German Research Foundation) - Project-ID 279384907 - SFB 1245.


\end{document}